\documentclass[12pt]{article}

\usepackage{arxiv}

\usepackage[utf8]{inputenc} % allow utf-8 input
\usepackage[T1]{fontenc}    % use 8-bit T1 fonts
\usepackage{booktabs}       % professional-quality tables
\usepackage{graphicx}
\usepackage{flushend}
\usepackage[nottoc]{tocbibind}
\usepackage[numbers,sort&compress]{natbib}

\usepackage[colorlinks,citecolor=blue,urlcolor=blue,linkcolor=blue]{hyperref}
\usepackage{marginnote}

\usepackage{bm,amssymb,amsbsy}
\usepackage{amsmath}
\usepackage{caption}
\usepackage{upgreek}
\usepackage{multirow}
\usepackage{enumitem}

\usepackage{lineno,hyperref}
\usepackage{appendix}
\newcommand*{\doi}[1]{\href{http://doi.org/#1}{#1}}
\newcommand{\vect}[1]{\boldsymbol{#1}}
\newcommand{\matr}[1]{\mathbf{#1}}
\newcommand{\matrgk}[1]{\mathsf{#1}}
\newcommand{\stoptocwriting}{\addtocontents{toc}{\protect\setcounter{tocdepth}{-5}}}
\newcommand{\resumetocwriting}{\addtocontents{toc}{\protect\setcounter{tocdepth}{\arabic{tocdepth}}}}
\newcommand{\appendixnumberline}[1]{Appendix\space}

\usepackage{tikz,xcolor,hyperref}

\definecolor{lime}{HTML}{A6CE39}
\DeclareRobustCommand{\orcidicon}{
        \begin{tikzpicture}
        \draw[lime, fill=lime] (0,0)
        circle [radius=0.16]
        node[white] {{\fontfamily{qag}\selectfont \tiny ID}};
        \draw[white, fill=white] (-0.0625,0.095)
        circle [radius=0.007];
        \end{tikzpicture}
        \hspace{-2mm}
}
\foreach \x in {A, ..., Z}{\expandafter\xdef\csname orcid\x\endcsname{\noexpand\href{https://orcid.org/\csname orcidauthor\x\endcsname}
                        {\noexpand\orcidicon}}
}
\title{\bf{Alignment of the Alpha Magnetic Spectrometer (AMS) in space}}

\usepackage{authblk}

\author[ \thanks{\href{mailto:qyan@cern.ch}{qyan@cern.ch}}]{Qi Yan}
\author[ ]{Vitaly Choutko}

\affil[ ]{Massachusetts Institute of Technology (MIT), Cambridge, Massachusetts 02139, USA}

\begin{document}
\maketitle

\makeatletter
\def\old@comma{,}
\catcode`\,=13
\def,{%
  \ifmmode%
    \old@comma\discretionary{}{}{}%
  \else%
    \old@comma%
  \fi%
}

\let\oldappendix\appendix
\makeatletter
\renewcommand{\appendix}{%
  \addtocontents{toc}{\let\protect\numberline\protect\appendixnumberline}%
  \renewcommand{\@seccntformat}[1]{Appendix~\csname the##1\endcsname\quad}%
  \oldappendix
}

\makeatother

\begin{abstract}
The Alpha Magnetic Spectrometer~(AMS) is a precision particle physics detector operating at an altitude of $\sim$410 km aboard the International Space Station. The AMS silicon tracker, together with the permanent magnet, measures the rigidity (momentum/charge) of cosmic rays in the range from  $\sim$0.5~GV to several TV.
In order to have accurate rigidity measurements, the positions of more than 2000 tracker modules have to be determined at the micron level by an alignment procedure. The tracker was first aligned using the 400 GeV/c proton test beam at CERN and then re-aligned using cosmic-ray events after being launched into space.
A unique method to align the permanent magnetic spectrometer for a space experiment is presented. The developed underlying mathematical algorithm is discussed in detail.
\end{abstract}


\vskip 1.2in
\begin{center}
\it{Submitted to The European Physical Journal C}
%{\scshape Submitted to The European Physical Journal C}
\end{center}
\newpage
\setcounter{tocdepth}{3}
\tableofcontents

%\linenumbers
\section{Introduction}
The Alpha Magnetic Spectrometer (AMS), operating aboard the International Space Station (ISS) since May 2011, is a unique large acceptance magnetic spectrometer in space. It aims to measure energy spectra of cosmic-ray charged particles, nuclei, antiparticles, antinuclei, and gamma-rays in the GeV-TeV region to understand Dark Matter, antimatter, and the origin of cosmic rays, as well as to explore new physics phenomena.
The AMS silicon tracker detector, together with the permanent magnet, determines the rigidity (momentum/charge) of charged cosmic rays by multiple measurements of the coordinates along the particle trajectory.
High performance of the tracker is crucial for the AMS mission and requires a sophisticated alignment to accurately determine the positions of the detector modules.

In August 2010, before AMS was launched, the complete detector 
was tested with a 400 GeV/c proton beam at the CERN Super Proton Synchrotron (SPS).
This data allows the precise alignment of the tracker with micron accuracy using the procedure described in this paper,
which aligns all the detector modules from different mechanical hierarchy levels in one step.
The strong accelerations and vibrations during launch, followed by the rapid outgassing of the support structure in vacuum, together with continuous temperature variations in space all change the positions of the tracker modules. The tracker is continuously re-aligned with cosmic-ray events to correct the resulting displacements. The unprecedented challenge in the alignment of the magnetic spectrometer in space is that the detector has to be aligned by using cosmic-ray events with unknown rigidities in the presence of the magnetic field. In this paper,
we report a unique mathematical approach which allows to overcome these difficulties and align the tracker to micron precision.

\section{AMS detector and the silicon tracker}
\begin{figure}[htb]
  \centering
  \includegraphics[width=0.65\textwidth]{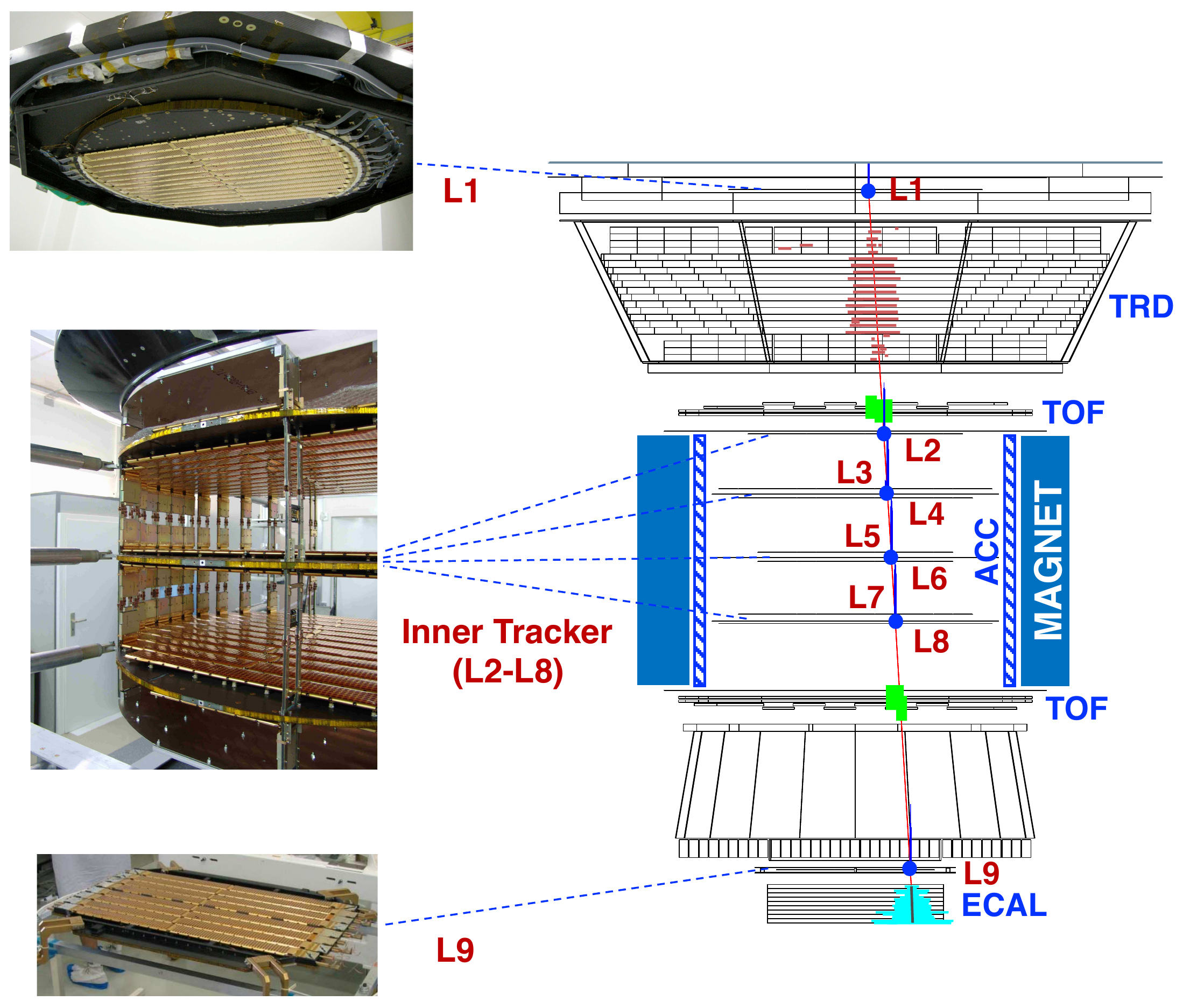}
  \caption{(right) Schematic view of a cosmic-ray fluorine nuclei event of 26~GV rigidity measured by AMS, with the signals in the TRD, TOF, silicon tracker, RICH, and ECAL. Also shown are the permanent magnet and ACC. (left) Layout of the tracker showing the 
upper external layer (L1), the inner tracker (L2-L8), and the lower external layer (L9) as well as their support planes.}
  \label{amsdet_fig}
\end{figure}
As shown in Fig.~\ref{amsdet_fig}, the AMS detector consists of a permanent magnet and an array of particle detectors to measure the velocity $\beta=v/c$, absolute charge $Q$, energy $E$, and rigidity $R$ of the passing particles.
Within the magnet bore and above and below the magnet are a total of 9 precision silicon tracker layers, L1 to L9. The tracker accurately measures $R$ and $Q$ of the particles. Above and below the magnet bore are the Upper and Lower Time of Flight (TOF) counters~\cite{TOF2014}. The TOF provides a charged particle trigger to AMS and determines $\beta$ and $Q$ of the incoming particles.
The Transition Radiation Detector (TRD)~\cite{TRD2013}, located above the Upper Time of Flight counters, identifies electrons and positrons.
The Ring Imaging Cherenkov detector (RICH)~\cite{RICH2010}, below the Lower Time of Flight counters, measures $\beta$ and $Q$ of passing particles.
The Electromagnetic Calorimeter (ECAL)~\cite{ECAL2013}, at the bottom of AMS,  measures $E$ of electromagnetic particles and separates protons from electrons and positrons.
The Anti-Coincidence Counters (ACC)~\cite{ANTI2009}, surrounding the inner tracker inside the magnet bore, reject cosmic rays entering AMS from the side.
The magnet~\cite{MAGNET} is made of 64 sectors of high-grade Nd-Fe-B assembled in a cylindrical shell.
The central field of the magnet is 1.4 kGauss. In 2010, the field was measured in 120 000 locations to an accuracy of better than 2 Gauss. Comparison with the measurements performed with the same magnet in 1997 shows that the field did not change within 1$\%$. On orbit, the magnet temperature varies from $-3$ to $+20^{\circ}$C. The field strength is corrected with a measured temperature dependence of $-0.09\%/^{\circ}$C~\cite{PROTONAMS2015}.

\begin{figure}[htb]
  \centering
  \includegraphics[width=0.45\textwidth]{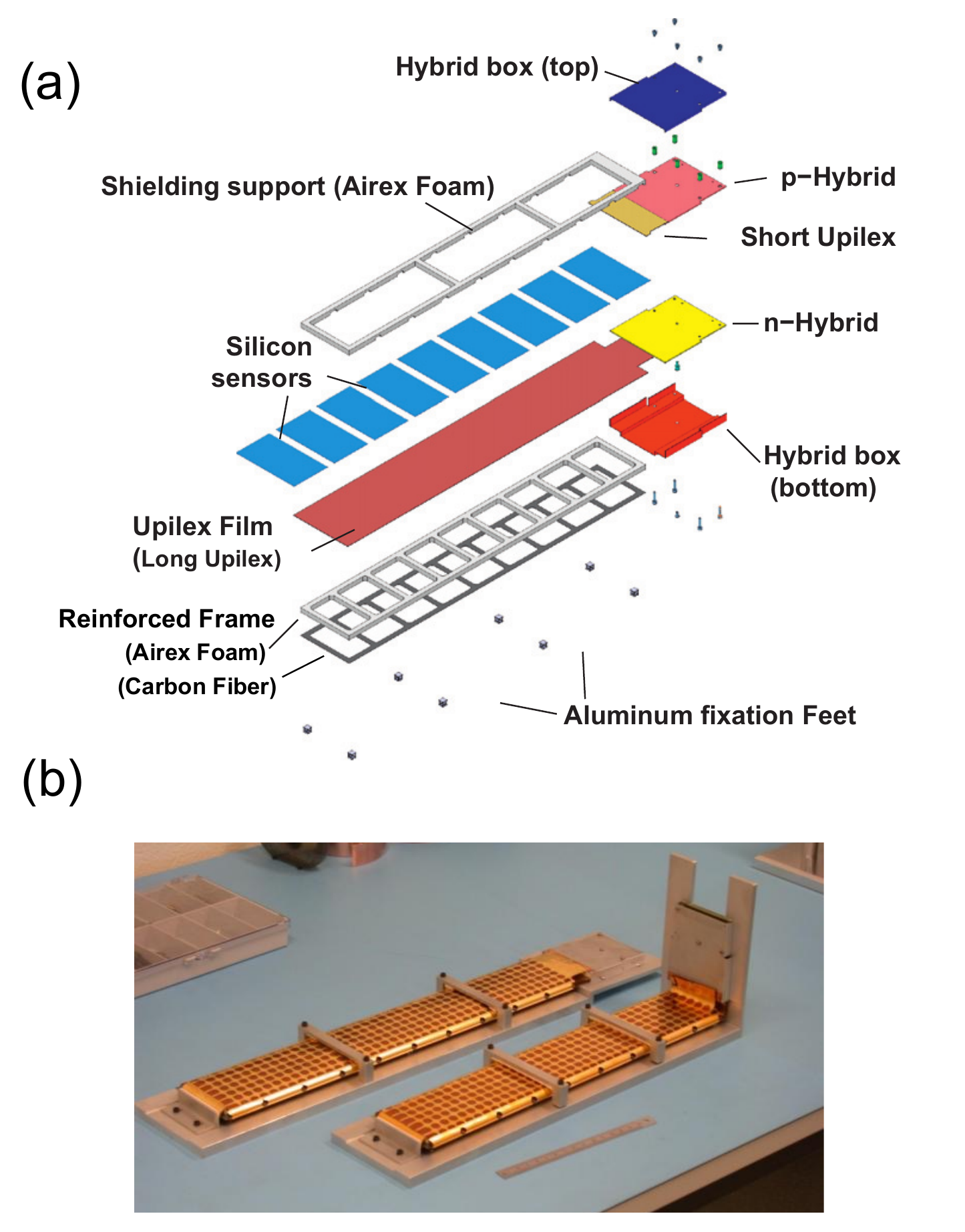}
  \caption{The AMS silicon tracker ladder: (a) the main components of the ladder and (b) two assembled ladders.}
  \label{amsladder_fig}
 \end{figure}
The AMS tracker comprises 2284 double-sided silicon micro-strip sensors each with a surface area of 41.360~$\times$~72.045 (active area of 39.832~$\times$~70.565)~mm$^{2}$ and thickness of 0.300~mm, 
assembled in 192 mechanical and electrical units called ladders~\cite{TRACKER2003}. 
Each ladder contains 9 to 15 sensors, see Fig.~\ref{amsladder_fig} (a).
The total active area is 6.42~m$^{2}$.
Both sides of a sensor are implanted with metallic strips running in orthogonal directions, 
providing a two-dimensional measurement of the particle position.
For the side with $p+$ doped strips ($p$-side), the implantation~(readout) strip pitch is 27.5~(110)~$\mathrm{\upmu m}$.
The opposite side ($n$-side) with $n+$ strips has an implantation~(readout) pitch of 104~(208)~$\mathrm{\upmu m}$.
The $p$-side ($n$-side) strips provide the measurement of the particle bending (non-bending) coordinate $y$ ($x$).
Combining the information from all signal strips in a sensor, the coordinate resolution in $y$ is $\sim$10~$\mathrm{\upmu m}$ for $Q=1$ and $\sim$5~$\mathrm{\upmu m}$ for $Q=6$ particles~\cite{TRACK2017}.
Sensors within a ladder are daisy-chained together through wire bonds on the $p$-side and are connected by a metalized Upilex film on the $n$-side which is then glued to a ladder reinforcement frame with layers of foam and carbon fiber (see Fig.~\ref{amsladder_fig} (a)). 

From 16 to 26 ladders are mounted onto one side of a support plane to form a layer.
As seen in Fig~\ref{amsdet_fig}, the tracker has 9 layers supported by 6 rigid planes.  Each plane is made of an aluminum honeycomb interior and carbon fiber skins.
The first layer (L1) is on plane 1 at the top of the detector,
the second (L2) is on plane 2 just above the magnet,
six (L3 to L8) are on 2 sides of planes 3, 4, and 5 within the bore of the magnet,
and the last (L9) is on plane 6 just above the ECAL.
The maximum lever arm from L1 to L9 is about 3~m.
L2 to L8 constitute the inner tracker.

The planes of the inner tracker are firmly held by a cylindrical carbon fiber structure which has near zero coefficient of thermal expansion and excellent mechanical strength~\cite{TRACKER2003}.
The material thickness of a plane, including 2 layers of ladders, represents $\sim$1$\%$ of a radiation length ($X_{0}$).
External plane 1 carrying L1 is bolted to another support sandwich plane (plane 1 NS) fastened to the top cover of the TRD.
External plane 6 carrying L9 is attached to the ECAL fixation blocks~\cite{MAGNET}. 
The deformation of the support structures of the TRD (M-Structure)~\cite{TRD2002} and ECAL (Unique Support Structure)~\cite{ECAL2007} due to gravity change or temperature variation (more than $\pm$10$^{\circ}$C in space) induce sizable displacements of L1 and L9 with respect to the position of the inner tracker.
The material thickness between L1 and L2, mostly the TRD and Upper TOF, is $\sim$0.3~$X_{0}$, and that between L8 and L9, mostly the Lower TOF and RICH, is $\sim$0.2~$X_{0}$~\cite{AMSNUCLEARXS}.

\section{Coordinate systems and composite alignment parameters}\label{compositealign}

\begin{figure}[htb]
  \centering 
  \includegraphics[width=0.65\textwidth]{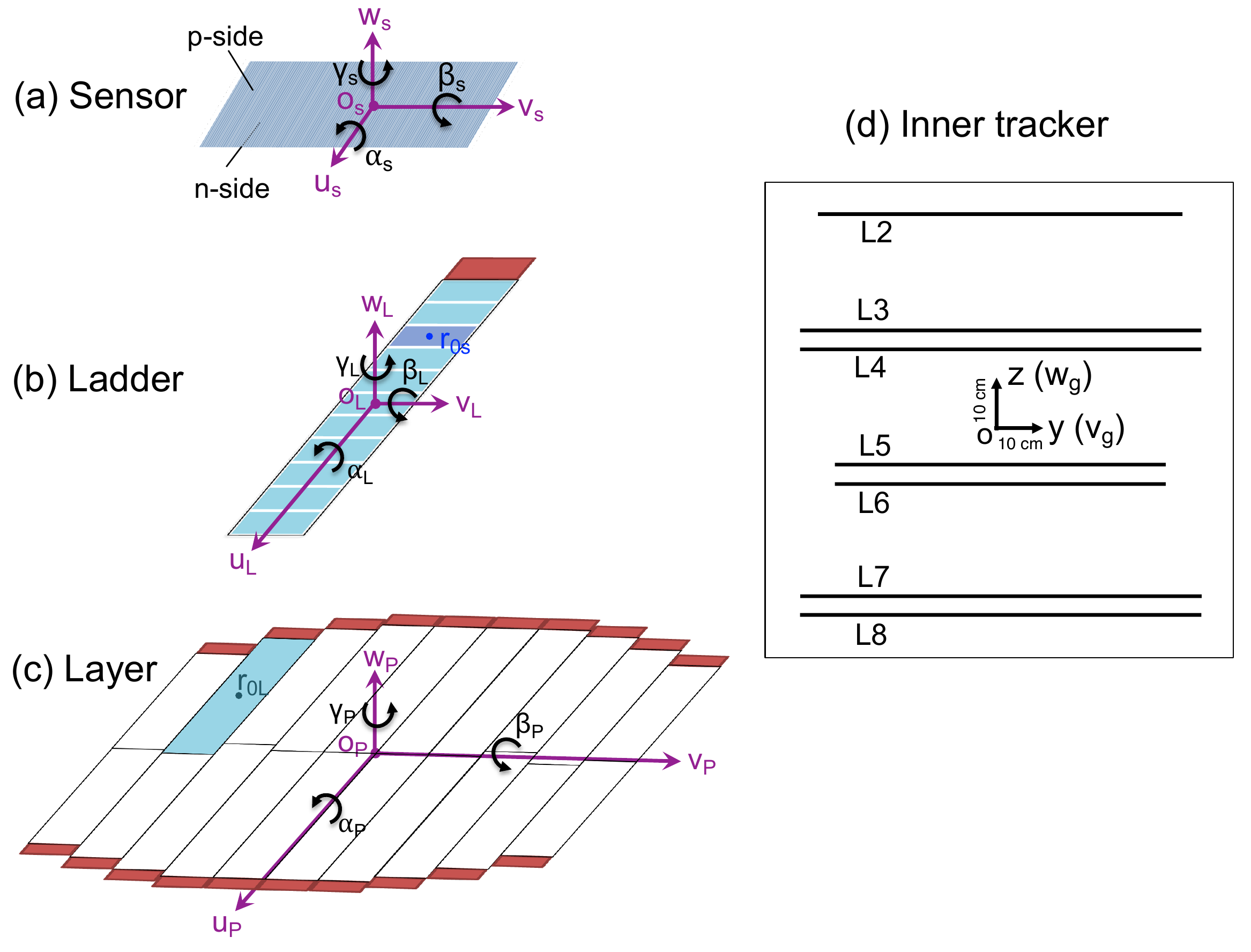}
  \caption{The components and coordinate systems of (a) a sensor, (b) a ladder, (c) a layer, and (d) the inner tracker.
The inner tracker coordinate system is also the global coordinate system.}
  \label{tkcoosystem}
\end{figure}
The AMS tracker modules (sensors, ladders, and layers) are assembled in a hierarchical support structure --- sensors in ladders, ladders on layers, and layers on planes into the tracker. Each module is positioned with respect to the next support structure by 6 degrees of freedom: 3 translations and 3 rotation angles.
Figure~\ref{tkcoosystem} (a) (b) (c) illustrates the local coordinate systems of a sensor, a ladder, and a layer where the geometric center of each module is defined as its origin point ($\vect{o}_{s}$, $\vect{o}_{L}$, or $\vect{o}_{P}$).
Taking the sensor coordinate system as an example, as shown in Fig.~\ref{tkcoosystem} (a),  the $u_{s}$-axis and the $v_{s}$-axis are defined along the coordinates measured by the strips of the $n$-side and the $p$-side respectively and the $w_{s}$-axis is normal to the sensor plane. 
Fig.~\ref{tkcoosystem} (d) shows the global coordinate system of the tracker where the geometric center of the inner tracker layers (L2-L8) is defined as its origin point ($\vect{o}_{g}$ or $\vect{o}$),
the $x$~($u_g$)-axis is along the coordinates measured by $n$-side strips parallel to the main component of the magnetic field, the $z$~($w_g$)-axis is pointing vertically perpendicular to the tracker layers, and the $y$~($v_g$)-axis completes to a right-handed orthogonal coordinate system.

In composite alignment, all detector modules from different hierarchy levels are aligned simultaneously. This approach was previously used in the CMS experiment~\cite{stoyecalibration}. In this section and section \ref{detectorcons}, we will introduce mathematical formulae for composite alignment. Specifically, section \ref{detectorcons} will address the implementation of constraints in composite alignment using our original numerical grid method.

\subsection{Coordinate transformation and alignment parameters}
The coordinates of the detector hit measured in the local sensor frame $\vect{q}=(u_{s},v_{s},w_{s})^{\mathsf{T}}$ can be transformed subsequently to the coordinates in the next reference frame, namely, in the ladder frame ($\vect{r}_L$), in the layer frame ($\vect{r}_P$), and in the global tracker frame ($\vect{r}_g$), as:
\begin{align}
\vect{r}_{L}&=\matr{R}^{\mathsf{T}}_{s}{\Delta}\matr{R}_{s}(\vect{q}+{\Delta}\vect{q}_{s})+\vect{r}_{0s} \label{f:frl} \\
\vect{r}_{P}&=\matr{R}^{\mathsf{T}}_{L}{\Delta}\matr{R}_{L}(\vect{r}_{L}+{\Delta}\vect{q}_{L})+\vect{r}_{0L} \label{f:frp} \\
\vect{r}_{g}&=\matr{R}^{\mathsf{T}}_{P}{\Delta}\matr{R}_{P}(\vect{r}_{P}+{\Delta}\vect{q}_{P})+\vect{r}_{0P} \label{f:frg}
\end{align}%
where $\vect{q}+{\Delta}\vect{q}_{s}$, $\vect{r}_{L}+{\Delta}\vect{q}_{L}$, and $\vect{r}_{P}+{\Delta}\vect{q}_{P}$ are the hit coordinates in the frames of the sensor, ladder, and layer respectively including small corrections on their individual position shifts of ${\Delta}\vect{q}_{s}$, ${\Delta}\vect{q}_{L}$, and ${\Delta}\vect{q}_{P}$;
$\matr{R}^{\mathsf{T}}_{s}$, $\matr{R}^{\mathsf{T}}_{L}$, and $\matr{R}^{\mathsf{T}}_{P}$ are the nominal rotation matrices from the sensor into the ladder, from the ladder into the layer, and from the layer into the tracker respectively and ${\Delta}\matr{R}_{s}$, ${\Delta}\matr{R}_{L}$, and ${\Delta}\matr{R}_{P}$ are their small individual corrections;
and $\vect{r}_{0s}$, $\vect{r}_{0L}$, and $\vect{r}_{0P}$ are the nominal positions of the sensor, ladder, and layer origin points in the next frame of the ladder, layer, and tracker respectively.
The corrections of each module displacement by an offset ${\Delta}\vect{q}_{i}=({\Delta}u,{\Delta}v,{\Delta}w)^{\mathsf{T}}$ and a rotation ${\Delta}\matr{R}_{i}={\Delta}\matr{R}_{i}^{\gamma}{\Delta}\matr{R}_{i}^{\beta}{\Delta}\matr{R}_{i}^{\alpha}$ have to be determined from the alignment procedure,
where ${\Delta}\matr{R}_{i}^{\alpha}$, ${\Delta}\matr{R}_{i}^{\beta}$, and ${\Delta}\matr{R}_{i}^{\gamma}$ are the decomposed rotation matrices defined by angles of rotation ${\alpha}$, ${\beta}$, and ${\gamma}$ around the $u$-axis, the new $v$-axis, and the new $w$-axis (Fig.~\ref{tkcoosystem}):
{\begin{equation*}
{\Delta}\matr{R}_{i}^{\alpha}= 
\begin{pmatrix}
1 &            0 & 0 \\
0 &  \mathrm{cos}{\alpha} & \mathrm{sin}{\alpha} \\
0 & -\mathrm{sin}{\alpha} & \mathrm{cos}{\alpha}
\end{pmatrix}
\
{\Delta}\matr{R}_{i}^{\beta}= 
\begin{pmatrix}
 \mathrm{cos}{\beta} &            0 & -\mathrm{sin}{\beta} \\
0                    &            1 & 0\\
\mathrm{sin}{\beta}  &            0 & \mathrm{cos}{\beta}
\end{pmatrix}
\end{equation*}}

{\begin{equation}
{\Delta}\matr{R}_{i}^{\gamma}= 
\begin{pmatrix}
\mathrm{cos}{\gamma} & \mathrm{sin}{\gamma} & 0 \\
-\mathrm{sin}{\gamma}& \mathrm{cos}{\gamma} & 0\\
0 &            0 & 1
\end{pmatrix}
\label{f:rotations}
\end{equation}}%
In the small-angle approximation, the correction matrix for rotation becomes:
{\begin{equation}
{\Delta}\matr{R}_{i}={\Delta}\matr{R}_{i}^{\gamma}{\Delta}\matr{R}_{i}^{\beta}{\Delta}\matr{R}_{i}^{\alpha}=
\begin{pmatrix}
1         & {\gamma} & -{\beta} \\
-{\gamma} & 1        & {\alpha} \\
{\beta}   &-{\alpha} & 1 
\end{pmatrix}
\label{f:rotation}
\end{equation}}

The transformation of a hit coordinate from the local sensor frame, $\vect{q}$, to the global tracker frame, $\vect{r}_{g}$, is given by: 
{{\begin{align}
\vect{r}_{g}{\simeq}&\matr{R}^\mathsf{T}(\vect{q}+\Delta\vect{q})+\vect{r}_{0} \label{f:frg3}
\end{align}}}%
where $\Delta\vect{q}$ is the total equivalent displacement correction in the local sensor frame including alignment parameters for all composite detector structures of the sensor, ladder, and layer,
$\matr{R}^\mathsf{T}$ is the nominal rotation matrix from the sensor into the global tracker frame,
and $\vect{r}_{0}$ is the nominal position of the sensor origin point in the global tracker frame. The definitions of $\Delta\vect{q}$, $\matr{R}^\mathsf{T}$, and $\vect{r}_{0}$ and the detailed calculation can be found in Appendix \ref{appA}.

\subsection{Coordinate measurement residual and its derivatives with respect to the alignment parameters}
The coordinate measurement (hit) residual $\vect{\varepsilon}$ is defined as the spatial difference between the predicted position of the track $\vect{q}_{p}$ and the measured position of the detector hit $\vect{q}$ in the sensor plane (local sensor frame), as:
\begin{equation}
\vect{\varepsilon}=\vect{q}_{p}-\vect{q}
\end{equation}%
The predicted position of the track in the sensor plane, $\vect{q}_{p}$,
is only sensitive to the sensor displacement, ${\Delta}\vect{q}$, along the ${w}_{s}$-axis (Fig.~\ref{tkcoosystem} (a)).
Such displacement will introduce a change of the track intersection position ${\Delta}\vect{q}_{p}$ as:
{\begin{equation}
{\Delta}\vect{q}_{p}=\matr{P}_{p}{\Delta}\vect{q}
\end{equation}}%
where
\begin{equation}
\matr{P}_{p}= 
\begin{pmatrix}
0 & 0 & \frac{du_{s}^{p}}{dw_{s}^{p}} \\
0 & 0 & \frac{dv_{s}^{p}}{dw_{s}^{p}} \\
0 & 0 & 1
\end{pmatrix}
\end{equation}%
The quantities ${du_{s}^{p}}/{dw_{s}^{p}}$ and ${dv_{s}^{p}}/{dw_{s}^{p}}$ are the track projected directions in the sensor $u_{s}w_{s}$-plane and $v_{s}w_{s}$-plane respectively. Hence, the total correction to the residual for the detector module displacements is: 
\begin{equation}
\Delta\vect{\varepsilon}=\Delta\vect{q}_{p}-\Delta\vect{q}=\matr{P}\Delta\vect{q} \label{f:residual}
\end{equation}%
where
\begin{equation}
\matr{P}=\matr{P}_{p}-\matr{E}= 
\begin{pmatrix}
-1 & 0 & \frac{du_{s}^{p}}{dw_{s}^{p}} \\
0 & -1 & \frac{dv_{s}^{p}}{dw_{s}^{p}} \\
0 & 0 & 0
\end{pmatrix}
\end{equation} 
and $\matr{E}$ is the unit matrix.

From Eq.(\ref{f:dq}) and Eq.(\ref{f:residual}), all the partial derivatives of the residual with respect to the alignment parameters can be calculated. Some examples are listed as follows:
{\begin{equation}
\begin{split}
\frac{\partial \vect{\varepsilon}}{\partial u_{s}}&=\matr{P}\vect{e}_{1} \\
\frac{\partial \vect{\varepsilon}}{\partial u_{L}}&=\matr{P}\matr{R}_{s}\vect{e}_{1} \\
\frac{\partial \vect{\varepsilon}}{\partial u_{P}}&=\matr{P}\matr{R}_{s}\matr{R}_{L}\vect{e}_{1} \\
\frac{\partial \vect{\varepsilon}}{\partial {\alpha}_{s}}&=\matr{P}\frac{\partial \Delta\matr{R}_{s}}{\partial {\alpha}_{s}}\vect{q} \\
\frac{\partial \vect{\varepsilon}}{\partial {\alpha}_{L}}&=\matr{P}\matr{R}_{s}\frac{\partial \Delta\matr{R}_{L}}{\partial {\alpha}_{L}}(\matr{R}_{s}^\mathsf{T}\vect{q}+\vect{r}_{0s})=\matr{P}\matr{R}_{s}\frac{\partial \Delta\matr{R}_{L}}{\partial {\alpha}_{L}}\hat{\vect{r}}_{L} \\
\frac{\partial \vect{\varepsilon}}{\partial {\alpha}_{P}}&=\matr{P}\matr{R}_{s}\matr{R}_{L}\frac{\partial \Delta\matr{R}_{P}}{\partial {\alpha}_{P}}\bigl[\matr{R}_{L}^\mathsf{T}(\matr{R}_{s}^\mathsf{T}\vect{q}+\vect{r}_{0s})+\vect{r}_{0L}\bigr] \\
   &=\matr{P}\matr{R}_{s}\matr{R}_{L}\frac{\partial \Delta\matr{R}_{P}}{\partial {\alpha}_{P}}\hat{\vect{r}}_{P}
\end{split}
\end{equation}}%
where $\vect{e}_{1}=(1,0,0)^\mathsf{T}$ is the unit vector of the $u$-axis, and $\hat{\vect{r}}_{L}=(\hat{u}_{L},\hat{v}_{L},\hat{w}_{L})^\mathsf{T}=\matr{R}_{s}^\mathsf{T}\vect{q}+\vect{r}_{0s}$ and $\hat{\vect{r}}_{P}=(\hat{u}_{P},\hat{v}_{P},\hat{w}_{P})^\mathsf{T}=\matr{R}_{L}^\mathsf{T}(\matr{R}_{s}^\mathsf{T}\vect{q}+\vect{r}_{0s})+\vect{r}_{0L}$ are the hit coordinates in the frames of the ladder and layer respectively without displacement.
Substituting the rotation (Eq.(\ref{f:rotation})) derivatives, the partial derivatives of the residual with respect to the alignment parameters of the sensor (${\partial \vect{\varepsilon}}/{\partial \vect{p}_{s}}$), ladder (${\partial \vect{\varepsilon}}/{\partial \vect{p}_{L}}$), and layer (${\partial \vect{\varepsilon}}/{\vect{p}_{P}}$) are obtained as:
{\begin{align}
\frac{\partial \vect{\varepsilon}}{\partial \vect{p}_{s}}&=\left(\frac{\partial \vect{\varepsilon}}{\partial u_{s}}, \frac{\partial \vect{\varepsilon}}{\partial v_{s}}, \frac{\partial \vect{\varepsilon}}{\partial w_{s}},
\frac{\partial \vect{\varepsilon}}{\partial {\alpha}_{s}},\frac{\partial \vect{\varepsilon}}{\partial {\beta}_{s}},\frac{\partial \vect{\varepsilon}}{\partial {\gamma}_{s}}\right)=\matr{P}\frac{\partial \vect{q}}{\partial \vect{p}_{s}} \nonumber\\
&=\matr{P}
\begin{pmatrix}
1 & 0 & 0 & 0      &-w_{s}=0& v_{s} \\
0 & 1 & 0 & w_{s}=0&       0&-u_{s} \\
0 & 0 & 1 &-v_{s}  &   u_{s}&    0
\end{pmatrix} \label{f:resps}\\
\frac{\partial \vect{\varepsilon}}{\partial \vect{p}_{L}}&=\left(\frac{\partial \vect{\varepsilon}}{\partial u_{L}}, \frac{\partial \vect{\varepsilon}}{\partial v_{L}}, \frac{\partial \vect{\varepsilon}}{\partial w_{L}},
\frac{\partial \vect{\varepsilon}}{\partial {\alpha}_{L}},\frac{\partial \vect{\varepsilon}}{\partial {\beta}_{L}},\frac{\partial \vect{\varepsilon}}{\partial {\gamma}_{L}}\right)=\matr{P}\frac{\partial \vect{q}}{\partial \vect{p}_{L}} \nonumber\\
&=\matr{P}\matr{R}_{s}
\begin{pmatrix}
1 & 0 & 0 & 0          &-\hat{w}_{L}& \hat{v}_{L} \\
0 & 1 & 0 & \hat{w}_{L}&           0&-\hat{u}_{L} \\
0 & 0 & 1 &-\hat{v}_{L}& \hat{u}_{L}&    0
\end{pmatrix} \label{f:respl}\\
\frac{\partial \vect{\varepsilon}}{\partial \vect{p}_{P}}&=\left(\frac{\partial \vect{\varepsilon}}{\partial u_{P}}, \frac{\partial \vect{\varepsilon}}{\partial v_{P}}, \frac{\partial \vect{\varepsilon}}{\partial w_{P}},
\frac{\partial \vect{\varepsilon}}{\partial {\alpha}_{P}},\frac{\partial \vect{\varepsilon}}{\partial {\beta}_{P}},\frac{\partial \vect{\varepsilon}}{\partial {\gamma}_{P}}\right)=\matr{P}\frac{\partial \vect{q}}{\partial \vect{p}_{P}} \nonumber\\
&=\matr{P}\matr{R}_{s}\matr{R}_{L}
\begin{pmatrix}
1 & 0 & 0 & 0          &-\hat{w}_{P}& \hat{v}_{P} \\
0 & 1 & 0 & \hat{w}_{P}&           0&-\hat{u}_{P} \\
0 & 0 & 1 &-\hat{v}_{P}& \hat{u}_{P}&    0
\end{pmatrix} \label{f:respp}
\end{align}}%
The alignment parameters of the sensor, ladder, and layer are ${\Delta}\vect{p}_{s}=(\Delta{u}_{s},\Delta{v}_{s},\Delta{w}_{s},\alpha_{s},\beta_{s},\gamma_{s})^\mathsf{T}$, ${\Delta}\vect{p}_{L}=(\Delta{u}_{L},\Delta{v}_{L},\Delta{w}_{L},\alpha_{L},\beta_{L},\gamma_{L})^\mathsf{T}$, and ${\Delta}\vect{p}_{P}=(\Delta{u}_{P},\Delta{v}_{P},\Delta{w}_{P},\alpha_{P},\beta_{P},\gamma_{P})^\mathsf{T}$ correspondingly.

\section{Constraints of the composite alignment parameters}\label{detectorcons}
For a composite detector which consists of several subcomponents, those modules on the same support structure are likely to have highly correlated displacements.
Applying the alignment directly on a single level of the hierarchy such as the sensors ignores the mechanical correlations and distorts the detector structure.
In the composite alignment, the alignment parameters in each level are defined relative to the next support structure as shown in Eqs.(\ref{f:frl})(\ref{f:frp})(\ref{f:frg}) and all the alignment parameters for all the detector modules (sensors, ladders, and layers) are aligned simultaneously.
In this way, all correlations are considered and the alignment accuracy is optimized.

If all composite modules are aligned at the same time without constraints, there will be no unique solution. For example, all the sensors in a ladder can move in one direction and the ladder can move in the opposite direction, which results in no movement of any sensors. To avoid this, 6 degrees of freedom must be constrained for every group of subcomponents on the same support structure.
The expressions of all the constraints are derived by our developed grid method described in the following sections \ref{csensorladder}, \ref{cladderlayer}, and \ref{clayertracker}.
\begin{figure}[htpb]
  \centering
  \includegraphics[width=0.6\textwidth]{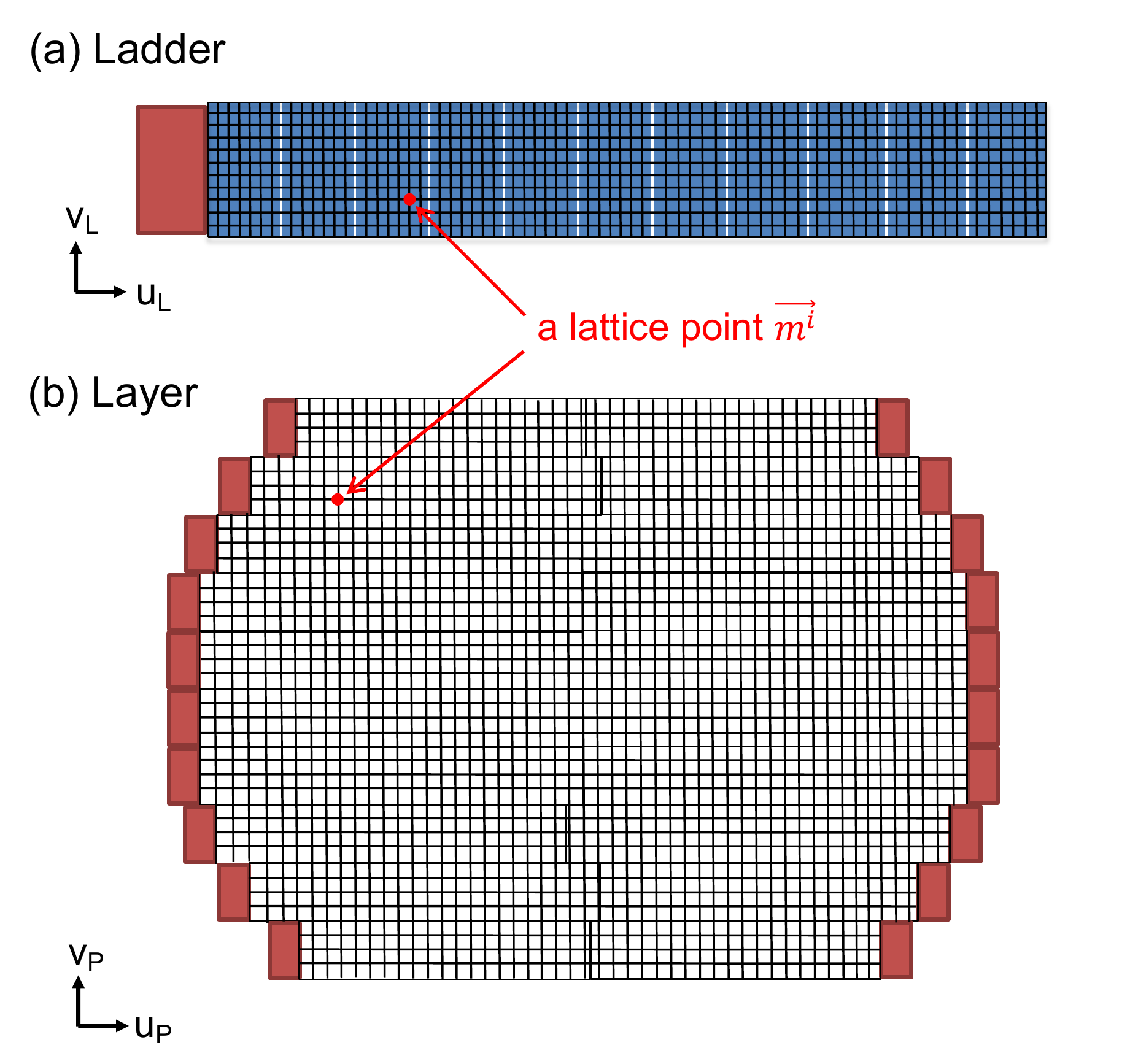}
  \caption{Schematics of (a) a ladder and (b) a layer divided into fine uniform grids.}
  \label{trackergrid}
\end{figure}

In addition, the stretching and shear deformations of the detector as subsets of the linear coordinate transformation will need specific constraints, as they are not sensed by the track alignment procedure.
In section \ref{cshearing}, we present our study to deal with this issue.

\subsection{Constraints of sensors in a ladder} \label{csensorladder}
Each sensor in a ladder can move individually. To investigate the displacements of the sensors with respect to the ladder, a ladder is divided into fine uniform grids spanning over all its sensors,  as illustrated in Fig.~\ref{trackergrid} (a). A lattice point $\vect{m}^{i}$ represents a movement in the $i$-th grid position induced by the displacement of the sensor.
The movement of the ladder ${\Delta}\vect{q}$ as a result of the displacements of all its sensors is estimated from all the lattices via ${\chi}^{2}$-minimization:
{{\begin{equation}
{\chi}^{2}=\sum_{i}|\vect{m}^{i}-{\Delta}\vect{q}|^{2}
\end{equation}}}%
The derivatives of the minimized ${\chi}^{2}$ with respect to the ladder movement parameters ${\Delta}\vect{p}_{L}$ are zero: 
{\begin{equation}
\frac{\partial {\chi}^{2} }{\partial \vect{p}_{L}}=\sum_{i}2\left(\frac{\partial \vect{q}}{\partial \vect{p}_{L}}\right)^\mathsf{T}_{i}(\vect{m}^{i}-\Delta\vect{q})=\vect{0} \label{f:dchisdpl}
\end{equation}}%
The displacements of sensors in a ladder are required to result in zero overall ladder displacement as ${\Delta}\vect{q}({\Delta}u_{L}^{s},{\Delta}v_{L}^{s},{\Delta}w_{L}^{s},\alpha_{L}^{s},\beta_{L}^{s},\gamma_{L}^{s})=\vect{0}$. 
Substituting ${\Delta}\vect{q}=\vect{0}$ and $\vect{m}^{i}=({\partial \vect{q}}/{\partial \vect{p}_{s}})_{i}{\Delta}\vect{p}_{s}^{i}$ (the first order approximation) into Eq.(\ref{f:dchisdpl}), 
6 constraints on the alignment parameters of sensors in a ladder are obtained by summing up all the lattice points, as:
{\begin{equation}
\sum_{i}\left(\frac{\partial \vect{q}}{\partial \vect{p}_{L}}\right)^\mathsf{T}_{i}\left(\frac{\partial \vect{q}}{\partial \vect{p}_{s}}\right)_{i}{\Delta}\vect{p}_{s}^{i}=\vect{0}
\label{f:senscons}
\end{equation}}%
where 
{\begin{equation}
\begin{split}
\left(\frac{\partial \vect{q}}{\partial \vect{p}_{L}}\right)^\mathsf{T}_{i}&=\left(\frac{\partial \vect{q}}{\partial u_{L}},\frac{\partial \vect{q}}{\partial v_{L}},\frac{\partial \vect{q}}{\partial w_{L}},\frac{\partial \vect{q}}{\partial \alpha_{L}},\frac{\partial \vect{q}}{\partial \beta_{L}},\frac{\partial \vect{q}}{\partial \gamma_{L}}\right)^\mathsf{T}_{i}\\
&=
\begin{pmatrix}
1 & 0 & 0 & 0     &-\hat{w}_{L}^{i}& \hat{v}_{L}^{i} \\
0 & 1 & 0 & \hat{w}_{L}^{i}&      0&-\hat{u}_{L}^{i} \\
0 & 0 & 1 &-\hat{v}_{L}^{i}& \hat{u}_{L}^{i}&    0
\end{pmatrix}
^\mathsf{T}\matr{R}_{s}^{i\mathsf{T}}\\
\end{split}
\end{equation}}%
is transposed from Eq.(\ref{f:respl}),
{\begin{equation}
\left(\frac{\partial \vect{q}}{\partial \vect{p}_{s}}\right)_{i}=
\begin{pmatrix}
1 & 0 & 0 & 0      &-w_{s}^{i}=0& v_{s}^{i} \\
0 & 1 & 0 & w_{s}^{i}=0&       0&-u_{s}^{i} \\
0 & 0 & 1 &-v_{s}^{i}  &   u_{s}^{i}&    0
\end{pmatrix}
\end{equation}}%
is from Eq.(\ref{f:resps}), and
{\begin{equation}
{\Delta}\vect{p}_{s}^{i}=(\Delta{u}_{s}^{i},\Delta{v}_{s}^{i},\Delta{w}_{s}^{i},\alpha_{s}^{i},\beta_{s}^{i},\gamma_{s}^{i})^\mathsf{T}
\end{equation}}

\subsection{Constraints of ladders in a layer} \label{cladderlayer}
Similarly, to study the displacements of the ladders with respect to the layer,  a layer is divided into fine uniform grids as illustrated in Fig.~\ref{trackergrid} (b). A lattice point $\vect{m}^{i}=({\partial \vect{q}}/{\partial \vect{p}_{L}})_{i}{\Delta}\vect{p}_{L}^{i}$ represents a movement in the $i$-th grid position induced by the displacement of the ladder. Using the same method, 6 constraints on the alignment parameters of ladders in a layer are derived as:
{{\begin{equation}
\sum_{i}\left(\frac{\partial \vect{q}}{\partial \vect{p}_{P}}\right)^\mathsf{T}_{i}\left(\frac{\partial \vect{q}}{\partial \vect{p}_{L}}\right)_{i}{\Delta}\vect{p}_{L}^{i}=\vect{0}
\label{f:laddercons}
\end{equation}}}%
where
{\begin{equation}
\begin{split}
\left(\frac{\partial \vect{q}}{\partial \vect{p}_{P}}\right)^\mathsf{T}_{i}&=\left(\frac{\partial \vect{q}}{\partial u_{P}},\frac{\partial \vect{q}}{\partial v_{P}},\frac{\partial \vect{q}}{\partial w_{P}},\frac{\partial \vect{q}}{\partial \alpha_{P}},\frac{\partial \vect{q}}{\partial \beta_{P}},\frac{\partial \vect{q}}{\partial \gamma_{P}}\right)^\mathsf{T}_{i}\\
&=
\begin{pmatrix}
1 & 0 & 0 & 0     &-\hat{w}_{P}^{i}& \hat{v}_{P}^{i} \\
0 & 1 & 0 & \hat{w}_{P}^{i}&      0&-\hat{u}_{P}^{i} \\
0 & 0 & 1 &-\hat{v}_{P}^{i}& \hat{u}_{P}^{i}&    0
\end{pmatrix}
^\mathsf{T}\matr{R}_{L}^{i\mathsf{T}}\matr{R}_{s}^{i\mathsf{T}}\\
\end{split}
\end{equation}}%
is transposed from Eq.(\ref{f:respp}),
{\begin{equation}
\left(\frac{\partial \vect{q}}{\partial \vect{p}_{L}}\right)_{i}=
\matr{R}_{s}^{i}
\begin{pmatrix}
1 & 0 & 0 & 0        &-\hat{w}_{L}^{i}     & \hat{v}_{L}^{i} \\
0 & 1 & 0 & \hat{w}_{L}^{i}&              0&-\hat{u}_{L}^{i} \\
0 & 0 & 1 &-\hat{v}_{L}^{i}&\hat{u}_{L}^{i}&    0
\end{pmatrix}
\end{equation}}%
is from Eq.(\ref{f:respl}), and
{\begin{equation}
{\Delta}\vect{p}_{L}^{i}=(\Delta{u}_{L}^{i},\Delta{v}_{L}^{i},\Delta{w}_{L}^{i},\alpha_{L}^{i},\beta_{L}^{i},\gamma_{L}^{i})^\mathsf{T}
\end{equation}}

\subsection{Constraints of layers in the tracker} \label{clayertracker}
The composite structure of layers in the tracker also has to be constrained to factor out the translations and rotations of the whole detector and to establish the basic position and orientation of AMS.
Considering mechanical and thermal stability, only the layers from the inner tracker (L2-L8), whose planes are firmly held by the carbon fiber cylinder, are used in the constraints.
All the inner tracker layers are divided into fine grids of equal size with each ($i$-th) lattice point representing the layer displacement at that position, see Fig.~\ref{trackergrid} (b).
By requiring the overall inner tracker to have neither translations nor rotations as ${\Delta}\vect{p}_{g}=(\Delta{x},\Delta{y},\Delta{z},\alpha,\beta,\gamma)^\mathsf{T}=\vect{0}$, 
the constraints on the alignment parameters of the inner tracker layers are obtained as:
{{\begin{equation}
\sum_{i}\left(\frac{\partial \vect{q}}{\partial \vect{p}_{g}}\right)^\mathsf{T}_{i}\left(\frac{\partial \vect{q}}{\partial \vect{p}_{P}}\right)_{i}{\Delta}\vect{p}_{P}^{i}=\vect{0}
\label{f:layercons}
\end{equation}}}%
where
{\begin{equation}
\left(\frac{\partial \vect{q}}{\partial \vect{p}_{g}}\right)^\mathsf{T}_{i}=
\begin{pmatrix}
1 & 0 & 0 & 0     &-\hat{z}^{i}& \hat{y}^{i} \\
0 & 1 & 0 & \hat{z}^{i}&      0&-\hat{x}^{i} \\
0 & 0 & 1 &-\hat{y}^{i}& \hat{x}^{i}&    0
\end{pmatrix}^\mathsf{T}
\matr{R}_{P}^{i\mathsf{T}}\matr{R}_{L}^{i\mathsf{T}}\matr{R}_{s}^{i\mathsf{T}}
\label{f:dqdpg}
\end{equation}}%
$\hat{\vect{r}}_{g}^{i}=(\hat{x}^{i},\hat{y}^{i},\hat{z}^{i})^\mathsf{T}$ is the $i$-th lattice point position in the global tracker frame without displacement,
{\begin{equation}
\left(\frac{\partial \vect{q}}{\partial \vect{p}_{P}}\right)_{i}=
\matr{R}_{s}^{i}\matr{R}_{L}^{i}
\begin{pmatrix}
1 & 0 & 0 & 0        &-\hat{w}_{P}^{i}     & \hat{v}_{P}^{i} \\
0 & 1 & 0 & \hat{w}_{P}^{i}&              0&-\hat{u}_{P}^{i} \\
0 & 0 & 1 &-\hat{v}_{P}^{i}&\hat{u}_{P}^{i}&    0
\end{pmatrix}
\label{f:dqdp}
\end{equation}}%
is from Eq.(\ref{f:respp}), and
{\begin{equation}
{\Delta}\vect{p}_{P}^{i}=(\Delta{u}_{P}^{i},\Delta{v}_{P}^{i},\Delta{w}_{P}^{i},\alpha_{P}^{i},\beta_{P}^{i},\gamma_{P}^{i})^\mathsf{T}
\end{equation}}

The grid density for calculation of Eq.(\ref{f:senscons}), Eq.(\ref{f:laddercons}), or Eq.(\ref{f:layercons}) is sufficiently large so that its contribution to the uncertainty of each constraint is negligible.

\subsection{Constraints of stretching and shear deformations}\label{cshearing}

\begin{figure}[htpb]
  \centering
  \includegraphics[width=0.65\textwidth]{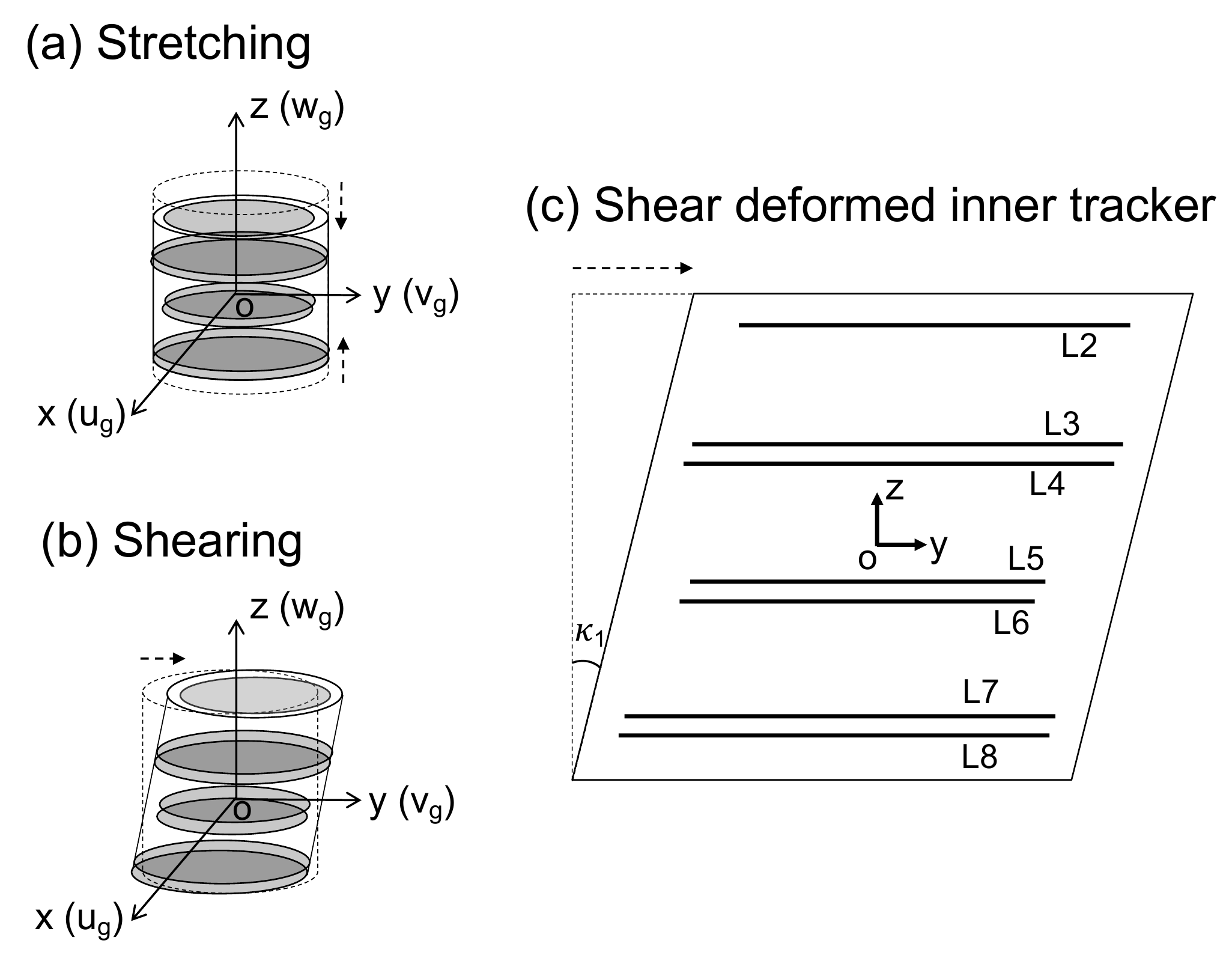}
  \caption{Schematics of the inner tracker deformations: (a) stretching, (b) shearing, and (c) shearing section view.}
  \label{sheartracker}
\end{figure}
The first alignment of the AMS tracker is based on the 400~GeV/c proton test beam, where the characteristics of tracks with given momenta in the magnetic field is equivalent to straight tracks. Any linear coordinate transformation will conserve the linearity of a straight track and hence not be sensed by the track alignment procedure. Conversely, without specific constraints, an unstable system of the alignment due to $\chi^{2}$ invariance could introduce this kind of transformation, manifested as an extra detector displacement or deformation.
A general linear transformation from a vector $\vect{r}=(x,y,z)^\mathsf{T}$ to a new vector $\vect{r}'=(x',y',z')^\mathsf{T}$ can be expressed by the matrix equation:
\begin{equation}
\vect{r}'=\matr{D}\vect{r}+\vect{d}
\end{equation}%
where $\matr{D}$ is a 3${\times}$3 matrix, called a transformation matrix, and $\vect{d}=(d_{1},{d}_2,d_{3})^\mathsf{T}$ is a vector representing a translation.
Clearly, in a linear transformation, there are a total of 12 free parameters (3 in $\vect{d}$ and 9 in $\matr{D}$), which can be categorized to describe the following decomposed transformations:
%\begin{enumerate}[label={[\alph*]}]
\begin{enumerate}[label=(\roman*)]
%\begin{enumerate}[label={[\roman*]}]
\item 3 translations represented by 3 elements in $\vect{d}$\label{transa}
\item 3 rotations whose matrix forms are shown in Eq.(\ref{f:rotations})\label{transb}
\item 3 stretchings with each leading to an expansion or shrinking of the object along the corresponding axis. As an example, Fig.~\ref{sheartracker} (a) shows the shrinking along the $z$-axis\label{transc} 
\item 3 shearings with each deforming the object shape on the corresponding projection plane as the one in Fig.~\ref{sheartracker} (b) shows the shearing on the $yz$-plane\label{transd}
\end{enumerate} 
The outcome of \ref{transa} translations and \ref{transb} rotations is a rigid-body displacement without changing the object shape or size.
The 3 translations and 3 rotations of the inner tracker ${\Delta}\vect{p}_{g}=(\Delta{x},\Delta{y},\Delta{z},\alpha,\beta,\gamma)^\mathsf{T}$ have already been constrained to be zero as previously discussed in section \ref{clayertracker}.
Next, we will focus on \ref{transc} stretchings and \ref{transd} shearings.

\subsubsection{Stretching}
The matrix of stretching $\matr{D}_{t}$ is diagonal:
{\begin{equation}
\matr{D}_{t}=
\begin{pmatrix}
\lambda_{1} & 0 & 0 \\
0 & \lambda_{2} & 0  \\
0 & 0 & \lambda_{3} 
\end{pmatrix}
\label{stretching}
\end{equation}}%
where $\lambda_{1}$, $\lambda_{2}$, and $\lambda_{3}$ are the extension-contraction coefficients along the $x$-, $y$-, and $z$-axes, respectively.

Stretching deformations are connected to the detector structure.
As shown in Fig.~\ref{tkcoosystem}, the silicon sensors through the ladder structure are tiled in the $xy$-plane to form all layers.
The stretching deformation in the $xy$-plane is constrained to some extent by the exactly known size of the sensors.
During exposure to the proton test beam or to cosmic rays in space, the incoming particles always enter the detector in various directions and positions. 
The distance between the neighboring sensors or ladders is well determined by many tracks, which are crossing them and the sensors from other layers in front and behind.
Therefore, during alignment, the extension-contraction coefficients $\lambda_{1}$ and $\lambda_{2}$ are naturally constrained by the intrinsic size of the sensors either themselves or the ones in front/behind and no external constrains are needed.

The extension-contraction coefficient along the $z$-axis, $\lambda_{3}$,  has no any sensor structure restriction (Fig.~\ref{sheartracker} (a)) and has to be defined.
According to Eq.(\ref{stretching}), the stretching length along the $z$-axis, ${\Delta}z$, is described as:
\begin{equation}
{\Delta}z=(\lambda_{3}-1)z=kz 
\end{equation}%
where no stretching deformation is $k\equiv(\lambda_{3}-1)=0$.
We can use the same grid method as in the previous section to derive the corresponding constraint on the alignment parameters.
As seen in Fig.~\ref{trackergrid} (b), the $i$-th lattice point of the inner tracker $m^{i}=\Delta{z}_{i}$  represents the $z$ position shift induced by the displacement of the layer at that position.
The stretching parameter $k$ is estimated from all the lattices via ${\chi}^{2}$-minimization, as:
{\begin{equation}
{\chi}^{2}=\sum_{i}(m^{i}-kz^{i})^{2}
\label{f:chisshear}
\end{equation}}%
where the derivative of ${\chi}^{2}$ with respect to $k$ is zero:
{\begin{equation}
\frac{\partial {\chi}^{2} }{\partial k}=\sum_{i}2z^{i}({m}^{i}-kz^{i})=0
\label{f:chisshear2}
\end{equation}}%
The constraint of $k=0$ leads to:
{\begin{equation}
\sum_{i}z^{i}{m}^{i}=\sum_{i}z^{i}{\Delta{z}}^{i}=0
\label{f:zshiftcons}
\end{equation}}%
$\Delta{z}^{i}$ in $\Delta\vect{p}_{g}^{i}=(\Delta{x}^{i},\Delta{y}^{i},\Delta{z}^{i},{\alpha}^{i},\beta,\gamma)^\mathsf{T}$ can be replaced by the layer alignment parameters of ${\Delta}\vect{p}_{P}^{i}=(\Delta{u}_{P}^{i},\Delta{v}_{P}^{i},\Delta{w}_{P}^{i},\alpha_{P}^{i},\beta_{P}^{i},\gamma_{P}^{i})^\mathsf{T}$ as:
{{\begin{equation}
\Delta\vect{p}_{g}^{i}=\left(\frac{\partial \vect{q}}{\partial \vect{p}_{g}}\right)^\mathsf{T}_{i}\left(\frac{\partial \vect{q}}{\partial \vect{p}_{P}}\right)_{i}{\Delta}\vect{p}_{P}^{i}
\label{f:dpg}
\end{equation}}}%
where $({\partial \vect{q}}/{\partial \vect{p}_{g}})_{i}^\mathsf{T}$ is from Eq.(\ref{f:dqdpg}) and $({\partial \vect{q}}/{\partial \vect{p}_{P}})_{i}$ is from Eq.(\ref{f:dqdp}). For the AMS inner tracker structure, the constraint of Eq.(\ref{f:zshiftcons}) can be simplified as:
\begin{equation}
\sum_{l=2}^{8}\matr{R}^{\mathsf{T}l}_{P}(3,3){\Delta}w_{P}^{l}z^{l}A^{l}=0
\label{f:zshiftcons2}
\end{equation}%
where $\matr{R}^{\mathsf{T}l}_{P}(3,3)$ is the (3,3) entry of the $l$-th layer rotation matrix,
${\Delta}w_{P}^{l}$ is the $l$-th layer alignment parameter on the translation along the $w_{P}$-axis,
and $z^{l}$ and $A^{l}$ are the $l$-th layer $z$ position and surface area respectively.

\subsubsection{Shearing}
Three individual matrices of pure shearing $\matr{D}_{h}^{\kappa1}$, $\matr{D}_{h}^{\kappa2}$, and $\matr{D}_{h}^{\kappa3}$ are given by:
{\begin{equation*}
\matr{D}_{h}^{\kappa1}=
\begin{pmatrix}
1 &  0 & 0 \\
0 &  1 & {\kappa}_{1}/2 \\
0 &  {\kappa}_{1}/2 & 1 
\end{pmatrix}
%\;
\quad
\matr{D}_{h}^{\kappa2}= 
\begin{pmatrix}
1 &  0 & {\kappa}_{2}/2 \\
0 &  1 & 0 \\
{\kappa}_{2}/2 &  0 & 1 
\end{pmatrix}
\end{equation*}}

{\begin{equation}
\matr{D}_{h}^{\kappa3}=
\begin{pmatrix}
1 &  {\kappa}_{3}/2 & 0 \\
{\kappa}_{3}/2 &  1 & 0 \\
0 &  0 & 1 
\end{pmatrix} \label{f:shearing}
\end{equation}}% 
where $\kappa_{1}$, $\kappa_{2}$, and $\kappa_{3}$ are the shear strains on the $yz$-, $xz$-, and $xy$-planes, respectively.
As seen, the pure shear matrices are symmetric in contrast with rotation matrices which are anti-symmetric as shown in Eq.(\ref{f:rotations}).

Using small angle and shear strain approximation, the product of matrices of shearing $\matr{D}_{h}^{\kappa1}$ and rotation ${\Delta}\matr{R}^{\alpha}$ is:
{\begin{equation}
\matr{D}_{h}^{\kappa1}{\Delta}\matr{R}^{\alpha}=
\begin{pmatrix}
1 &  0 & 0 \\
0 &  1 & {\kappa}_{1}/2+\alpha \\
0 &  {\kappa}_{1}/2-\alpha & 1
\end{pmatrix}
\label{f:rotateshear}
\end{equation}}%
When $\alpha={\kappa}_{1}/2$, $\matr{D}_{h}^{\kappa1}{\Delta}\matr{R}^{\alpha}$ becomes a simple shearing~\cite{SHEARPAPER} along the $y$-axis on the $yz$-plane as illustrated in Fig.~\ref{sheartracker} (c).
In the presence of both shearing and rotation in the $yz$-plane, the change of the object position along the $y$-axis, $\Delta{y}$, obtained from Eq.(\ref{f:rotateshear}) is:
{\begin{equation}
\Delta{y}=({\kappa}_{1}/2+\alpha)z=k_{1}z
\label{f:dyshearing} 
\end{equation}}%
where the requirment of the object to have neither rotation $\alpha=0$ nor shear deformation ${\kappa}_{1}=0$, defines $k_{1}=0$.
Repeating ${\chi}^{2}$ minimization to Eq.(\ref{f:dyshearing}) together with the constraint $k_{1}=0$ leads to:
{\begin{equation}
\sum_{i}z^{i}{\Delta{y}}^{i}=0
\label{f:yzshearcons}
\end{equation}}%
where ${\Delta{y}}^{i}$ can be replaced by the layer alignment parameters as shown in Eq.(\ref{f:dpg}).
For the AMS inner tracker structure, the corresponding constraint is simplified to be:
\begin{equation}
\sum_{l=2}^{8}\matr{R}^{\mathsf{T}l}_{P}(2,2){\Delta}v_{P}^{l}z^{l}A^{l}=0
\label{f:yzshearcons2}
\end{equation}%
where $\matr{R}^{\mathsf{T}l}_{P}(2,2)$ is the (2,2) entry of the $l$-th layer rotation matrix,
${\Delta}v_{P}^{l}$ is the $l$-th layer alignment parameter on the translation along the $v_{P}$-axis, and $z^{l}$ and $A^{l}$ are the $l$-th layer $z$ position and surface area respectively.
From Eq.(\ref{f:rotateshear}), we can also study the change of the object position along the $z$-axis instead of the $y$-axis to derive another constraint on the $yz$-plane as:
{\begin{equation}
\Delta{z}=({\kappa}_{1}/2-\alpha)y=k'_{1}y
\label{f:dzshearing}
\end{equation}}% 
Nevertheless, given a rotation constraint on $\alpha$, the constraints on $k'_{1}$ of Eq.(\ref{f:dzshearing}) and $k_{1}$ of Eq.(\ref{f:dyshearing}) are not independent as $k'_{1}=k_{1}-2\alpha$, which means the $k'_{1}$ constraint is just a linear combination of the ${k}_1$ constraint and the $\alpha$ constraint.
To restrict both rotation and shearing on the $yz$-plane, a pair of constraints on any of ($\alpha$, $k_{1}$), ($\alpha$, $k'_{1}$), or ($k_{1}$, $k'_{1}$) are sufficient and they are equivalent to each other.   

For the $xz$-plane, which is similar to the $yz$-plane (Fig.~\ref{tkcoosystem}), the requirment of the object to have neither rotation $\beta=0$ nor shearing ${\kappa}_2=0$, leads to:
{\begin{equation}
\sum_{i}z^{i}{\Delta{x}}^{i}=0
\label{f:xzshearcons}
\end{equation}}%
where ${\Delta{x}}^{i}$ can be replaced by the layer alignment parameters as shown in Eq.(\ref{f:dpg}).
For the AMS inner tracker structure, the corresponding constraint can be simplified as:
\begin{equation}
\sum_{l=2}^{8}\matr{R}^{\mathsf{T}l}_{P}(1,1){\Delta}u_{P}^{l}z^{l}A^{l}=0
\label{f:xzshearcons2}
\end{equation}%
where $\matr{R}^{\mathsf{T}l}_{P}(1,1)$ is the (1,1) entry of the $l$-th layer rotation matrix and ${\Delta}u_{P}^{l}$ is the $l$-th layer alignment parameter on the translation along the $u_{P}$-axis.

The detector structure of the $xy$-plane, where the sensors are tiled, is completely different from the $yz$- and $xz$-planes.
The essence of shear deformation is a symmetric strain tensor that results in a change in angle.
So, the shearing on the $xy$-plane to a sensor will shear the sensor surface and break the orthogonal system of the strips on the opposite sides, which is mechanically not allowed. In this sense, the pure shearing on a $xy$-plane or a layer, which leads to a homogeneous deformation of all detector microscopic components, is practically non-existent. 
Another kind of pseudo-shearing of a layer with only shifting the positions of its ladders along the $x$-axis ($u_{P}$-axis in Fig.~\ref{trackergrid} (b)) without deforming the ladders' shape, is also constrained by the intrinsic structure of the sensors in the track alignment procedure, where the relative position between neighboring ladders in a layer is well defined by many tracks crossing them and the sensors from other layers in front and behind.
Accordingly, similar to $\lambda_{1}$ and $\lambda_{2}$ in the stretching deformation, the shearing strain $\kappa_{3}$ also does not need external constraint. 

In this section, we have studied the 12 degrees of freedom in the linear transformation with each of them corresponding to a kind of detector displacement or deformation. They were all constrained:
\begin{enumerate}[label={(\alph*)}]
\item 3 translations and 3 rotations by Eq.(\ref{f:layercons}), 
\item 2 stretchings and 1 shearing by the intrinsic size and shape of the sensors during track alignment,
\item 1 stretching by Eq.(\ref{f:zshiftcons}) or Eq.(\ref{f:zshiftcons2}),
\item 2 shearings, one by Eq.(\ref{f:yzshearcons}) or Eq.(\ref{f:yzshearcons2}) and the other by Eq.(\ref{f:xzshearcons}) or Eq.(\ref{f:xzshearcons2}).
\end{enumerate}
 
\section{Global track alignment}\label{globalalignment}
The global alignment method was first introduced in Ref.~\cite{Softwarealignment}. It is widely used in HEP and other fields~\cite{flucke2008cms}\cite{lhcbalignment}\cite{alice2010alignment}.
In addition to this method, there are also other alignment methods, such as the one presented in Ref.~\cite{altas2020alignment}.

In magnetic field, each track trajectory is characterized by a number of parameters (5 for a helix without multiple-scattering) which has to be determined from the track fitting procedure.
Besides the position measurements, multiple scattering due to Coulomb interaction of the particle with the detector materials also impacts the accurate determination of the track. 
Taking into account the scattering angles being extra measurement quantities, for a given track $i$, the track parameters $\Delta\vect{q}_{i}$ are determined via ${\chi}^2$ minimization~\cite{BGLfitting}: 
{\begin{equation}
{\chi}^{2}_{i}=\sum_{j=1}^{n_{meas}}\vect{\varepsilon}_{j}(\vect{q}_{i})^\mathsf{T}\matr{V}_{j}^{-1}\vect{\varepsilon}_{j}(\vect{q}_{i})+\sum_{j=2}^{n_{scat}-1}\vect{\beta}_{j}(\vect{q}_{i})^\mathsf{T}\matr{W}_{j}^{-1}\vect{\beta}_{j}(\vect{q}_{i})
\end{equation}}%
where $\vect{\varepsilon}_{j}$ is the $j$-th hit residual with the position measurement covariance matrix $\matr{V}_{j}$, and $\vect{\beta}_{j}$ is the $j$-th scattering angle with the covariance matrix $\matr{W}_{j}$~\cite{multiplescattering}\cite{GBLmultiplescattering}. 

In the AMS global alignment, the global detector alignment parameters, $\Delta\vect{p}$, and the local track parameters, $\Delta\vect{q}$, of all tracks are determined simultaneously through a vast ${\chi}^2$ minimization, taking account of both residual measurements and multiple-scattering effects:
{\begin{equation}
\begin{split}
{\chi}^{2}(\vect{q},\vect{p})=\sum_{i=1}^{N_{track}}~\Biggl[\sum_{j=1}^{n_{meas}}{\vect{\varepsilon}}_{ij}(\vect{q}_{i},\vect{p})^\mathsf{T}\matr{V}_{ij}^{-1}\vect{\varepsilon}_{ij}(\vect{q}_{i},\vect{p}) 
+\sum_{j=2}^{n_{scat}-1}{\vect{\beta}}_{ij}(\vect{q}_{i})^\mathsf{T}\matr{W}_{ij}^{-1}{\vect{\beta}}_{ij}(\vect{q}_{i})\Biggr]
\end{split}
\label{f:alignchis}
\end{equation}}

Setting the partial derivatives of the ${\chi}^2$ of Eq.(\ref{f:alignchis}) with respect to each global parameter and each local track parameter equal to zero leads to the matrix equation:
{ \begin{equation}
\begin{pmatrix}
\sum_{i}\matr{C}^{i} &  \matr{G}^{1} &  \ldots &\matr{G}^{j} & \ldots & \matr{G}^{N} \\
(\matr{G}^{1})^\mathsf{T} & \matrgk{\Gamma}^{1} & \ldots & \matr{0} & \ldots & \matr{0}  \\
\vdots  & \vdots & \ddots & \vdots &\ddots & \vdots\\
(\matr{G}^{j})^\mathsf{T} & \matr{0} & \ldots & \matrgk{\Gamma}^{j} &  \ldots & \matr{0} \\
\vdots & \vdots & \ddots & \vdots & \ddots & \vdots\\
(\matr{G}^{N})^\mathsf{T} & \matr{0} & \ldots& \matr{0} & \ldots & \matrgk{\Gamma}^{N} \\
\end{pmatrix}
\begin{pmatrix}
\Delta\vect{p} \\
\Delta\vect{q}_{1} \\
\vdots\\
\Delta\vect{q}_{j} \\
\vdots\\
\Delta\vect{q}_{N} \\
\end{pmatrix}
=\\
\begin{pmatrix}
\sum_{i}\vect{d}^{i} \\
\vect{b}^{1} \\
\vdots \\
\vect{b}^{j} \\
\vdots \\
\vect{b}^{N} \\
\end{pmatrix}
\label{f:globalmatrix}
\end{equation}}%
see Appendix \ref{appB} for the definitions of matrices $\matr{C}$, $\matr{G}$, $\matrgk{\Gamma}$ and vectors $\vect{d}$, $\vect{b}$ as well as the detailed calculation.
The solution requires the inversion of the matrix of dimension $(n_{g}+N{\cdot}n_{l})^{2}$, where $n_{g}$ is the number of global alignment parameters (up to $\sim$15 000 for the AMS tracker), $N$ is the number of tracks used for the alignment (e.g. $\sim$10$^{9}$ tracks for the alignment with cosmic rays collected in flight), and $n_{l}$ is the number of local parameters per track (e.g. up to 27 for the General Broken Lines algorithm~\cite{BGLfitting} with 13 equivalent thin scatterers representing the AMS materials).
The dimension of the inversion matrix for solving the global alignment parameters can be reduced to $n_{g}^{2}$ by partitioning~\cite{MILLEPEDE}.
The constraints discussed in section \ref{detectorcons} are added into the matrix via Lagrange multipliers.
The matrix inversion is handled by the Pede program~\cite{MILLEPEDE2}.
A presigma, which can be interpreted as an initial detector mounting precision, can be assigned to the diagonal matrix element of each alignment parameter to optimize the matrix solution in the program.

In principle, the matrix inversion for solving the global alignment parameters only needs to be performed once and no iterations are required. 
However due to potential inaccuracies in the solution of the large linear system and due to a required outlier (large residual events) treatment, a few internal iterations for the matrix inversion may be necessary.
For the "Inversion" solution method in the Pede, 3 internal iterations are more than enough.
The presigmas are always defined with respect to the previous iteration, hence alignment corrections significantly larger than the presigmas can still occur after iterations.
In this sense, the presigmas are considered not to bias the result if enough iterations are performed but will impact the choice of the preferred solution among all possible candidates with similar ${\chi}^{2}$, which will be discussed in more detail in the next section. 

Recently, the version of the Pede written in Fortran has been implemented to be compatible with multi-threading~\cite{millpedesoft}.
But it is still deficient in dealing with massive local parameters of billions of tracks ($N{\cdot}n_{l}\sim10^9\times20$) and a sizable number of global parameters ($n_g\sim15~000$). This version of Pede is extended by the AMS collaboration to become fully parallelized using the OpenMP platform~\cite{OPENMP}, which allows much faster I$/$O and computational processing. In particular, the most restricted I$/$O part is improved by replacement with the parallelized ROOT~\cite{brun1997root} I$/$O.  Using CERN 64-CPU machines and the EOS storage system~\cite{peters2015eos}, it takes $\sim$30 hours to process the matrix inversion for 1 billion tracks with 3 internal iterations.

\section{Alignment based on the 400~GeV/c proton test beam}\label{testbeamsec}
Each module of the AMS tracker has its own initial mechanical mounting precision varying from a few microns to thousands of microns:
the assembly accuracy for a sensor in the ladder is ${\sim}6~\mathrm{\upmu m}$,
the mounting accuracy for a ladder on the layer is ${\sim}70~\mathrm{\upmu m}$,
the installation accuracy for an inner tracker layer is ${\sim}40~\mathrm{\upmu m}$ along $x$ and $y$ and ${\sim}200~\mathrm{\upmu m}$ along $z$ while for an external layer it is ${\sim}1000~\mathrm{\upmu m}$ for $x$, $y$, and $z$.
A summary of the initial mounting precision can be found in Table \ref{T:presigma} (a). 
The test-beam track alignment aims to reduce the module misalignment from all these sources down to a micron level for the rigidity measurement.

\begin{figure}[htpb]
  \centering
  \includegraphics[width=0.6\textwidth]{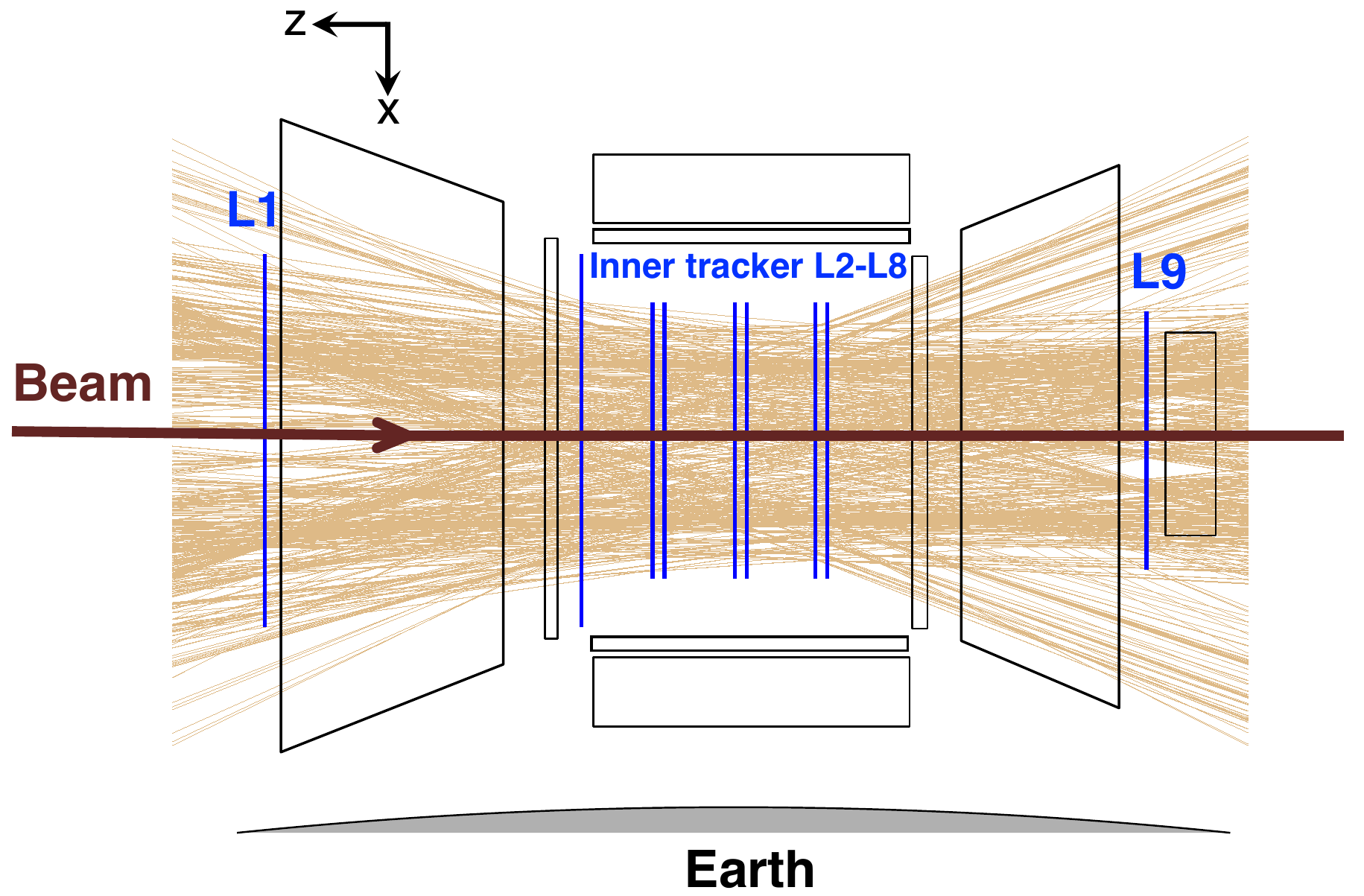}
  \caption{Schematics of the nominal attitude of the AMS in the beam test: the $z$-axis of the AMS against the beam direction, the $x$-axis to the nadir, and the $y$-axis (pointing out of the page) parallel to the Earth. The densely packed lines represent the 886 directions of the primary 400 GeV/c proton test beam passing through AMS.}
  \label{testbeamtracker}
\end{figure}
\subsection{Setup of the test beam}
During the beam test, AMS was installed on a rotation stand  which allows the detector to be exposed to particles from different positions and directions. To minimize the potential deformation of the tracker planes as well as the contraction of the support structures due to gravity, the nominal attitude of the AMS illustrated in Fig.~\ref{testbeamtracker} was pointing to be the $z$-axis against the beam direction, the $x$-axis to the nadir (down), and the $y$-axis parallel to the Earth (horizontal), hence the positions of the tracker modules along the $y$-axis, i.e. the particle bending direction, was the least deformed.

The track alignment is performed based on the primary 400 GeV/c proton beam, where the positions and orientations of the detector were adjusted 886 times to collect events in the full acceptance of the tracker as illustrated in Fig.~\ref{testbeamtracker}. 
The beam spot size, defined as the spot radius to include 68$\%$ of events at each position, was rather narrow at $\sim$3.5~mm.
With ${\sim}10^{4}$ events per position, the total collected number of events for the alignment was ${\sim}10^{7}$.

\begin{figure*}[htpb]
  \includegraphics[width=1\textwidth]{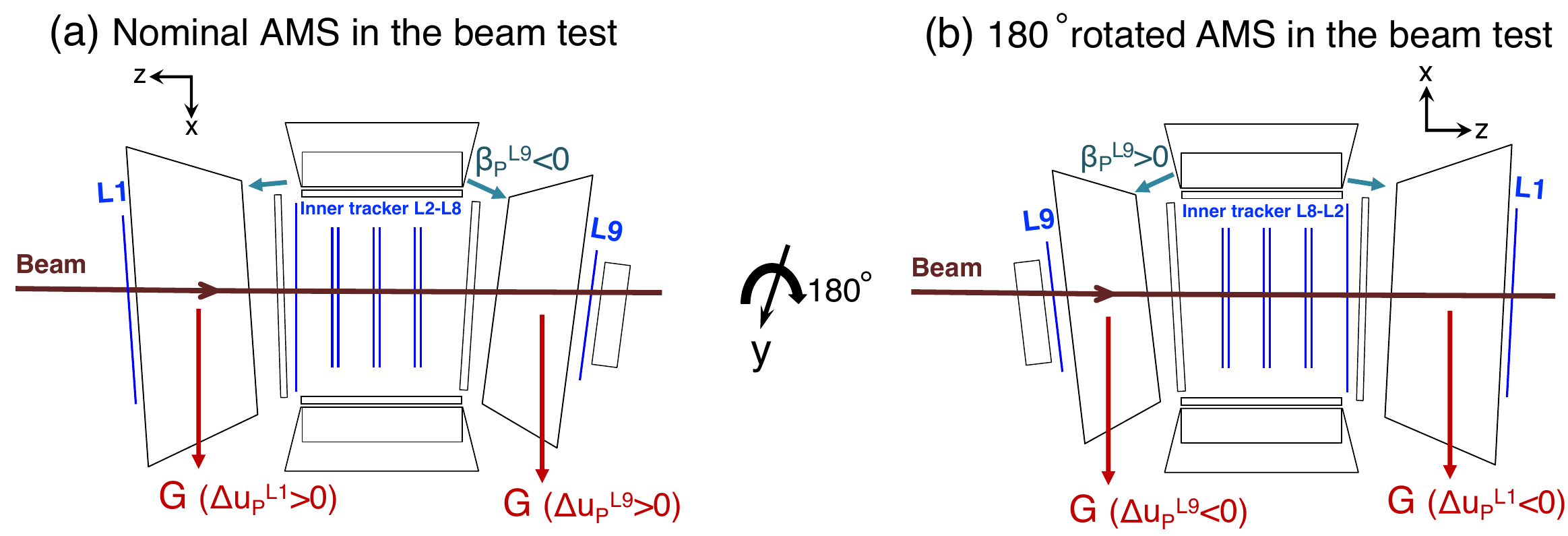}
  \caption{Schematics of the detector deformations due to gravity for (a) nominal AMS and for (b) 180$^{\circ}$ rotated AMS in the beam test.}
  \label{horizontalams}
\end{figure*}

Besides the normal data collection, AMS also collected a special dataset of the 400 GeV/c proton beam, in which the whole detector was rotated around the $y$-axis by 180$^{\circ}$ to examine the mechanical stability of the tracker, as illustrated in Fig.~\ref{horizontalams}. 
There were 60 assigned beam positions for this configuration and the total number of the collected events was ${\sim}10^{6}$.
This data is only used for the alignment verification purpose instead of being directly used in the test-beam alignment. 

\subsection{Alignment procedure}
In the test-beam alignment, all the composite tracker modules are aligned simultaneously using the global composite alignment approach as discussed in sections \ref{compositealign}, \ref{detectorcons}, and \ref{globalalignment}.
The General Broken Lines (GBL) algorithm with fixed curvature ($1/R=1/400$~GV$^{-1}$) track fitting is imposed to derive the residuals $\vect{\varepsilon}_{ij}(\vect{q}_{i}^{0},\vect{p}^{0})$, the partial derivatives with respect to the local track parameters of the residuals ${\partial \vect{\varepsilon}_{ij}}/{\partial \vect{q}_{i}}$ and the scattering angles ${\partial \vect{\beta}_{ij}}/{\partial \vect{q}_{i}}$, for Eq.(\ref{f:globalmatrix}), see also Eqs.(\ref{f:dg})(\ref{f:Gmatrix})(\ref{f:bii})(\ref{f:gamma}). 
The 400 GeV/c proton Monte Carlo sample produced by Geant4~\cite{GEANT42003} is used for the alignment optimization. 

\begin{table}[htb]
\begin{center}
{ \
\begin{tabular}{ccccccc}
\multicolumn{7}{c}{(a) Initial mechanical mounting precision} \\
\hline
\multirow{2}{*}{Precision} & \multicolumn{3}{c}{Translation ($\mathrm{\upmu m}$)}  & \multicolumn{3}{c}{Rotation (mrad)}\\
\cline{2-7}
& $\Delta{u}$ & $\Delta{v}$ & $\Delta{w}$ & $\alpha$   & $\beta$ &  $\gamma$ \\
\hline
L1/L9 &1000 & 1000 & 1500 & 0.1 & 0.1 & 0.5 \\
L2-L8 & 50 & 30 & 200 & 0.06 & 0.06 & 0.05 \\
Ladder& 70 & 100 & 40 & 0.3 & 0.3 & 0.2 \\
Sensor& 6   & 6   & 6  & 0.1 & 0.1 & 0.1 \\
\hline
\multicolumn{7}{c}{}\\
\multicolumn{7}{c}{(b) Presigmas used in the test-beam alignment} \\
\hline
\multirow{2}{*}{Presigma} & \multicolumn{3}{c}{Translation ($\mathrm{\upmu m}$)}  & \multicolumn{3}{c}{Rotation (mrad)}\\
\cline{2-7}
& $\Delta{u}$ & $\Delta{v}$ & $\Delta{w}$ & $\alpha$   & $\beta$ &  $\gamma$ \\
\hline
L1/L9 &5000$^\dagger$ & 5000$^\dagger$ & 5000$^\dagger$ & 10$^\dagger$ & 10$^\dagger$ & 10$^\dagger$ \\
L2-L8 & 200$^\dagger$ & 200$^\dagger$ & 200 & 0.25$^\dagger$ & 0.25$^\dagger$ & 0.2$^\dagger$ \\
Ladder& 100 & 100 & 50 & 0.3 & 0.3 & 0.2 \\
Sensor& 6   & 6   & -- & -- & -- & 0.1 \\
\hline
\\
\end{tabular}
}
\caption{(a) The initial mechanical mounting precision of the tracker modules and (b) the presigmas of the alignment parameters used in the test-beam alignment.
The presigmas labeled "--" indicate the parameters that cannot be precisely determined by the alignment due to the limited beam directions per sensor and therefore are fixed to 0.
The presigmas labeled "$^\dagger$" are significantly increased to approach the preferred solution.} 
\label{T:presigma}
\end{center}
\end{table}
\subsubsection{Presigmas in the alignment}
The external layers, L1 and L9, have much worse mounting accuracy than the inner tracker (L2-L8). At the small scale, the assembly accuracy of the sensors-in-ladders or ladders-on-layers for L1 and L9 are similar to that of the inner tracker. It means that the positions of the external layers in the sensor or ladder level can be treated equally as the inner tracker in the alignment and help to reduce the overall bias. But this can only be achieved by the composite alignment, where all the modules are defined relative to the next support structures and all the modules from the inner tracker and external layers are aligned together taking into account all the correlations.
In the composite alignment, the presigmas of the layer alignment parameters for L1 and L9 are set to be more than 20 times larger than the inner tracker (see Table \ref{T:presigma} (b)), while the presigmas of the sensor/ladder alignment parameters are assigned to be the same for every layer, so that the preferred alignment solution tends to correct the displacements of the whole external layers with reference to the position of the inner tracker.

On the other hand, under similar conditions or ${\chi}^{2}$, the solutions with displacements of the larger modules are preferred to the solutions with displacements of the smaller components.
Presigmas of the alignment parameters can be properly adjusted to favor displacements of the larger modules.
As seen in Table \ref{T:presigma}, the presigmas of the layer alignment parameters labeled "$^\dagger$" are significantly increased compared with the layer mounting precision to strengthen the preference of the corrections on the layers rather than on the ladders.

\subsubsection{Fixed parameters in the alignment}
With a total of 886 beam spots distributed over $\sim$250 sensors per layer, the average number of beam spots per sensor is $\sim$3.  Due to the limited beam positions and directions, $\sim$75$\%$ of the sensors with the crucial sensor parameters of $\Delta{u}_{s}$, $\Delta{v}_{s}$, and $\gamma_{s}$ can be aligned: 
for a sensor with a small number of passing events, $<$2000, $\Delta{u}_{s}$, $\Delta{v}_{s}$, and $\gamma_{s}$ are fixed to 0;
for a sensor with the passing beam spots close together, such as $\sigma({u}_{s})<$10 mm and $\sigma({v}_{s})<$12 mm, $\gamma_{s}$ cannot be precisely determined and is fixed to 0, where $\sigma$ represents the standard deviation.

One ladder in L3 is completely inactive and its alignment parameters are fixed as $\Delta{u}_{L}=\Delta{v}_{L}=\Delta{w}_{L}=\alpha_{L}=\beta_{L}=\gamma_{L}=0$. Another ladder in L4 is inactive on the $n$-side and its $\Delta{u}_{L}$  is fixed to 0. 
For a ladder with the passing beams at small inclination angles and small position spanning along the ${v}_{L}$-axis, both $\sigma(du_{L}^{p}/dw_{L}^{p}{\cdot}{v}_{L})<2.2$ mm and $\sigma(dv_{L}^{p}/dw_{L}^{p}{\cdot}{v}_{L})<2.2$ mm, ${\alpha}_{L}$ cannot be precisely obtained from the alignment and is fixed to 0, where $du_{L}^{p}/dw_{L}^{p}$ and $dv_{L}^{p}/dw_{L}^{p}$ are the beam projected directions in the ladder $u_{L}w_{L}$-plane and $v_{L}w_{L}$-plane respectively (see Fig.~\ref{tkcoosystem} (b)).
Similarly, for a ladder with the passing beams of small inclination angles and small position spanning along the ${u}_{L}$-axis, both $\sigma(du_{L}^{p}/dw_{L}^{p}{\cdot}{u}_{L})<7$ mm and $\sigma(dv_{L}^{p}/dw_{L}^{p}{\cdot}{u}_{L})<7$ mm, ${\beta}_{L}$ is fixed to 0.

\begin{table}[htb]
\begin{center}
{ \
\begin{tabular}{cccccc}
%\hline
\multicolumn{6}{c}{(a) Number of fixed ladder alignment parameters}\\
\hline
$\Delta{u}_{L}$ & $\Delta{v}_{L}$ & $\Delta{w}_{L}$ & $\alpha_{L}$ & $\beta_{L}$ & $\gamma_{L}$ \\
\hline
2 & 1 & 1 & 39 & 2 & 1 \\
\hline
\multicolumn{6}{c}{}\\
%\hline
\multicolumn{6}{c}{(b) Number of fixed sensor alignment parameters} \\
\hline
$\Delta{u}_{s}$ & $\Delta{v}_{s}$ & $\Delta{w}_{s}$ & $\alpha_{s}$ & $\beta_{s}$ & $\gamma_{s}$ \\
\hline
734 & 572 & 2284 & 2284 & 2284 & 1276 \\
\hline
\\
\end{tabular}
}
\caption{The number of ladders (a) and sensors (b) with fixed parameters in the test-beam alignment.
Note that the AMS tracker has 192 ladders  and 2284 sensors.} 
\label{T:fixparTB}
\end{center}
\end{table}
Table \ref{T:fixparTB} summarizes the number of ladders and sensors with fixed alignment parameters.
As seen, 39 ladders --- out of a total 192 ladders --- have the alignment parameter $\alpha_{L}$ fixed.
From Eq.(\ref{f:respl}),  we can derive that the $\alpha_{L}$ equivalent alignment corrections on the ladder hit position are $du_{L}^{p}/dw_{L}^{p}{\cdot}{v}_{L}{\cdot}\alpha_{L}$ and $dv_{L}^{p}/dw_{L}^{p}{\cdot}{v}_{L}{\cdot}\alpha_{L}$ for the $u_{L}$- and $v_{L}$-projections respectively.
Assuming the particle incident angle $du_{L}^{p}/dw_{L}^{p}~(\mathrm{or}~dv_{L}^{p}/dw_{L}^{p})=0.3$, for the hit with the largest ${v}_{L}=35$~mm at the ladder edge, a typical mounting precision of $\sigma({\alpha}_{L})=0.3$ mrad (see Table \ref{T:presigma} (a)) or a fixed $\alpha_{L}=0$ will introduce a misalignment of 3.15~$\mathrm{\upmu m}$, which is a small inaccuracy.
This is also the case for sensor alignment parameters of ${\Delta}w_{s}$, ${\alpha}_{s}$, and ${\beta}_{s}$ fixing them in the alignment will not result in a sizable misalignment.
Owing to a good sensor assembly precision of $\sigma(\gamma_{s})=0.1$ mrad, a fixed $\gamma_{s}=0$ for part of sensors will also give a small misalignment of up to $|v_{s}\gamma_{s}|=3.5~\mathrm{\upmu m}$ (${v}_{L}=35$~mm) and  $|u_{s}\gamma_{s}|=1.9~\mathrm{\upmu m}$ (${u}_{L}=19$~mm) for the $u_{L}$- and $v_{L}$-projections respectively. 

\subsection{Alignment results}

\begin{figure*}[htpb]
  \includegraphics[width=1.\textwidth]{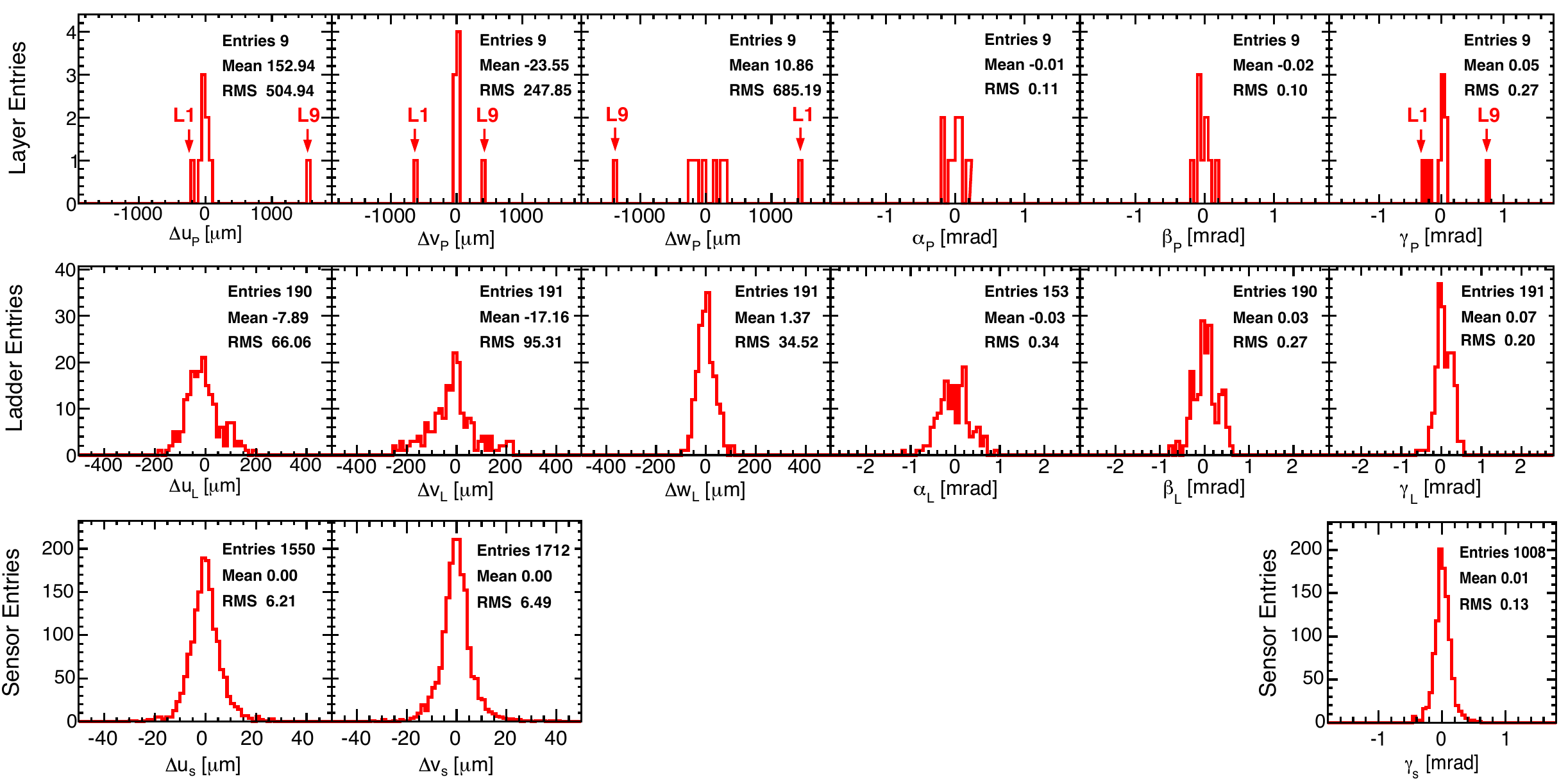}
  \caption{The distributions of the alignment parameters of layers (top row), ladders (middle row), and sensors (bottom row) obtained from the test-beam alignment. The fixed alignment parameters are not included.}
  \label{alignpartestbeam}
\end{figure*}
\begin{figure}[htpb]
  \centering
  \includegraphics[width=0.75\textwidth]{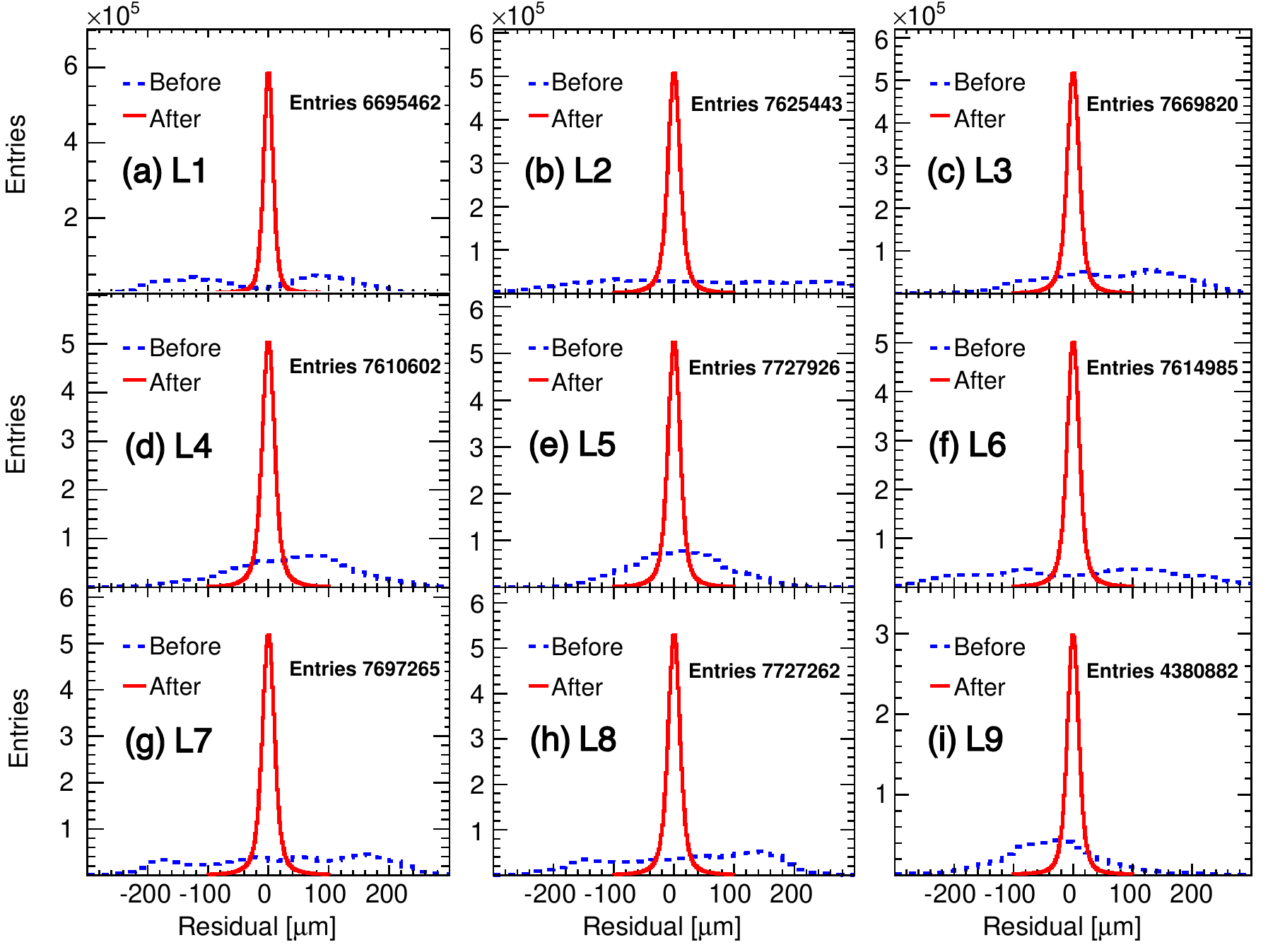}
  \caption{The residual distributions of the individual layers in the sensor $v_{s}$ direction before (dashed histograms) and after (solid histograms) the test-beam alignment.} 
  \label{alignrestestbeam}
\end{figure}
\begin{figure}[htpb]
  \centering
  \includegraphics[width=0.75\textwidth]{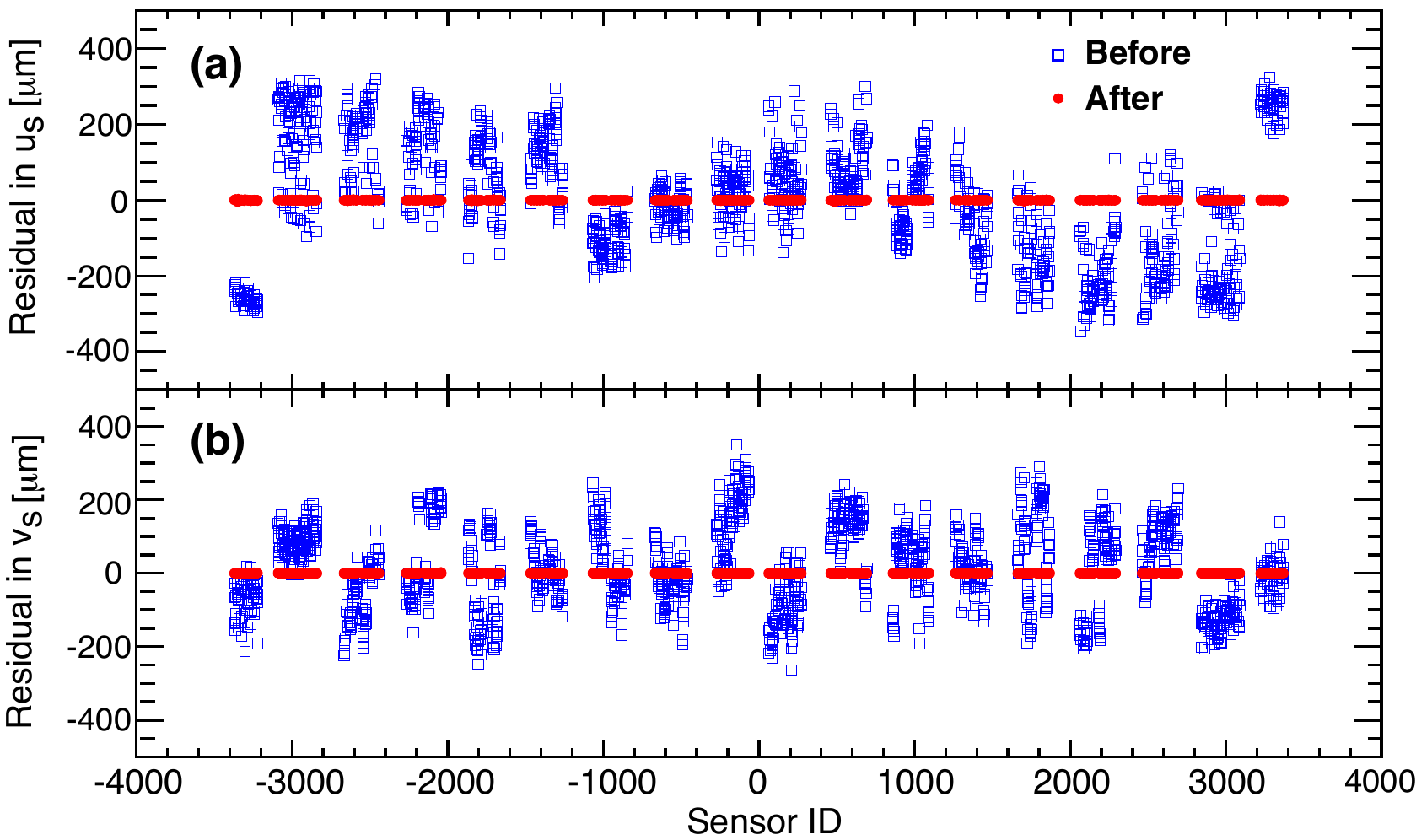}
  \caption{The residual biases of the individual sensors in (a) the $u_{s}$ direction and (b) the $v_{s}$ direction before (open squares) and after (full circles) the test-beam alignment. A circle or square represents the residual bias of each sensor. The circles or squares of a common group are the sensors from the same half of a tracker layer.
The sensor ID is defined as $(sensor+20{\times}ladder+400{\times}layer){\times}half$,
where $sensor$ is the sensor number $[1...15]$, $ladder$ is the ladder number $[1...13]$, $layer$ is the layer number $[0...8]$,
and $half$ is $-1$ for the ladders located on the negative half ($u_{0L}<0$) and $+1$ on the positive half ($u_{0L}>0$) of a layer.
}
  \label{alignressensortb}
\end{figure}

The alignment parameters obtained from the test-beam alignment are shown in Fig.~\ref{alignpartestbeam}. As seen, the external layers, L1 and L9, have much larger layer-biases both in translations and rotations compared with the layers of the inner tracker. Other than that, no significant large outliers on the alignment parameters occur.
Figure~\ref{alignrestestbeam} shows the residual distributions of the 9 layers in the sensor $v_{s}$ direction before and after the test-beam alignment.
A large improvement of the residual distributions is obvious.
Figure~\ref{alignressensortb} shows the residual biases of all sensors before and after the alignment. As seen, there is no bias in each sensor after the alignment.
Even taking into account the limited beam positions and directions, the overall misalignment in the $v_{s}$ direction for the rigidity measurement is 1-2~$\mathrm{\upmu m}$. 

\subsection{Mechanical stability study with the 180$^{\circ}$ runs}
\begin{figure}[htpb]
  \centering
  \includegraphics[width=0.75\textwidth]{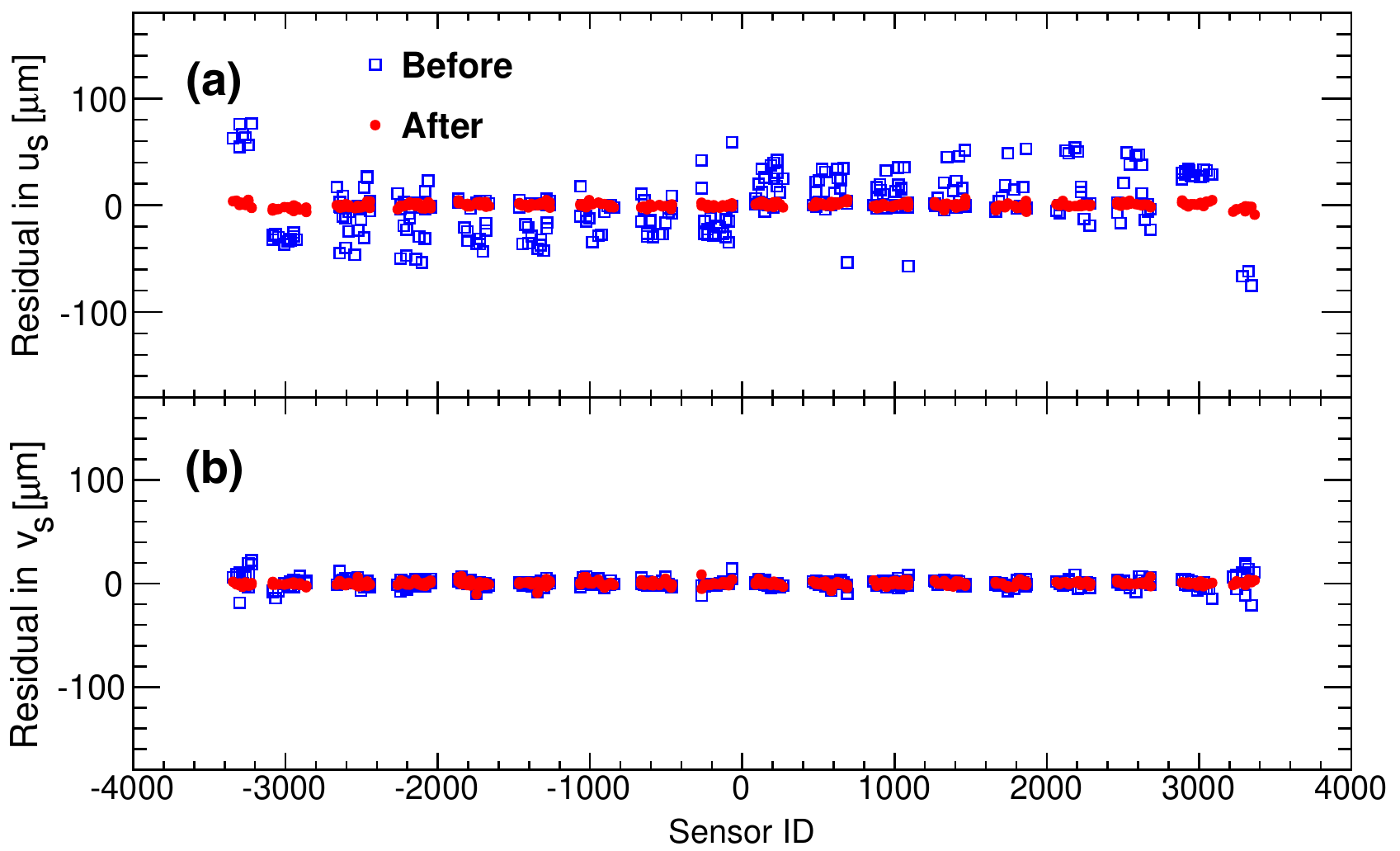}
  \caption{The residual biases in (a) the $u_{s}$ and (b) the $v_{s}$ directions of the individual sensors of the 180$^{\circ}$ test-beam runs using the alignment corrections from the nominal runs but before (open squares) and after (full circles) the additional alignment on the external layers.}
  \label{horizontalruns}
\end{figure}
The test-beam alignment is done based on the nominal data where the AMS $z$-axis is against the beam direction and the $x$-axis is to the nadir as illustrated in Fig.~\ref{horizontalams} (a).
The obtained alignment corrections are then applied to the data collected with the whole detector rotated around the $y$-axis by 180$^{\circ}$ where now the $z$-axis is along the beam direction and the $x$-axis is pointing to the zenith as illustrated in Fig.~\ref{horizontalams} (b).
After rotation, as seen in Fig.~\ref{horizontalruns} (a), there is a significant bias of each sensor in the sensor $u_{s}$ direction (along or opposite to the $x$-axis), while the residual bias of each sensor in the sensor $v_{s}$ direction (along or opposite to the $y$-axis) is tiny as shown in Fig.~\ref{horizontalruns} (b). This clearly indicates the displacement induced by gravity whose direction is parallel to the $x$-axis.

Compared with the inner tracker support structure, a carbon fiber cylinder, the support structures of the external layers, the TRD M-Structure and the Unique Support Structure, are made from aluminum, which is much less stiff.
The resulting detector deformations due to gravity before and after the detector rotation are illustrated in Fig.~\ref{horizontalams} (a) and (b) respectively.
As seen, when the direction of gravity was switched from along to opposite to the AMS $x$-axis, the most prominent changes of the external layer positions in the tracker frame are expected to be the layer translation along the $x$-axis (e.g. from ${\Delta}u_{P}^\mathrm{L1}>0$ to ${\Delta}u_{P}^\mathrm{L1}<0$ for L1) and the layer rotation around the $y$-axis (e.g. from $\beta_{P}^\mathrm{L9}<0$ to $\beta_{P}^\mathrm{L9}>0$ for L9).

\begin{table}[htb]
\begin{center}
{ \
\begin{tabular}{ccccccc}
\hline
\multirow{2}{*}{Displacement} & \multicolumn{3}{c}{Translation ($\mathrm{\upmu m}$)}  & \multicolumn{3}{c}{Rotation (mrad)}\\
\cline{2-7}
& $\Delta{u}_{P}$ & $\Delta{v}_{P}$ & $\Delta{w}_{P}$ & $\alpha_{P}$   & $\beta_{P}$ &  $\gamma_{P}$ \\
\hline
L1 & -200 & -1 & -42 & -0.016 & 0.297 & 0.002 \\
L9 & -580 & -2 &  96 & 0.038  & 1.253 & -0.007 \\
\hline
\\
\end{tabular}
}
\caption{The displacements of L1 and L9 introduced by 180$^{\circ}$ detector rotation obtained from the test-beam alignment.}
\label{T:horitontaldp}
\end{center}
\end{table}
To justify this reasoning, an additional alignment to correct the displacements of the external layers is performed to the 180$^{\circ}$ runs, where all the alignment parameters on the sensors and ladders as well as the layers of the inner tracker are fixed to be the same as the nominal runs except the layer alignment parameters of L1 and L9 which are left free.
The obtained relative changes (180$^{\circ}$ with respect to the nominal) of the layer alignment parameters of L1 and L9 are shown in Table~\ref{T:horitontaldp}.
As seen, when reversing the gravity load in the $x$-direction, the largest translation displacements are along the $x$-axis, $-200~\mathrm{\upmu m}$ and $-580~\mathrm{\upmu m}$ for L1 and L9 respectively, and the largest rotation displacements are around the $y$-axis, 0.297 mrad and 1.253 mrad for L1 and L9 respectively. The translation displacement along the $y$-axis, which is the most critical direction, namely the particle bending direction, is the smallest, $-1~\mathrm{\upmu m}$ and $-2~\mathrm{\upmu m}$ for L1 and L9 respectively. 
Most strikingly, with the alignment only on the external layers, all the major structures of the sensor residual biases disappear and the remaining deviations are within $2~\mathrm{\upmu m}$ as shown in Fig.~\ref{horizontalams}.
This demonstrates that the major outcome of the tracker deformation due to gravity in the beam test is the rigid-body displacement of the external layers.

With the 180$^{\circ}$ runs, the alignment has been verified, the inner tracker support structure has been proved to be rigid, and significant movements induced by gravity of the external layers as rigid bodies have been observed. 

\section {Dynamic alignment of the external tracker layers in space}
After AMS was launched into space, we found that the positions of ladders and sensors were permanently changed up to tens of microns compared to their positions on the ground. 
In addition, the continuous temperature variations on orbit, through the thermal deformation of the support structures, cause the periodic movements of the whole external layers at hundreds of microns per half-obit ($\sim$46 min).
The first kind of displacement is corrected by the static alignment with billions of cosmic-ray events, which will be discussed in section \ref{staticalign}. The second kind of displacement is corrected by the dynamic alignment with instantaneously collected cosmic-ray events and will be reported in this section.
Prior to the static alignment, the dynamic alignment should be applied to remove large periodic movements of the external layers and decrease the inaccuracy of the external tracker layers to the same level as that of the inner tracker.

\subsection{Thermal environment and data collection on orbit}
The ISS orbits the Earth every 93 minutes with an orbital inclination of 52$^{\circ}$.
The thermal environment of AMS on the ISS has both short-term and long-term variations.
The regular short-term variation is the periodic temperature cycle that follows orbital day and night transition.
The long-term variation is mainly due to the change of the angle between the ISS orbital plane and the direction to the Sun, or solar beta angle, which has a precession period of 60 days and can reach up to $\pm75^{\circ}$.
Other thermal variables such as the positions of the ISS radiators and solar arrays, ISS attitude changes for visiting vehicles and reboosts, and shading of AMS by adjacent payloads can also have a big influence on the temperature changes at different time scales, from minutes to months. 

The sensor positions with respect to the carbon fiber reinforced ladders should not change over time, as carbon fiber has near zero coefficient of thermal expansion.
Likewise, the positions of ladders on the carbon fiber skinned planes is stable.
The positions of the inner tracker layers should also not change as their planes are firmly embedded in the carbon fiber cylinder. 
However, the variation of the temperature and gradients across the aluminum mechanical structures (mainly the TRD M-Structure and the Unique Support Structure) lead to continuous periodic movements of the external layers, which are corrected by the dynamic alignment using the concurrently collected cosmic-ray events, mainly protons and helium. 

In flight, the AMS event trigger rates vary from 200 Hz near the equator to $\sim$2000 Hz near the Earth’s magnetic poles. The average event acquisition rate is $\sim$700 Hz.
The events from each quarter of the ISS orbit (from near the pole to the equator or vice versa), about 23 minutes, are arranged in sequence as one run. Detector hardware calibrations are done between runs and last up to two minutes.

\subsection{Alignment procedure}
In the dynamic alignment, only the rigid-body movements of the external layers are considered. 
In this case, there are a total of 12 alignment parameters, 6 for L1 of $({\Delta}u_{P}^\mathrm{L1},{\Delta}v_{P}^\mathrm{L1},{\Delta}w_{P}^\mathrm{L1},\alpha_{P}^\mathrm{L1},\beta_{P}^\mathrm{L1},\gamma_{P}^\mathrm{L1})^\mathsf{T}$ and 6 for L9 of $({\Delta}u_{P}^\mathrm{L9}$,${\Delta}v_{P}^\mathrm{L9},{\Delta}w_{P}^\mathrm{L9},\alpha_{P}^\mathrm{L9},\beta_{P}^\mathrm{L9},\gamma_{P}^\mathrm{L9})^\mathsf{T}$.
For a short time interval with a finite number of cosmic-ray events which are mostly at low rigidities~\cite{PROTONAMS2015}\cite{HELIUMAMS2015}, the main constraint on the alignment precision of an external layer comes from the multiple scattering due to the materials of between L1 and L2, $\sim$0.3~$X_{0}$, or between L8 and L9, $\sim$0.2~$X_{0}$ (see Fig.~\ref{amsdet_fig}). 
As an example, for a particle with 10~GV rigidity, the average scattering angle between L1 and L2  is $\sim$0.7 mrad, which corresponds to ${\sim}700~\mathrm{\upmu m}$ smearing on the L1 position using the 1~m extrapolation from the inner tracker. 
Since multiple scattering and the resulting position smearing decreases linearly with increasing rigidity~\cite{multiplescattering},
the efficient usage of cosmic-ray events and particularly those at high rigidities is critical for the precision of the dynamic alignment. 

\subsubsection{Dynamic alignment in a short-time window}
The developed global alignment approach as discussed in sections \ref{compositealign} and \ref{globalalignment} is applied for the dynamic alignment.
The GBL algorithm with free curvature (inverse rigidity, $1/R$) track fitting is used to derive the residuals $\vect{\varepsilon}_{ij}(\vect{q}_{i}^{0},\vect{p}^{0})$, the partial derivatives with respect to the local track parameters of the residuals ${\partial \vect{\varepsilon}_{ij}}/{\partial \vect{q}_{i}}$ and the scattering angles ${\partial \vect{\beta}_{ij}}/{\partial \vect{q}_{i}}$, for Eq.(\ref{f:globalmatrix}), see also Eqs.(\ref{f:dg})(\ref{f:Gmatrix})(\ref{f:bii})(\ref{f:gamma}). 
The event sample used for the dynamic alignment is required to have a reconstructed track and hits on the external layers. 
The crucial ingredient for the dynamic alignment accuracy, the covariance matrix of the scattering angle, $\matr{W}_{ij}~{\propto}~1/R_{i}^{2}$ in Eq.(\ref{f:alignchis}), can be calculated iteratively event by event using the measured rigidity with the following alignment procedures:
%\begin{enumerate}[label={[\arabic*]}]
\begin{enumerate}[label={(\roman*)}]
\item Initialize $\matr{W}_{ij}(R_{i})$ event by event using the rigidity measured from the inner tracker.
\item Determine the alignment parameters of L1 and L9 by minimization of Eq.(\ref{f:alignchis}).\label{dynamticiter2}
\item Recalculate $\matr{W}_{ij}(R_{i})$ event by event by replacing the rigidity with the new one meausured from both the inner tracker and external layers including the latest alignment corrections from step \ref{dynamticiter2}.\label{dynamticiter3}
\item Repeat steps \ref{dynamticiter2} \ref{dynamticiter3} until all the alignment parameters converge.
\end{enumerate}

\begin{figure}[htpb]
  \centering
  \includegraphics[width=0.75\textwidth]{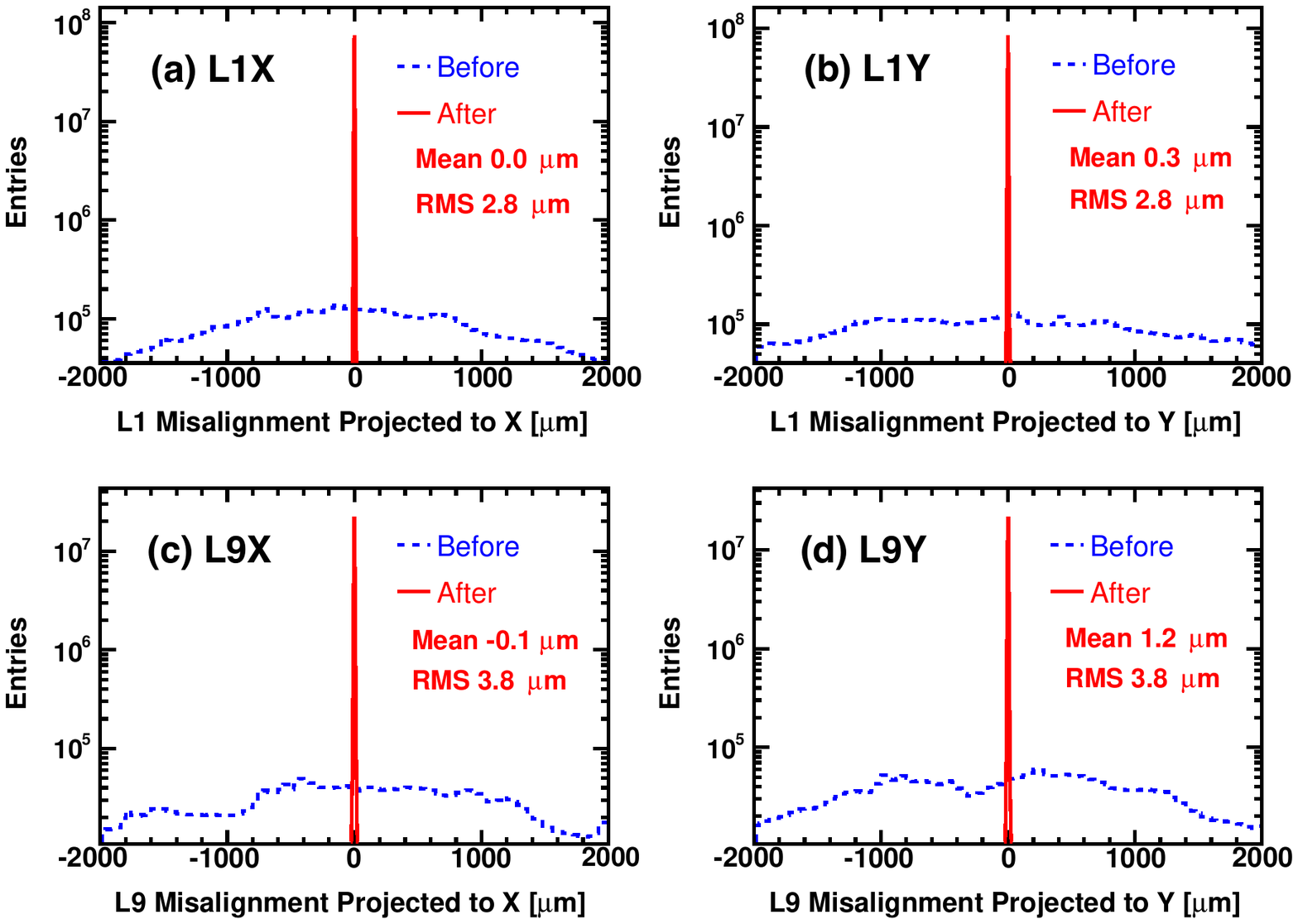}
  \caption{Misalignments of L1 (a, b) and L9 (c, d) before (dashed histograms) and after (solid histograms) the dynamic alignment derived from the MC simulation. For every 100 000 simulated events passing through either L1 or L9, the positions of the external layers in the MC are randomly displaced, which is then followed by a dynamic alignment. One entry of misalignment in each histogram corresponds to one set of displacements of the external layers. With the alignment, the misalignments projected to $x$ (a, c) and $y$ (b, d) coordinates are reduced from more than a thousand microns down to 2.8~$\mathrm{\upmu m}$ for L1 and 3.8~$\mathrm{\upmu m}$ for L9.}
  \label{dynamicalignmc}
\end{figure}

An isotropic cosmic-ray Monte Carlo (MC) sample produced by Geant4~\cite{GEANT42003}\cite{PROTONAMS2015}\cite{HELIUMAMS2015} is used for validation of the alignment.
For every 100 000 events, the positions of the external layers in the MC are randomly displaced, which is then followed by a dynamic alignment. 
Figure~\ref{dynamicalignmc} shows the misalignments of L1 and L9 before and after the alignment derived directly from the MC.
As seen, with the alignment, the misalignments are reduced from more than a thousand microns down to a few microns.

For the flight data, this alignment is performed in time-slices of $\Delta_{t}\approx90$~sec.
The set of alignment parameters obtained in each time-slice have significant statistical errors, which are further reduced by combining the alignment results from the nearby time-slices via a custom developed smoothing procedure described in the next section. 

\subsubsection{Alignment smoothing for long time period}
\begin{figure}[htpb]
  \centering
  \includegraphics[width=0.7\textwidth]{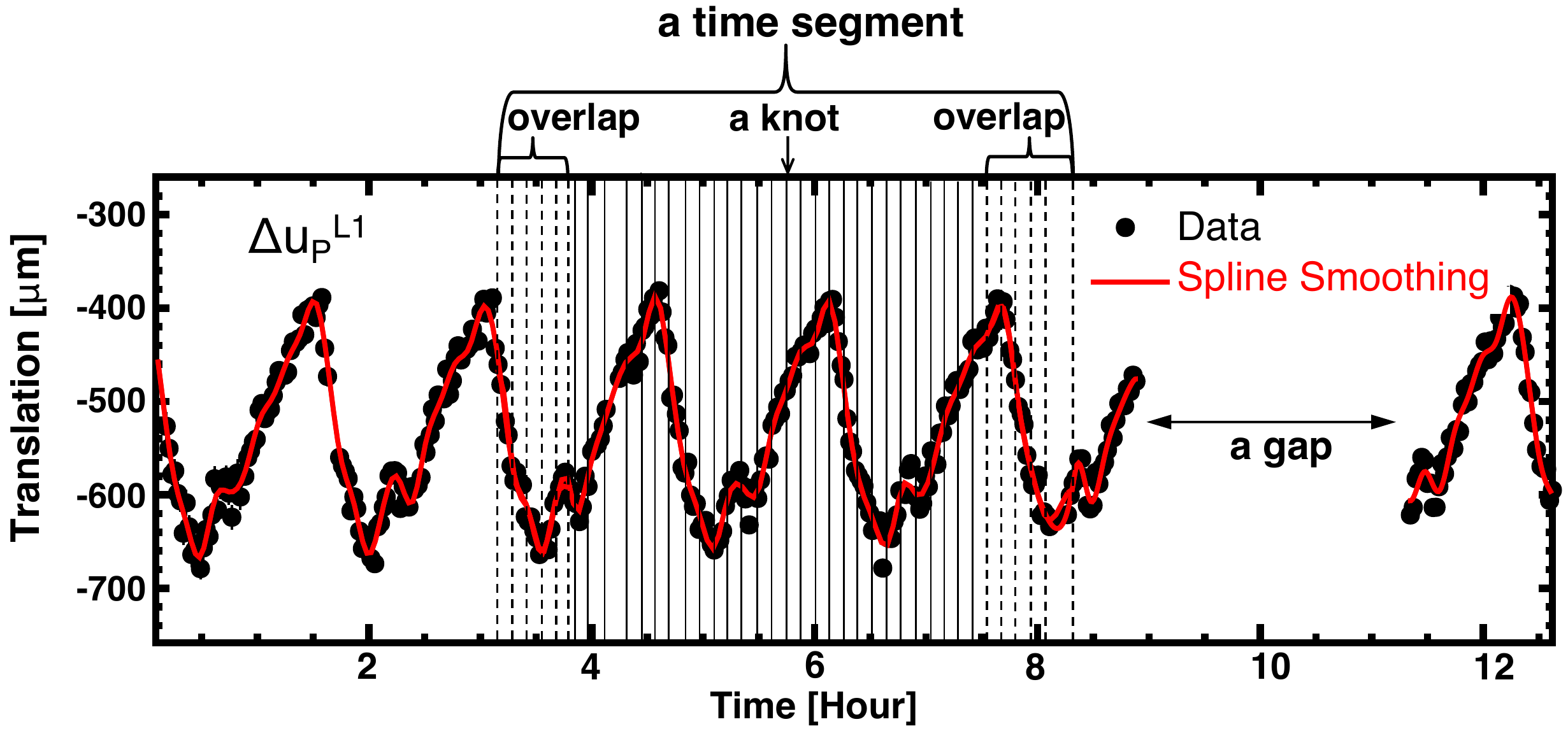}
  \caption{Illustration of the spline smoothing for describing the variation of the dynamic alignment parameter $\Delta{u}_{P}^{L1}$ over time. The entire time block is divided into several smaller overlapping time segments, where the alignment data (points) in each segment are smoothed by a spline function (curve) as indicated.
The distribution of knots of the spline is indicated by the vertical lines including those knots shared with the neighboring splines (dashed vertical lines).}
  \label{splinesmoothing}
\end{figure}

After the short-time dynamic alignment, for a given time period of $N\Delta_{t}$, $N$ sets of alignment parameters are smoothed as functions of time to describe the differential movements of the external layers. 
To fully exploit the alignment information, the time period for each smoothing should be as long as possible, but that introduces too many fitting parameters to solve.
Instead, in our approach, the entire 10 years is divided into small overlapping time segments of a few hours, where the alignment data in each time segment are smoothed by a spline function~\cite{splinesmoothing}, as illustrated in Fig.~\ref{splinesmoothing}:
%\begin{enumerate}[label={(\roman*)}]
\begin{enumerate}
\item Each spline function has up to 40 knots (indicated as vertical lines in Fig.~\ref{splinesmoothing}) which are distributed over time with an equal number of data points per knot. 
\item The neighboring segments overlap (share) 6 knots of the alignment data (Fig.~\ref{splinesmoothing} dashed vertical lines' region).
\item If there is a data gap (more than 1 hour), the new segment will restart once the next alignment data appears. 
\end{enumerate}
To achieve the minimal alignment error, the assignment of the knots is critical:
the more alignment-data points per knot, the smaller the statistical error but the larger the systematic error;
while the fewer data points per knot, the smaller the systematic error but the larger the statistical error.

In the short-time dynamic alignment, the error of each alignment parameter for each time slice ($\Delta_{t}$), $\sigma_{t}$, is estimated from error propagation,
which has a small bias depending on (a) the track fitting model along with the assessment of errors on the multiple scattering and coordinate resolution and (b) the intrinsic correlation among the alignment parameters.
A correction factor $k$, which scales the alignment parameter error to the true one as $k\sigma_{t}$, can be derived from the alignment data over a long time period ($N_{0}\Delta_{t}$) by bootstrapping:
{\begin{equation}
k=\sqrt{\frac{{\chi}_0^2}{n_0}}=\sqrt{\frac{{\chi}_0^2}{N_{0}-m_{0}}}
\end{equation}}% 
where ${\chi}_0^{2}$, $m_{0}$, and $n_{0}=N_{0}-m_{0}$ are the fitting chi-square, number of knots, and degrees of freedom, for the spline fitting to $N_{0}$ data points with a sufficient number of knots to reach a negligible systematic error.
In view of the observed rate of the external-layer movement, every 2 data points or 180 sec per knot ($m_{0}=N_{0}/2$) is enough to derive $k$.

For a spline fit to $N$ data points with a given number of data points per knot, the total alignment error after smoothing is the sum in quadrature of the statistical and systematic errors:
\begin{equation}
\sigma_{tot}=\sqrt{\sigma_{stat}^{2}+\sigma_{sys}^{2}}=\sqrt{k^{2}\sigma_{fit}^{2}+\Bigl(\frac{{\chi}^2}{n}-k^{2}\Bigr)\sigma_{t}^{2}}
\label{f:errdatapernode}
\end{equation}%
where $\sigma_{fit}$, ${\chi}^{2}$, and $n$ are the fitting error, chi-square, and degrees of freedom respectively, $\sigma_{stat}=k\sigma_{fit}$ is the statistical error which decreases as increasing data points per knot, and $\sigma_{sys}=\sigma_{t}\sqrt{{{\chi}^2}/{n}-k^{2}}$ is the systematic error which increases as increasing data points per knot. 

The smoothing of the external layer movement is optimized by assigning the knots with the minimal total error of Eq.({\ref{f:errdatapernode}}) for every alignment parameter.

\subsection{Alignment results}
\begin{figure*}[htpb]
  \includegraphics[width=1.\textwidth]{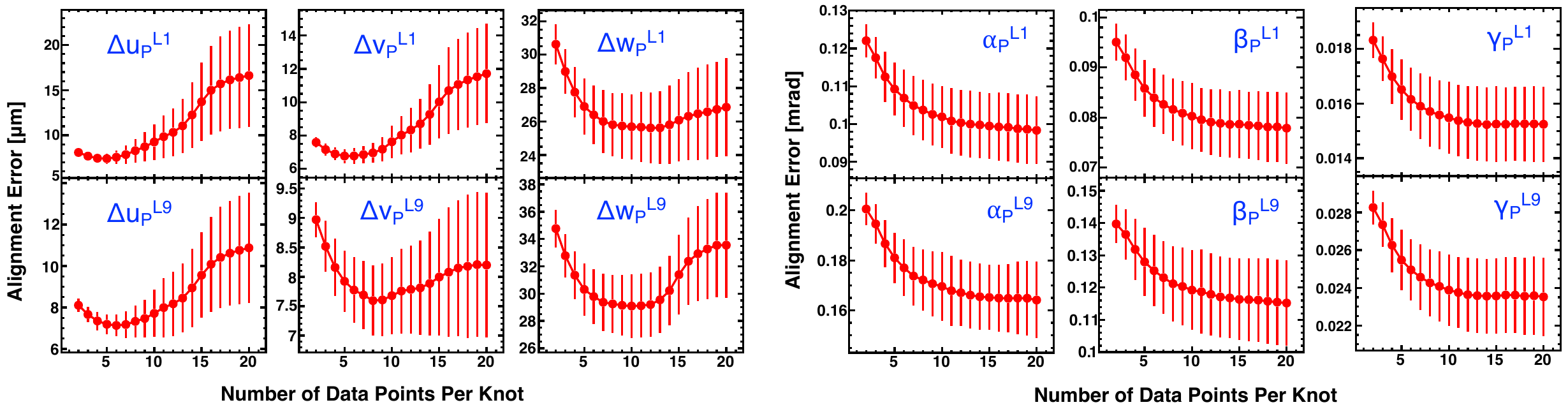}
  \caption{The total errors of the dynamic alignment parameters as functions of number of data points per knot, calculated over 10 years from Eq.(\ref{f:errdatapernode}). The error bars in each plot represent the standard deviations of the alignment errors arising from the time dependence.}
  \label{dyalignerr}
\end{figure*}
The total errors of the individual alignment parameters as functions of number of data points per knot calculated over 10 years from Eq.(\ref{f:errdatapernode}) are shown in Fig.~\ref{dyalignerr}.
Accordingly, the time intervals between adjacent knots for the spline smoothings with the minimal alignment errors are summarized in Table~\ref{T:fitwindow} (a).
As seen, compared with rotations, translations need more dense knots to trace their variations, indicating that the movements of the external layers in terms of translations are more rapid than in terms of rotations.

\begin{figure*}[htpb]
  \includegraphics[width=1.\textwidth]{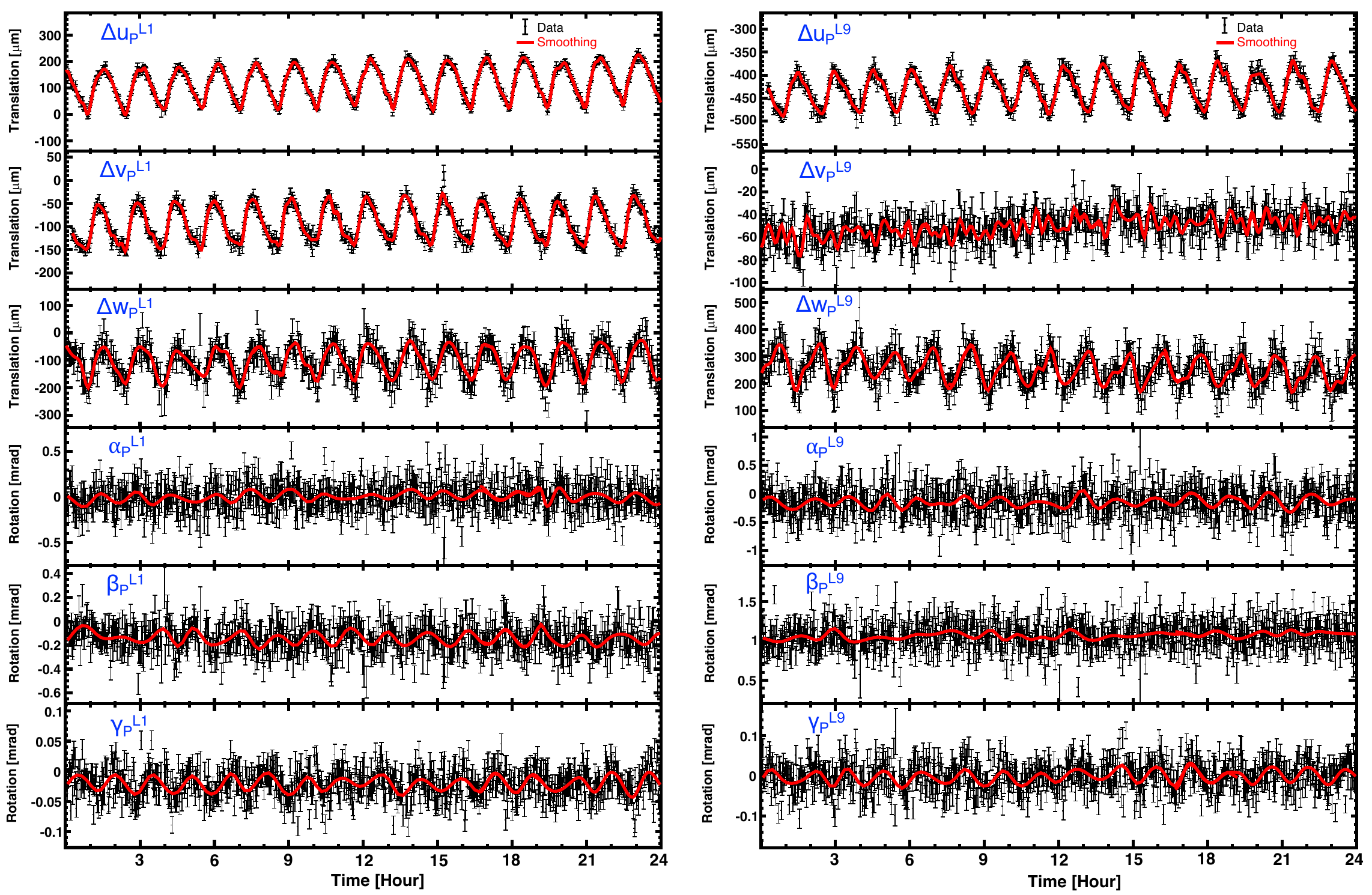}
  \caption{The variations of the dynamic alignment parameters of L1 (left column) and L9 (right column) over 24 hours on Dec. 17, 2015.}
  \label{smoothingday}
\end{figure*}
\begin{figure*}[htpb]
  \includegraphics[width=1.\textwidth]{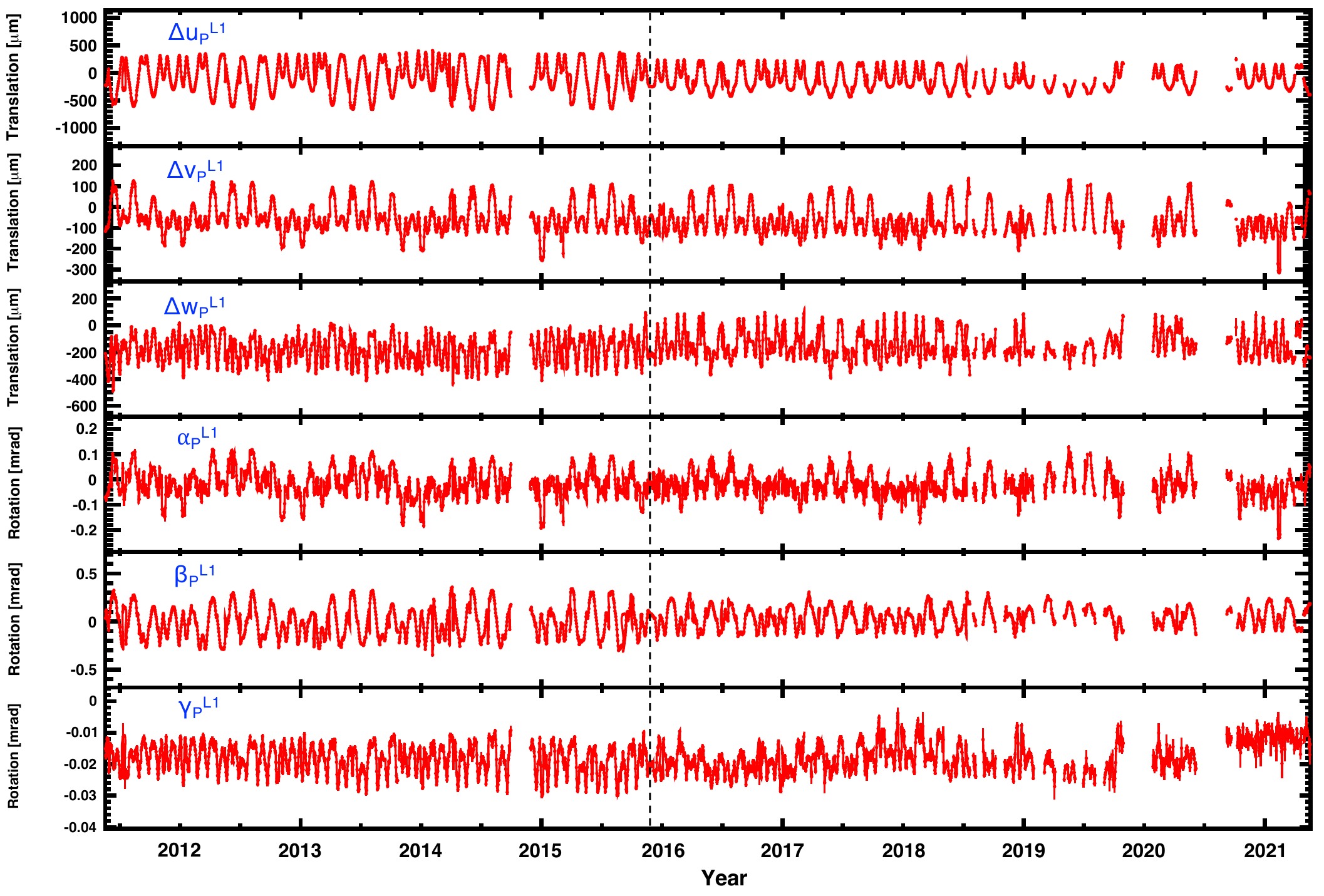}
  \caption{The variations of the dynamic alignment parameters of L1 over 10 years from May 20, 2011 to May 20, 2021. Note the change in behavior starting from the end of 2015, which is due to the installation of a thermal blanket on the port ($-x$) side of AMS on Oct 28, 2015 (indicated by the vertical dashed line).}
  \label{smoothingmonthl1}
\end{figure*}
\begin{figure*}[htpb]
  \includegraphics[width=1.\textwidth]{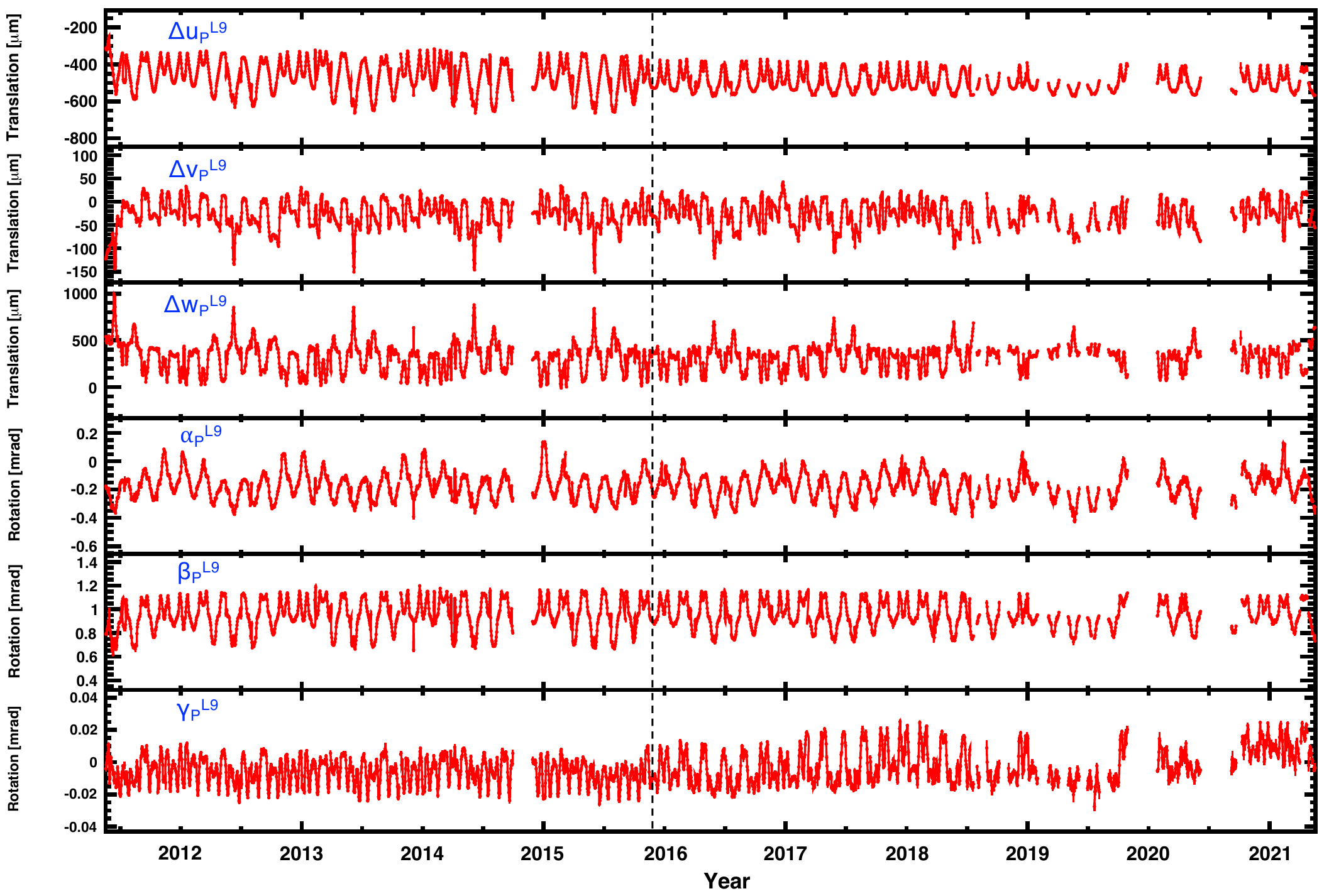}
  \caption{The variations of the dynamic alignment parameters of L9 over 10 years from May 20, 2011 to May 20, 2021.  Note the change in behavior starting from the end of 2015, which is due to the installation of a thermal blanket on the port ($-x$) side of AMS on Oct 28, 2015 (indicated by the vertical dashed line).}
  \label{smoothingmonthl9}
\end{figure*}

Typical variations of the individual alignment parameters over a day together with the smoothings are shown in Fig.~\ref{smoothingday}.
The orbital period of $\sim$93 minutes can be clearly seen.
As shown in the figure, the movement of L1 (L9) in terms of translation is $\sim$200~$\mathrm{\upmu m}$, $\sim$100~$\mathrm{\upmu m}$, and $\sim$200~$\mathrm{\upmu m}$ ($\sim$100~$\mathrm{\upmu m}$, $\sim$20~$\mathrm{\upmu m}$, and $\sim$200~$\mathrm{\upmu m}$) per half orbit in the $x$-, $y$-, and $z$-directions (strictly the $u_{P}$-, $v_{P}$-, and $w_{P}$-directions) respectively and of rotation is $\sim$0.2~mrad, $\sim$0.2~mrad, and $\sim$0.03~mrad ($\sim$0.2~mrad, $\sim$0.1~mrad, and $\sim$0.05~mrad) per half orbit around the $x$-, $y$-, and $z$-axes (strictly the $u_{P}$-, $v_{P}$-, and $w_{P}$-axes) respectively.
In addition to the orbital movements, the external layers also display the long-term movements with a cycle of about 2 months --- the period of the solar beta angle. Figures~\ref{smoothingmonthl1} and \ref{smoothingmonthl9} show the variations of the individual alignment parameters of L1 and L9 respectively, over 10 years from May 20, 2011 to May 20, 2021,
 where each data point represents the alignment parameter averaged over a day.
As seen, the long-term movements of L1 (L9) translations are up to $\sim$1000~$\mathrm{\upmu m}$, $\sim$200~$\mathrm{\upmu m}$, and $\sim$300~$\mathrm{\upmu m}$ ($\sim$300~$\mathrm{\upmu m}$, $\sim$100~$\mathrm{\upmu m}$, and $\sim$700~$\mathrm{\upmu m}$) per month in the $x$-, $y$-, and $z$-directions respectively and the rotations can reach $\sim$0.2~mrad, $\sim$0.6~mrad, and $\sim$0.02~mrad ($\sim$0.4~mrad, $\sim$0.5~mrad, and $\sim$0.03~mrad) per month around the $x$-, $y$-, and $z$-axes respectively. 

\begin{table}[htb]
\begin{center}
{ \
\begin{tabular}{ccccccc}
\multicolumn{7}{c}{(a) The time intervals between adjacent knots (sec)}\\
\hline
Parameter & $\Delta{u}_{P}$ & $\Delta{v}_{P}$ & $\Delta{w}_{P}$ & $\alpha_{P}$   & $\beta_{P}$ &  $\gamma_{P}$ \\
\hline
L1 & 450 & 540  & 990 & 1800 & 1800 & 1800 \\
L9 & 540 & 720 &  900 & 1800 & 1800 & 1800 \\
\hline
\multicolumn{7}{c}{}\\
\multicolumn{7}{c}{(b) External layer dynamic alignment precision}\\
\hline
\multirow{2}{*}{Parameter} & \multicolumn{3}{c}{Translation ($\mathrm{\upmu m}$)}  & \multicolumn{3}{c}{Rotation (mrad)}\\
\cline{2-7}
& $\Delta{u}_{P}$ & $\Delta{v}_{P}$ & $\Delta{w}_{P}$ & $\alpha_{P}$   & $\beta_{P}$ &  $\gamma_{P}$ \\
\hline
L1 & 7.4 & 6.8 & 25.6 & 0.098 & 0.078 & 0.015 \\
L9 & 7.2 & 7.6 & 29.1 & 0.164 & 0.115 & 0.023 \\
\hline
\\
\end{tabular}
}
\caption{(a) The time intervals between adjacent knots used for the spline smoothings of the individual alignment parameters that provide (b) the best dynamic alignment precision.}
\label{T:fitwindow}
\end{center}
\end{table}

\begin{figure}[htpb]
  \centering 
  \includegraphics[width=0.65\textwidth]{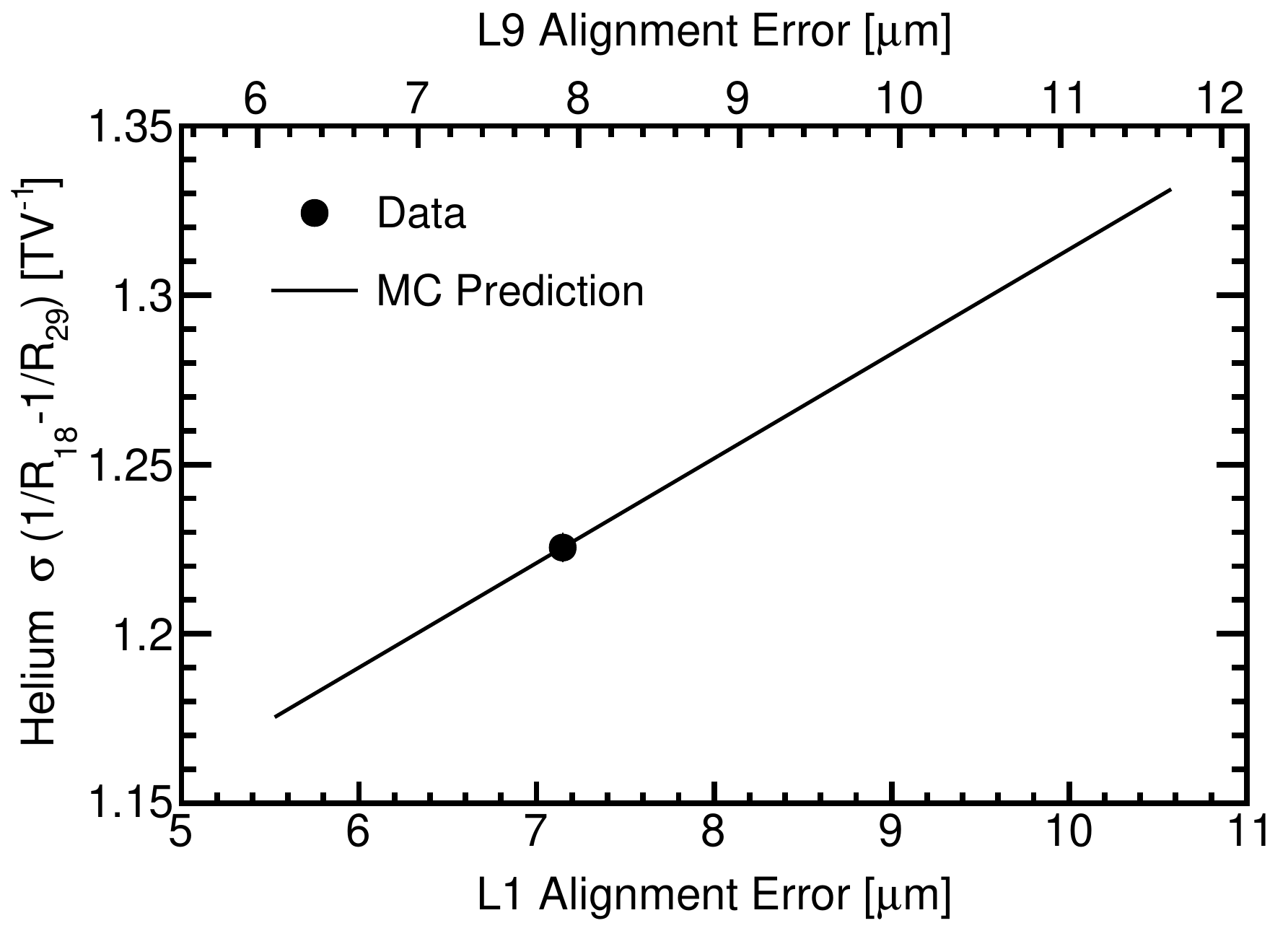}
  \caption{The standard deviation of the difference in the inverse rigidities measured using the upper span (L1–L8) and using the lower span (L2–L9) of the tracker, $\sigma(1/R_{18}-1/R_{29})$, for cosmic-ray helium data with the alignment corrections (full circle) and for the Monte Carlo prediction based on the alignment errors of L1 and L9 (line) in the rigidity range $R_{19}>570$~GV. 
As seen, the data point best matches the Monte Carlo prediction at the alignment errors of 7.1~$\mathrm{\upmu m}$ and 7.9~$\mathrm{\upmu m}$ for L1 and L9 respectively.}
  \label{l1l9alignerr}
\end{figure}

The final achieved alignment precision for all the alignment parameters derived from Fig.~\ref{dyalignerr} is summarized in Table~\ref{T:fitwindow} (b).
As seen, for example, with the dynamic alignment, the translational movement in the $y$-direction (${\Delta}v_{P}$) is aligned to a precision of 6.8~$\mathrm{\upmu m}$ for L1 and 7.6~$\mathrm{\upmu m}$ for L9. To evaluate the total residual misalignments of the external layers in the particle bending direction which is connected to the rigidity resolution, the rigidity measured using the upper span of the tracker, namely from L1 to L8 ($R_{18}$), are compared to the rigidity measured using the lower span, namely from L2 to L9 ($R_{29}$), for a helium sample with the full-span rigidity (measured from L1 to L9) $R_{19}>570$~GV.
Figure~\ref{l1l9alignerr} shows the Gaussian sigma of the $1/R_{18}-1/R_{29}$ distribution derived from the flight data (full circle) and its fit to the prediction from the MC simulation (line).
As seen, with the dynamic alignment, the total residual misalignments (alignment errors) on the rigidity measurement are estimated to be 7.1~$\mathrm{\upmu m}$ for L1 and 7.9~$\mathrm{\upmu m}$ for L9. 

\section{Static alignment of the tracker in space}\label{staticalign}
Before launch, AMS has been aligned based on the primary 400 GeV/c proton test beam as discussed in section \ref{testbeamsec}. However, the strong accelerations and vibrations during launch, followed by the rapid outgassing of the support structure in vacuum 
permanently changed the positions of all the tracker modules. Therefore, the entire tracker has to be aligned again with cosmic-ray events to correct the resulting displacements. The most challenging part of this alignment is the unknown curvatures ($1/R$) of the incoming particles in the presence of the magnetic field.
A track alignment approach similar to the test-beam alignment but with free curvature track fitting (see Eq.(\ref{f:alignchis})) is not enough for such an alignment as the curvatures of the tracks can be biased by any value without changing the alignment $\chi^{2}$.
The development of a new mathematical description is required for such an alignment.

\subsection{Global track alignment with curvature constraints}\label{globalalignment2}
For the alignment with a magnetic field and with particles whose rigidities are unknown, a new term, ${{\rho}_{i}^{2}(\vect{p})}/{Z_{i}}$, is introduced in the global alignment $\chi^{2}$ to constrain the curvature change:
{ \begin{equation}
\begin{split}
{\chi}^{2}(\vect{q},\vect{p})=\sum_{i=1}^{N_{track}}~\Biggl[\sum_{j=1}^{n_{meas}}{\vect{\varepsilon}}_{ij}(\vect{q}_{i},\vect{p})^\mathsf{T}\matr{V}_{ij}^{-1}{\vect{\varepsilon}}_{ij}(\vect{q}_{i},\vect{p}) 
+\sum_{j=2}^{n_{scat}-1}{\vect{\beta}}_{ij}(\vect{q}_{i})^\mathsf{T}\matr{W}_{ij}^{-1}\vect{\beta}_{ij}(\vect{q}_{i})+\frac{{\rho}_{i}^{2}(\vect{p})}{Z_{i}}\Biggr]
\end{split}
\label{f:alignchisiss}
\end{equation}}%
where ${\rho}_{i}(\vect{p})={\rho}_{i}(\vect{p}^{0})+\sum_{g'}{\frac{\partial {\rho}_{i}}{\partial p_{g'}}}{\Delta}p_{g'}$ is the curvature bias ($\Delta{R}^{-1}$) for the $i$-th track, that depends on the global alignment parameters ${\Delta}\vect{p}$, and is equal to ${\rho}_{i}(\vect{p}^{0})$ before the alignment, namely ${\Delta}\vect{p}=\vect{0}$; and $Z_{i}$ is its variance.
$Z{\rightarrow}0$ will impose no change of the curvature measurement before and after the alignment. Conversely, $Z{\rightarrow}{\infty}$ means no curvature constraints in the alignment, making Eq.(\ref{f:alignchisiss}) the same as Eq.(\ref{f:alignchis}).
In the absence of a curvature reference, the measured curvature of a track is supposed to have no bias before the alignment, as ${\rho}_{i}(\vect{p}^{0})=0$, with an uncertainty represented by the variance (squared error) $Z_{i}$.

Setting the partial derivative of the ${\chi}^2$ of Eq.(\ref{f:alignchisiss}) with respect to each global parameter $\Delta{p}_g$ equal to zero, we can derive a matrix equation similar to Eq.(\ref{f:di}), as:
{ \begin{equation}
\sum_{i=1}^{N_{track}}\vect{d'}^{i}=\Bigl(\sum_{i=1}^{N_{track}}\matr{C'}^{i}\Bigr){\Delta}\vect{p}+\sum_{i=1}^{N_{track}}\matr{G}^{i}{\Delta}\vect{q}_{i}
\label{f:di2}
\end{equation}}%
where $\vect{d'}^{i}$ is a vector whose $g$-th element is given by:
{ \begin{equation}
{d'}_{g}^{i}=-\sum_{j=1}^{n_{meas}}\Bigl({\frac{\partial \vect{\varepsilon}_{ij}}{\partial p_{g}}}\Bigr)^\mathsf{T}\matr{V}_{ij}^{-1}\vect{\varepsilon}_{ij}(\vect{q}_{i}^{0},\vect{p}^{0})-\frac{\partial {\rho}_{i}}{\partial p_{g}}{Z}_{i}^{-1}{\rho}_{i}(\vect{p}^{0})
\label{f:dg2}
\end{equation}}%
$\matr{C'}^{i}$ is a matrix whose $(g,g')$ entry is given by:
{ \begin{equation}
{C'}_{gg'}^{i}=\sum_{j=1}^{n_{meas}}\Bigl({\frac{\partial \vect{\varepsilon}_{ij}}{\partial p_{g}}}\Bigr)^\mathsf{T}\matr{V}_{ij}^{-1}{\frac{\partial \vect{\varepsilon}_{ij}}{\partial p_{g'}}}+\frac{\partial \rho_{i}}{\partial p_{g}}{Z}_{i}^{-1}{\frac{\partial \rho_{i}}{\partial p_{g'}}}
\label{f:Cmatrix2}
\end{equation}}%
and $\matr{G}^{i}$ is the matrix whose entry has been defined in Eq.(\ref{f:Gmatrix}).

Setting the partial derivative of the ${\chi}^2$ of Eq.(\ref{f:alignchisiss}) with respect to each local track parameter of each track equal to zero, we obtain the same matrix equation as Eq.(\ref{f:bi}). 

Combining Eq.(\ref{f:bi}) and Eq.(\ref{f:di2}), all the global alignment parameters, $\Delta\vect{p}$, and all the local track parameters, $\Delta\vect{q}$, can be solved simultaneously as in Eq.(\ref{f:globalmatrix}) with the replacement of $\vect{d}^{i}\rightarrow\vect{d'}^{i}$ and $\matr{C}^{i}\rightarrow\matr{C'}^{i}$.

The partial derivatives of the curvature change with respect to the global alignment parameters, ${\partial {\rho}_{i}}/{\partial \vect{p}}$, present in both $\vect{d'}^{i}$ of Eq.(\ref{f:dg2}) and $\matr{C'}^{i}$ of Eq.(\ref{f:Cmatrix2}), are needed for the alignment.
For the $i$-th track, the alignment corrections $\Delta\vect{p}$ will change the ($j$-th) hit residual by: 
{ \begin{equation}
{\vect{\widetilde{\varepsilon}}}_{ij}^{0}=\sum_{g'}\frac{\partial \vect{\varepsilon}_{ij}}{\partial p_{g'}}{\Delta}p_{g'}
\end{equation}}%
A track fitting is performed on ${\vect{\widetilde{\varepsilon}}}_{i}^{0}$ from all the hits to derive the local track parameters, $\Delta\widetilde{\vect{q}}_{i}$, which represent the alignment corrections on the $i$-th track trajectory. 
Minimization of the fitting $\widetilde{\chi}^{2}$ leads to the partial derivative with respect to each local track parameter, $\widetilde{q}_{il}$, equal to zero: 
{ \begin{equation}
\begin{split}
0=\frac{\partial {\widetilde{\chi}}^{2}}{\partial \widetilde{q}_{il}}~{\simeq}~2\sum_{j=1}^{n_{meas}}\Bigl({\frac{\partial \vect{\varepsilon}_{ij}}{\partial \widetilde{q}_{il}}}\Bigr)^\mathsf{T}\Bigl({\vect{\widetilde{\varepsilon}}}_{ij}^{0}+\sum_{l'}{\frac{\partial \vect{\varepsilon}_{ij}}{\partial \widetilde{q}_{il'}}}{\Delta}\widetilde{q}_{il'}\Bigr) 
=2\sum_{j=1}^{n_{meas}}\Bigl({\frac{\partial \vect{\varepsilon}_{ij}}{\partial \widetilde{q}_{il}}}\Bigr)^\mathsf{T}\Bigl(\sum_{g'}\frac{\partial \vect{\varepsilon}_{ij}}{\partial p_{g'}}{\Delta}p_{g'}+\sum_{l'}{\frac{\partial \vect{\varepsilon}_{ij}}{\partial \widetilde{q}_{il'}}}{\Delta}\widetilde{q}_{il'}\Bigr)
\end{split}
\label{dchisdqtilde}
\end{equation}}%
Eq.(\ref{dchisdqtilde}) can be simplified in matrix form as:
{ \begin{equation}
 \vect{0}=(\matr{\widetilde{G}}^{i})^\mathsf{T}{\Delta}\vect{p}+\matrgk{\widetilde{\Gamma}}^{i}{\Delta}\vect{\widetilde{q}}_{i}
\label{f:bitilde}
\end{equation}}%
where $\matr{\widetilde{G}}^{i}$ is a matrix whose $(g,l')$ entry is given by:
{ \begin{equation}
\widetilde{G}_{gl'}^{i}=\sum_{j=1}^{n_{meas}}\Bigl({\frac{\partial \vect{\varepsilon}_{ij}}{\partial p_{g}}}\Bigr)^\mathsf{T}{\frac{\partial \vect{\varepsilon}_{ij}}{\partial \widetilde{q}_{il'}}}
\label{f:Gmatrixtilde}
\end{equation}}%
and $\matrgk{\widetilde{\Gamma}}^{i}$ is a matrix whose $(l,l')$ entry is given by:
{ \begin{equation}
\widetilde{\Gamma}_{ll'}^{i}=\sum_{j=1}^{n_{meas}}\Bigl({\frac{\partial \vect{\varepsilon}_{ij}}{\partial \widetilde{q}_{il}}}\Bigr)^\mathsf{T}\frac{\partial \vect{\varepsilon}_{ij}}{\partial \widetilde{q}_{il'}}
\label{f:gammatilde}
\end{equation}}%
The partial derivatives of the residual with respect to the local track parameters, ${\partial \vect{\varepsilon}_{ij}}/{\partial \vect{\widetilde{q}}_{i}}$, for Eqs.(\ref{f:Gmatrixtilde}) (\ref{f:gammatilde}), are derived from the track fitting algorithm (e.g. the GBL algorithm) without multiple scattering.
Hence, the local track parameters, ${\Delta}\vect{\widetilde{q}}_{i}$, which represent the $i$-th track trajectory change by the alignment, are obtained from Eq.(\ref{f:bitilde}) as:
{ \begin{equation}
{\Delta}\vect{\widetilde{q}}_{i}=\bigl[-(\matrgk{\widetilde{\Gamma}}^{i})^{-1}(\matr{\widetilde{G}}^{i})^\mathsf{T}\bigr]{\Delta}\vect{p}=\matr{\widetilde{H}}^{i}{\Delta}\vect{p}
\label{dqtilde}
\end{equation}}%
where ${\Delta}\vect{\widetilde{q}}_{i}=\bigl(\Delta{\rho}_{i}=\Delta\widetilde{R}_{i}^{-1},{\Delta}\widetilde{q}_{i2},{\Delta}\widetilde{q}_{i3},{\Delta}\widetilde{q}_{i4},{\Delta}\widetilde{q}_{i5}\bigr)^\mathsf{T}$ has only 5 parameters, much fewer than ${\Delta}\vect{q}_{i}$ with multiple scattering appearing in Eqs.(\ref{f:bi}) (\ref{f:di2}), and the matrix $\matr{\widetilde{H}}^{i}$ is given by $\matr{\widetilde{H}}^{i}=-(\matrgk{\widetilde{\Gamma}}^{i})^{-1}(\matr{\widetilde{G}}^{i})^\mathsf{T}$.
As $\Delta{\rho}_{i}={\Delta}\widetilde{q}_{i1}$ in Eq.(\ref{dqtilde}),  the partial derivative of the curvature change with respect to the $g$-th global alignment parameter, ${\partial {\rho}_{i}}/{\partial {p}_{g}}$, is the $(1,g)$ entry of $\matr{\widetilde{H}}^{i}$:
{ \begin{equation}
\frac{\partial {\rho}_{i}}{\partial {p}_{g}}=\matr{\widetilde{H}}^{i}(1,g)
\end{equation}} 

The variance of ${\rho}_{i}$, namely $Z_{i}$, present in both $\vect{d'}^{i}$ of Eq.(\ref{f:dg2}) and $\matr{C'}^{i}$ of Eq.(\ref{f:Cmatrix2}), is also needed for the alignment.
As inferred from Eq.(\ref{dqtilde}), $Z_{i}$ can be interpreted as the error propagation from a given covariance matrix of $\Delta\vect{p}$ denoted by $\matr{\widetilde{V}}^{\Delta\vect{p}}$, as:
{ \begin{equation}
Z_{i}=\bigl[\matr{\widetilde{H}}^{i}\matr{\widetilde{V}}^{\Delta\vect{p}}(\matr{\widetilde{H}}^{i})^\mathsf{T}\bigr](1,1)
\end{equation}}%
Each layer alignment translation parameter can be assigned an error, $\widetilde{\sigma}$, for the calculation of $\matr{\widetilde{V}}^{\Delta\vect{p}}$ as $\matr{\widetilde{V}}^{\Delta\vect{p}}=\matr{\widetilde{V}}(\widetilde{\sigma})$ and propagated to  $Z_{i}(\widetilde{\sigma})$ as:
{ \begin{equation}
Z_{i}(\widetilde{\sigma})=\bigl[\matr{\widetilde{H}}^{i}\matr{\widetilde{V}}(\widetilde{\sigma})(\matr{\widetilde{H}}^{i})^\mathsf{T}\bigr](1,1)
\label{f:zisigma}
\end{equation}}%
Note that $Z_{i}$ is set via $\widetilde{\sigma}$ rather than itself merely for the sake of understanding: for instance, $Z_{i}(\widetilde{\sigma})$ with $\widetilde{\sigma}=10~\mathrm{\upmu m}$ is equal to the curvature variance (squared error) arising from a position uncertainty of $10~\mathrm{\upmu m}$ on each tracker layer.
The assignment of $\widetilde{\sigma}$ passing to $Z_{i}(\widetilde{\sigma})$ should be optimized to attain the best alignment precision as discussed below in section \ref{staticalignstatergy}. 

\subsection{Alignment data sample}
Most of the collected cosmic-ray events are at low rigidities, below 10~GV~\cite{PROTONAMS2015}\cite{HELIUMAMS2015}. To achieve micron level alignment accuracy for each sensor, the alignment will require billions of cosmic-ray events to overcome the multiple scattering arising from the detector materials, especially the large amounts between the external layers and inner tracker ($\sim$0.3~$X_{0}$ between L1 and L2 and $\sim$0.2~$X_{0}$ between L8 and L9).

Multiple scattering decreases linearly with increasing rigidity~\cite{multiplescattering}.
By selecting the latitude and longitude where the minimal geomagnetic cutoff~\cite{geomagneticcutoff}\cite{IGRF12paper} in the AMS field of view is greater than $7.6$~GV, the number of events at rigidities below 10~GV is reduced to 5$\%$; while $\sim$40$\%$ of the high rigidity ($>30$~GV) events are kept for the alignment.

In the static alignment data sample, there are 1.6 billion cosmic-ray events, which corresponds to the full AMS dataset from May 2011 to January 2015 (over 3.5 years period).
The track information from all those events is filled into one matrix to solve all the alignment parameters in one step (see sections \ref{globalalignment} and \ref{globalalignment2}).
Owing to the massive amount of data used, the statistical error in the alignment is negligible.

\subsection{Alignment procedure}\label{staticalignstatergy}
After the previous dynamic alignment, the external tracker layers have been aligned with respect to the inner tracker. Next, the modules from the external layers and inner tracker can be aligned together to reduce the overall misalignment. In particular, the positions of the external layers in the ladder or sensor level can help to improve the alignment precision of the inner tracker.

The developed global alignment approach as discussed in sections \ref{compositealign}, \ref{detectorcons}, and \ref{globalalignment2} is applied for the static alignment.
The GBL algorithm with multiple scattering and with free curvature ($1/R$) track fitting is used to derive the residuals $\vect{\varepsilon}_{ij}(\vect{q}_{i}^{0},\vect{p}^{0})$, the partial derivatives with respect to the local track parameters of the residuals ${\partial \vect{\varepsilon}_{ij}}/{\partial \vect{q}_{i}}$ and the scattering angles ${\partial \vect{\beta}_{ij}}/{\partial \vect{q}_{i}}$, for Eqs.(\ref{f:Gmatrix})(\ref{f:bii})(\ref{f:gamma})(\ref{f:dg2}).
The GBL algorithm without multiple scattering and with free curvature track fitting is used to derive the partial derivatives of the residuals with respect to the local track parameters ${\partial \vect{\varepsilon}_{ij}}/{\partial \vect{\widetilde{q}}_{i}}$ for Eqs.(\ref{f:Gmatrixtilde})(\ref{f:gammatilde}). 

\subsubsection{Alignment validation with Monte Carlo}\label{issalignvalid}
An isotropic cosmic-ray Monte Carlo sample produced by Geant4~\cite{GEANT42003}\cite{PROTONAMS2015}\cite{HELIUMAMS2015} is used to validate the static alignment.
All the tracker modules in the MC are randomly displaced by Gaussian sampling using the displacement parameters similar to the flight data.
The static alignment (see Eq.(\ref{f:alignchisiss})) accuracy is optimized by varying the curvature constraint, namely $\widetilde{\sigma}$, which defines the curvature variance $Z_{i}(\widetilde{\sigma})$ for the alignment as shown in Eq.(\ref{f:zisigma}). 

\begin{figure}[htpb]
  \centering
  \includegraphics[width=0.6\textwidth]{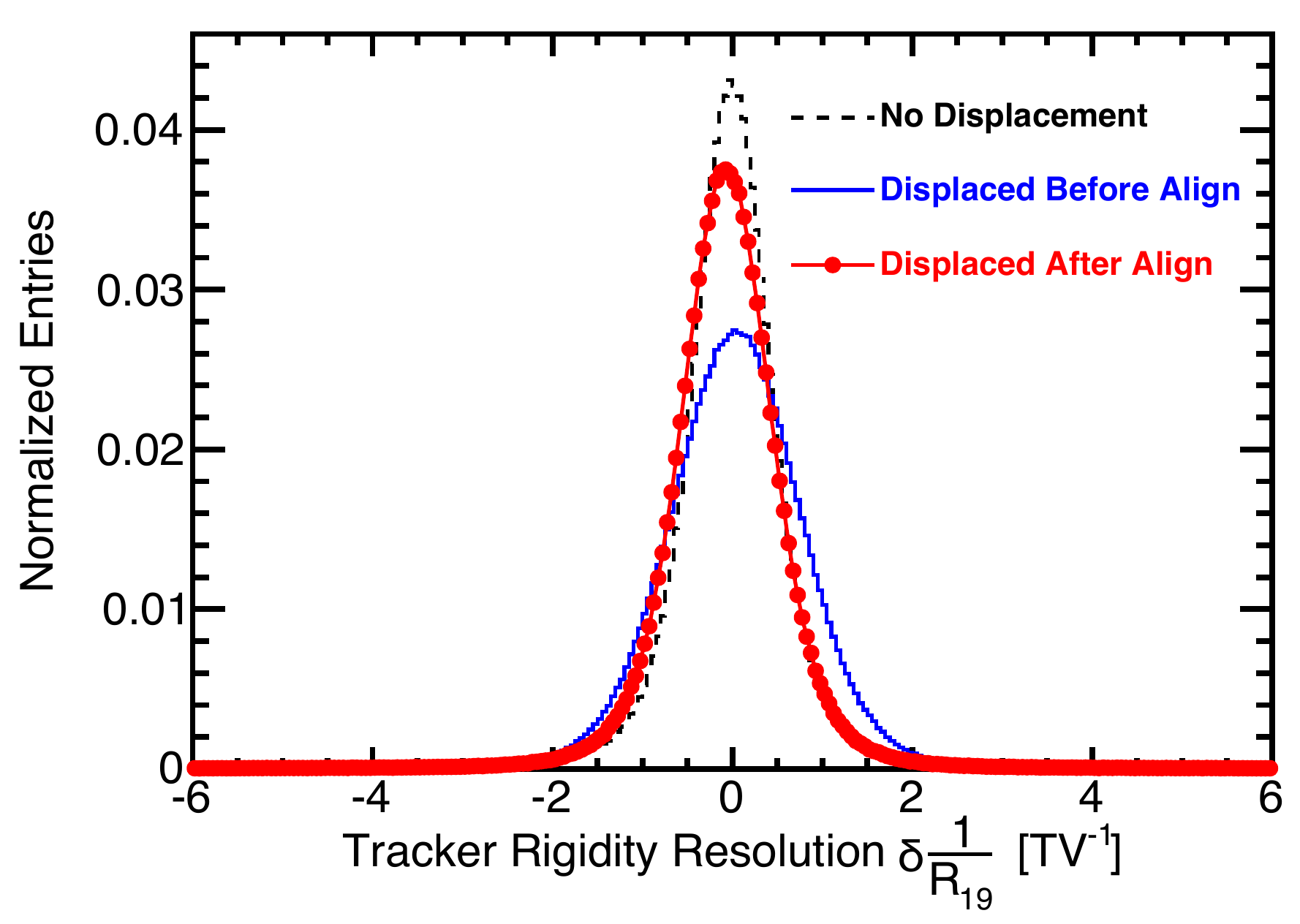}
  \caption{The distributions of the proton full-span rigidity resolution (${\delta}R_{19}^{-1}$) at 1.5~TV for the MC samples with no tracker module displacement (dashed histogram), displaced modules before alignment (solid histogram), and displaced modules after alignment using $\widetilde{\sigma}=200~\mathrm{\upmu m}$ (full circle histogram).}
  \label{rigresomc}
\end{figure}
\begin{figure}[htpb]
  \centering
  \includegraphics[width=0.55\textwidth]{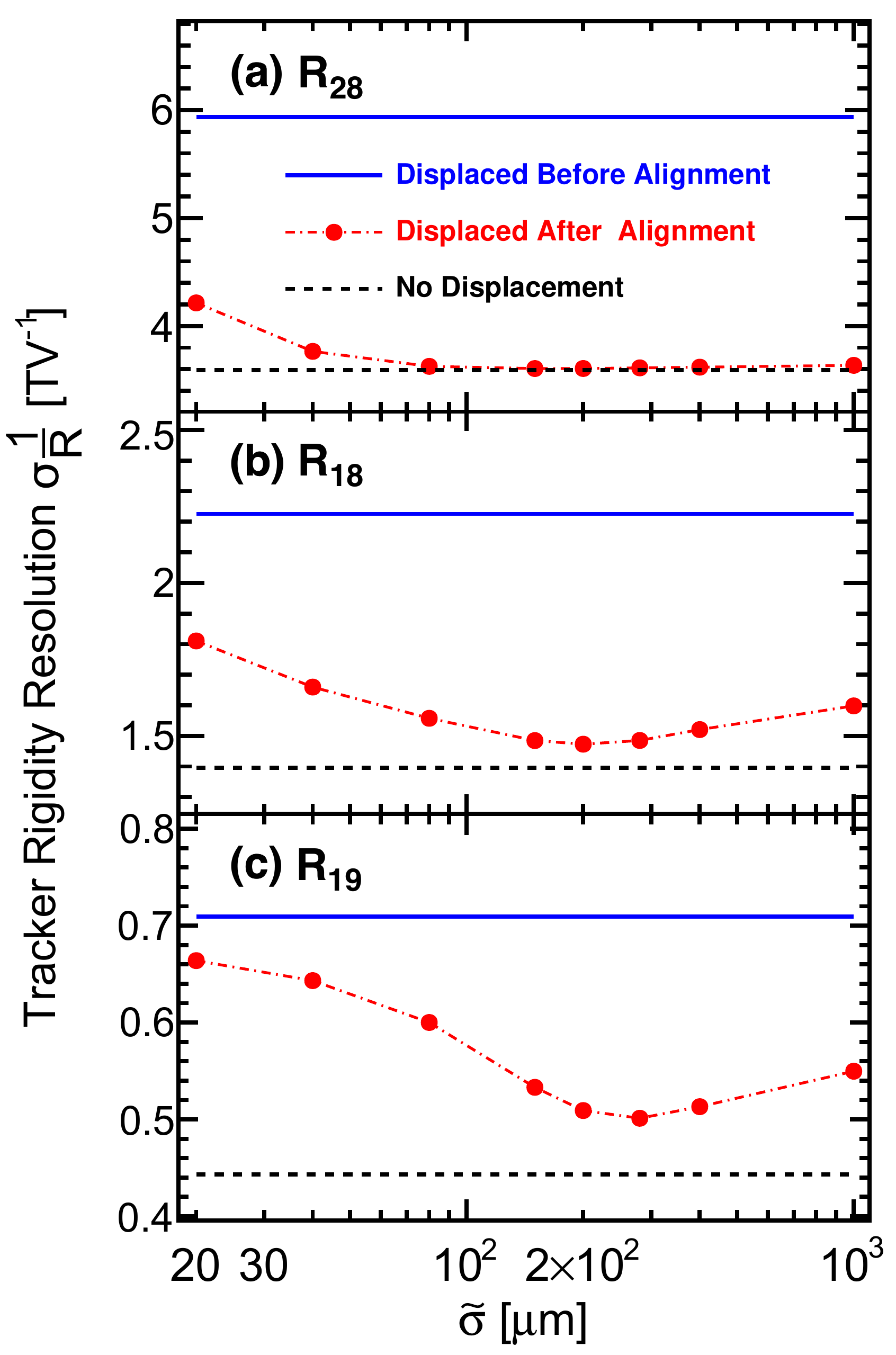}
  \caption{The proton rigidity resolutions $\sigma(1/R)$ of (a) the inner tracker ($R=R_{28}$), (b) L1 and the inner tracker ($R=R_{18}$), and (c) the full-span tracker ($R=R_{19}$) at 1.5~TV as functions of the curvature constraint $\widetilde{\sigma}$, obtained from the alignment on the MC with the tracker modules displaced (full circles and dot-dashed curves). The rigidity resolutions for no module displacement (dashed lines) and displaced modules before alignment (solid lines) are also shown.} 
  \label{rigresosigmamc}
\end{figure}

Figure~\ref{rigresomc} shows the distributions of the proton full-span rigidity resolution (${\delta}R_{19}^{-1}$) at 1.5~TV for no module displacement (dashed histogram), displaced modules before alignment (solid histogram), and displaced modules after alignment with $\widetilde{\sigma}=200~\mathrm{\upmu m}$ (full circle histogram).
As seen, the developed alignment procedure is capable of restoring most of the smeared rigidity resolution.
Note that the small shift in the mean of the measured rigidity will be precisely corrected by using the rigidity-scale determination procedure in section \ref{scaledetermine}. 
Figure~\ref{rigresosigmamc} shows the proton rigidity resolutions of (a) the inner tracker ($R_{28}$), (b) L1 and the inner tracker ($R_{18}$), and (c) the full-span tracker ($R_{19}$) as functions of the curvature constraint $\widetilde{\sigma}$ (full circles and dot-dashed curves).
As seen, in the alignment, the optimal values of $\widetilde{\sigma}$ that derive the best rigidity resolutions, are $\sim$150~$\mathrm{\upmu m}$ for $R_{28}$, $\sim$200~$\mathrm{\upmu m}$ for $R_{18}$, and $\sim$280~$\mathrm{\upmu m}$ for $R_{19}$.
It is clear that the curvature constraint $\widetilde{\sigma}$ should be neither too tight as that will force no track curvature change before and after alignment, nor too loose as that will result in arbitrary change of the curvature in the alignment. Compared with a typical tracker intrinsic position resolution of $\sim$10$~\mathrm{\upmu m}$, $Z_{i}(\widetilde{\sigma})$ with $\widetilde{\sigma}\sim200~\mathrm{\upmu m}$ is a rather loose variance, which is equal to a curvature variance transformed from a position uncertainty of $\sim$200$~\mathrm{\upmu m}$ on each tracker layer.

\subsubsection{Alignment optimization for the flight data}
As shown in the MC study (section \ref{issalignvalid}), the alignment precision is sensitive to the curvature variance, $Z_{i}(\widetilde{\sigma})$, used in Eq.(\ref{f:alignchisiss}), which also needs to be derived from the flight data.
The primary goal for the alignment is to improve the track curvature ($1/R$) measurement precision, i.e. to reduce the curvature bias.
Residuals cannot be used for the study of the curvature misalignment as the curvature bias cannot be seen from the residuals.
However, the curvature bias or rigidity bias is very sensitive to the cosmic-ray flux measurement --- or, more precisely, the rigidity dependence of the cosmic-ray flux measured at high rigidities~\cite{PROTONAMS2015}\cite{HELIUMAMS2015}\cite{TRACKScale}. 
This feature can be exploited to probe the curvature misalignment.

\begin{figure}[htpb]
  \centering
  \includegraphics[width=0.7\textwidth]{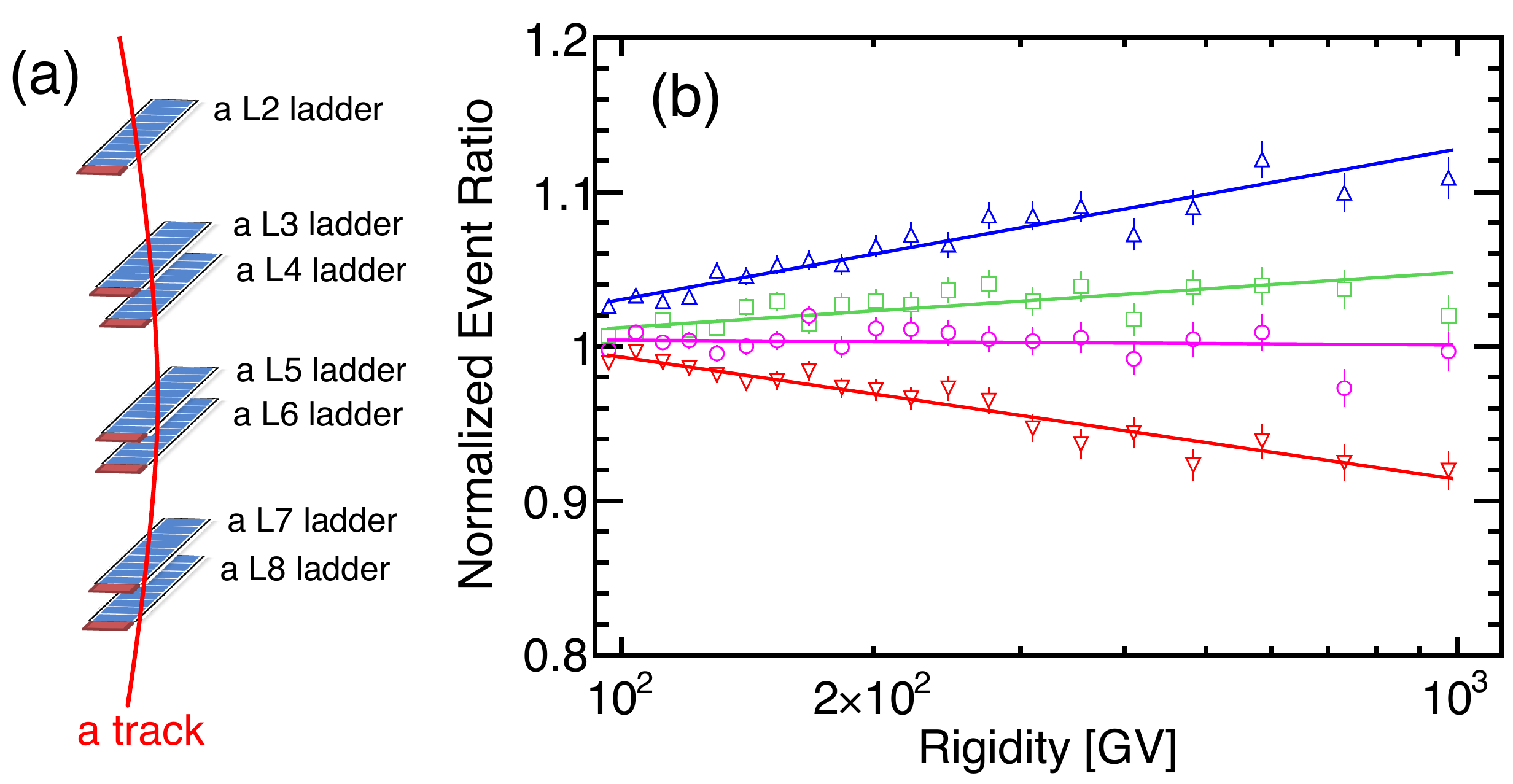}
  \caption{(a) Schematic of a ladder combination of the inner tracker and (b) the slope fits (lines) to the normalized event ratios (${\widehat{n}_{i}}/{\widehat{n}}$) of 4 different ladder combinations with each specified by a set of symbols (up triangles, squares, circles, or down triangles). The different rigidity dependences of the ratios, or the deviation among the slopes, induced by the different curvature biases, are clearly seen.} 
  \label{ladderslopefit}
\end{figure}

As cosmic rays are isotropic, for an ideal tracker without misalignment, the cosmic-ray fluxes measured with a similar pattern in layers but different detector-module combinations, such as different ladder combinations (see one ladder combination illustrated in Fig.~\ref{ladderslopefit} (a)), are expected to be the same.
Therefore, as a result of the differential curvature bias, the deviation of the fluxes or the rigidity dependencies of the event rates (the number of the collected events per second) obtained from different ladder combinations, is used as an estimator of the tracker misalignment.

To display the relative rigidity dependence, the cosmic-ray event rates measured from the $i$-th ladder combination are divided by the event rates measured with the total tracker, denoted by ${n}_{i}/n$.
Then the obtained ${n}_{i}/n$ is normalized by its acceptance fraction ${A}_{i}/A$, as:
{ \begin{equation}
\frac{\widehat{n}_{i}}{\widehat{n}}=\frac{{n}_{i}/n}{{A}_{i}/A}=\frac{{n}_{i}/n}{{\aleph}_{i}/\aleph}
\label{f:evratios}
\end{equation}}%
where ${\aleph}_{i}$ is the total number of events for the $i$-th ladder combination, which sums up all the passing events above 30~GV --- the rigidity region that has no influence from the geomagnetic field~\cite{AMSONISSPART2}; and ${\aleph}_{i}/\aleph$ is the ratio of the total events between the $i$-th ladder combination and the full tracker, which is used to calculate the acceptance fraction as ${A}_{i}/A={\aleph}_{i}/{\aleph}$.

For the $i$-th ladder combination, the normalized event ratio, ${\widehat{n}_{i}}/{\widehat{n}}$,  is fitted over the high rigidity range 90-1000~GV to derive the event-ratio slope $k_{i}$, with:
{ \begin{equation}
\frac{\widehat{n}_{i}}{\widehat{n}}=k_{i}\mathrm{log}(R)+b_{i}
\label{f:evratioslope}
\end{equation}}%
where the slope $k_{i}$ and the intercept $b_{i}$ are the two fitting parameters.

As an illustration, Fig.~\ref{ladderslopefit} (b) shows the slope fits to the normalized event ratios of 4 different ladder combinations.
A clear deviation among the event-ratio slopes of different ladder combinations is seen.
For each track pattern of L2-L8, L1-L8, or L1-L9, the standard deviation of the event-ratio slopes from the 1000 most populated ladder combinations (i.e. with the largest number of passing events) is used as a gauge to evaluate the misalignment.

\begin{figure}[htpb]
  \centering
  \includegraphics[width=0.55\textwidth]{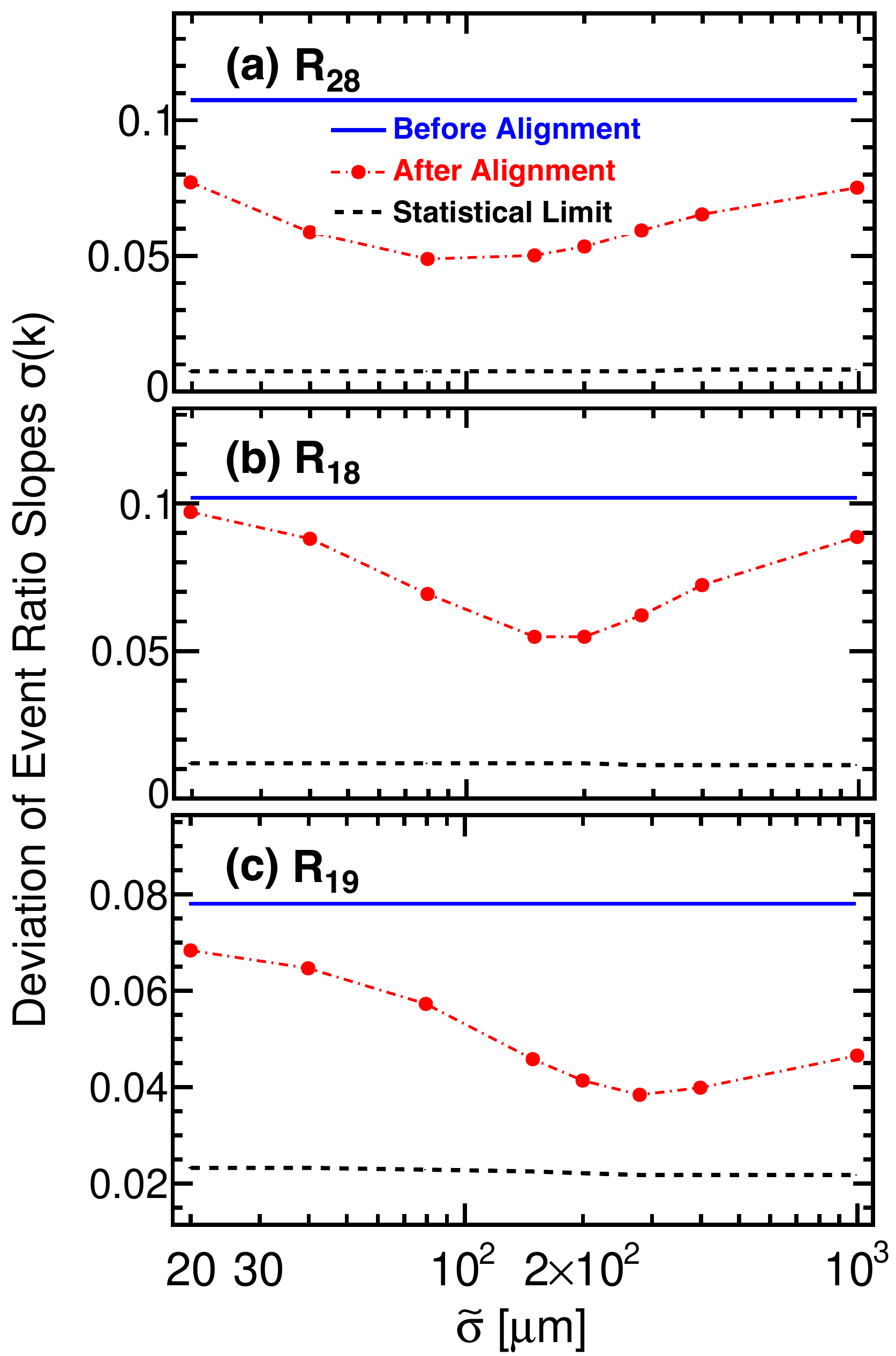}
  \caption{The standard deviations of the cosmic-ray event-ratio slopes (flux rigidity dependences), $\sigma({k})$, as functions of the curvature constraint $\widetilde{\sigma}$ for the ladder combinations of (a) the inner tracker ($R_{28}$), (b) L1 and the inner tracker ($R_{18}$), and (c) the full-span tracker ($R_{19}$), obtained from the static alignment on the flight data (full circles and dot-dashed curves). 
The deviations of the slopes before the static alignment (solid lines) and the statistical limits due to the slope uncertainties arising from the limited number of cosmic-ray events at high rigidities (dashed lines) are also shown. }
  \label{slopesigiss}
\end{figure}

Figure~\ref{slopesigiss} shows the standard deviations of the event-ratio slopes, $\sigma({k})$, as functions of the alignment used $\widetilde{\sigma}$ for the ladder combinations of (a) the inner tracker ($R_{28}$), (b) L1 and the inner tracker ($R_{18}$), and (c) the full-span tracker ($R_{19}$).
As seen, the optimal values of $\widetilde{\sigma}$ for the flight data that have the minimal curvature misalignment, are $80-150~\mathrm{\upmu m}$ for $R_{28}$, $150-200~\mathrm{\upmu m}$ for $R_{18}$, and $\sim$280~$\mathrm{\upmu m}$ for $R_{19}$, which are consistent with the previous estimation from the MC (section \ref{issalignvalid}).
Taking all the track patterns (L2-L8, L1-L8, and L1-L9) into account, the curvature constraint of $\widetilde{\sigma}=200~\mathrm{\upmu m}$ is chosen for the static alignment.
As shown in the figure, after the static alignment, the quality of the rigidity measurement or the rigidity resolution has been significantly improved. There is also no misalignment of the residuals after this step. 
However, a small remaining misalignment of the curvature still exists and is further reduced by the 2nd static alignment performed afterwards using the curvature alignment approach introduced below. 

\subsubsection{Refinement with the curvature alignment}\label{curvatirealign}
In the 2nd static alignment, the alignment corrections obtained from the 1st static alignment are applied.
Different from the 1st static alignment, which was using zero mean for the curvature constraint as ${\rho}_{i}(\vect{p})={\rho}_{i}(\vect{p}^{0})+\sum_{g'}{\frac{\partial {\rho}_{i}}{\partial p_{g'}}}{\Delta}p_{g'}$ with ${\rho}_{i}(\vect{p}^{0})=0$ in Eq.(\ref{f:alignchisiss}), the 2nd static alignment, namely the curvature alignment, uses the curvature bias ${\rho}_{i}(\vect{p}^{0})$ estimated from the data to further improve the result.
The method to obtain ${\rho}_{i}(\vect{p}^{0})$ is based on the isotropic property of cosmic-ray fluxes, i.e. the same rigidity dependence of the cosmic-ray event rates measured with the different detector-module combinations.

In the $j$-th rigidity bin $[R_{j},R_{j+1}]$, the event rate, $n_{j}={N_{j}}/{T}$ (the number of the events per second), measured from the total tracker which has a small curvature misalignment of $\rho$, can be described by:
{ \begin{equation}
n_{j}(\rho)=\int_{R_{j}}^{R_{j+1}}\frac{d{R}}{R^{2}}\int_{0}^{\infty} \Phi(R_{0})A(R_{0})M\Bigl(R_{0},\frac{1}{R}-\frac{1}{R_{0}}+\rho\Bigr)d{R_{0}}
\end{equation}}% 
where $1/R+\rho$ and $1/R$ are the measured inverse rigidities with and without the curvature bias respectively, $R_{0}$ is the true rigidity before detector resolution smearing, $\Phi(R_{0})$ is the cosmic-ray flux, $A(R_{0})$ is the acceptance of the tracker, and $M(R_{0},{1}/{R}-{1}/{R_{0}}+\rho)$ is the probability density function of the tracker rigidity resolution for a given true rigidity $R_{0}$ expressed as a function of $1/R-1/R_{0}+\rho$.
The total tracker is assumed to have no curvature bias as $\rho=0$.
With $A$ and $M$ parameterized from the MC simulation, the parameterization of $\Phi$ is obtained from the fit to the event rates measured with the total tracker.

In the $j$-th rigidity bin, the ratio of the event rate of the $i$-th detector-module combination, $n_{ij}$, to the total event rate, $n_{j}$, is:
{\begin{align}
\frac{n_{ij}}{n_{j}}=\frac{f_{i}n_{j}(\rho=\rho_{i})}{n_{j}(\rho=0)}
=\frac{f_{i}\int_{R_{j}}^{R_{j+1}}\frac{d{R}}{R^{2}}\int_{0}^{\infty} \Phi(R_{0})A(R_{0})M\bigl(R_{0},\frac{1}{R}-\frac{1}{R_{0}}+\rho_{i}\bigr)d{R_{0}}}{\int_{R_{j}}^{R_{j+1}}\frac{d{R}}{R^{2}}\int_{0}^{\infty} \Phi(R_{0})A(R_{0})M\bigl(R_{0},\frac{1}{R}-\frac{1}{R_{0}}\bigr)d{R_{0}}} \label{f:curvealign}
\end{align}}%
where $\rho_{i}$ is the curvature bias of the $i$-th detector-module combination,
$f_{i}=A_{i}/A$ is the constant acceptance ratio of the $i$-th detector-module combination to the total tracker,
and $n_{ij}=A_{i}/A{\cdot}n_{j}(\rho=\rho_{i})=f_{i}n_{j}(\rho=\rho_{i})$.  
From the fit of Eq.(\ref{f:curvealign}) to the event-rate ratio at high rigidity bins (90-1000~GV as in Fig.~\ref{ladderslopefit} (b)), the curvature bias $\rho_{i}$ is obtained.
 
After the 1st static alignment, 2500 ladder combinations of the inner tracker (86$\%$ of the total sample), 4000 ladder combinations of L1 and the inner tracker (88$\%$ of the total sample),
and 2000 ladder combinations of the full-span tracker (77$\%$ of the total sample) are estimated for their remaining curvature biases, which are used as the curvature reference ${\rho}_{i}(\vect{p}^{0})$ in Eq.(\ref{f:alignchisiss}) for the 2nd static alignment. 

\begin{figure}[htpb]
  \centering
  \includegraphics[width=0.75\textwidth]{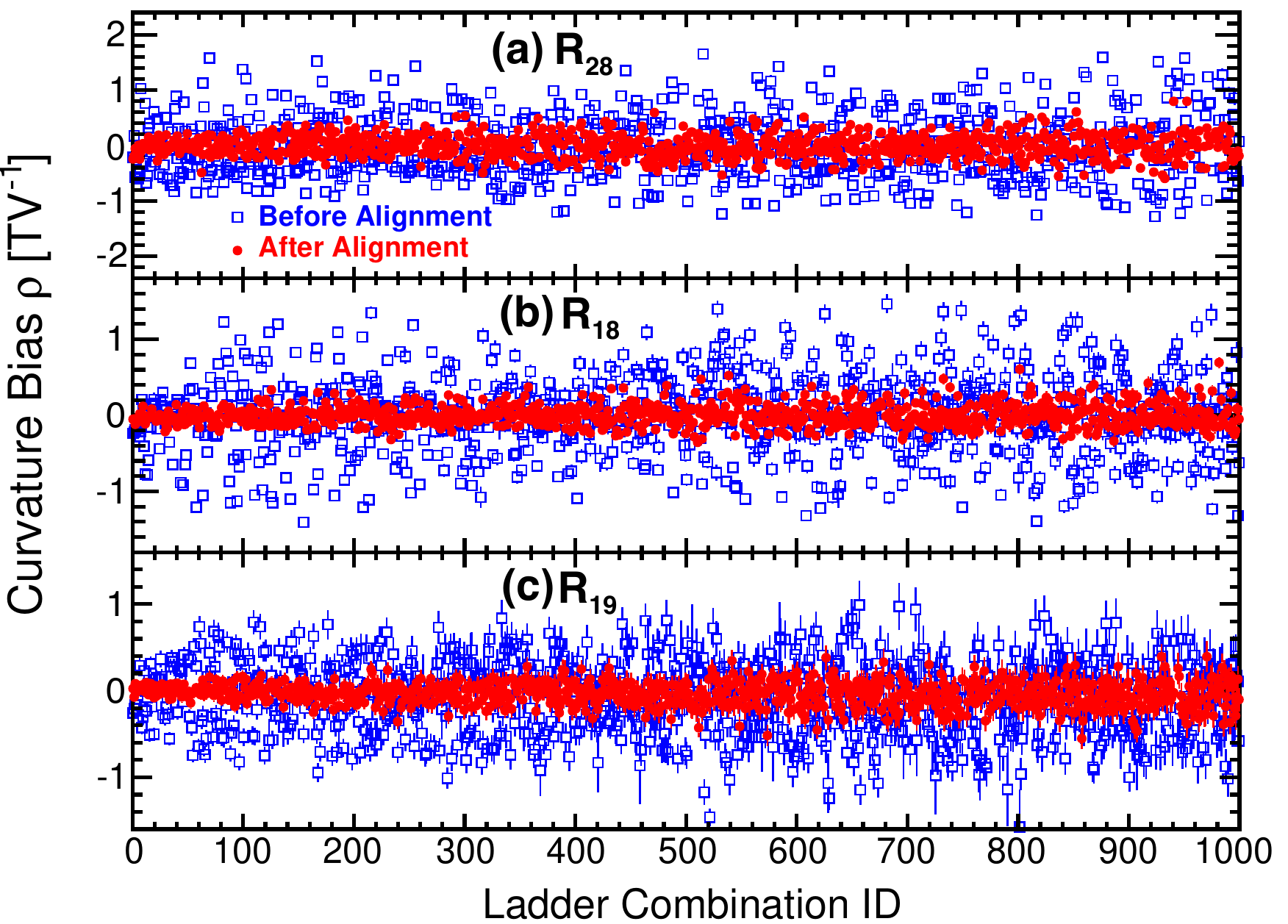}
  \caption{The curvature biases of the 1000 most populated ladder combinations for the individual track patterns of (a) L2-L8 ($R_{28}$), (b) L1-L8 ($R_{18}$), and (c) L1-L9 ($R_{19}$) before the static alignment (open squares) and after the full static alignment (full circles).}
  \label{curvbaisladder}
\end{figure}
\begin{figure}[htpb]
  \centering
  \includegraphics[width=0.65\textwidth]{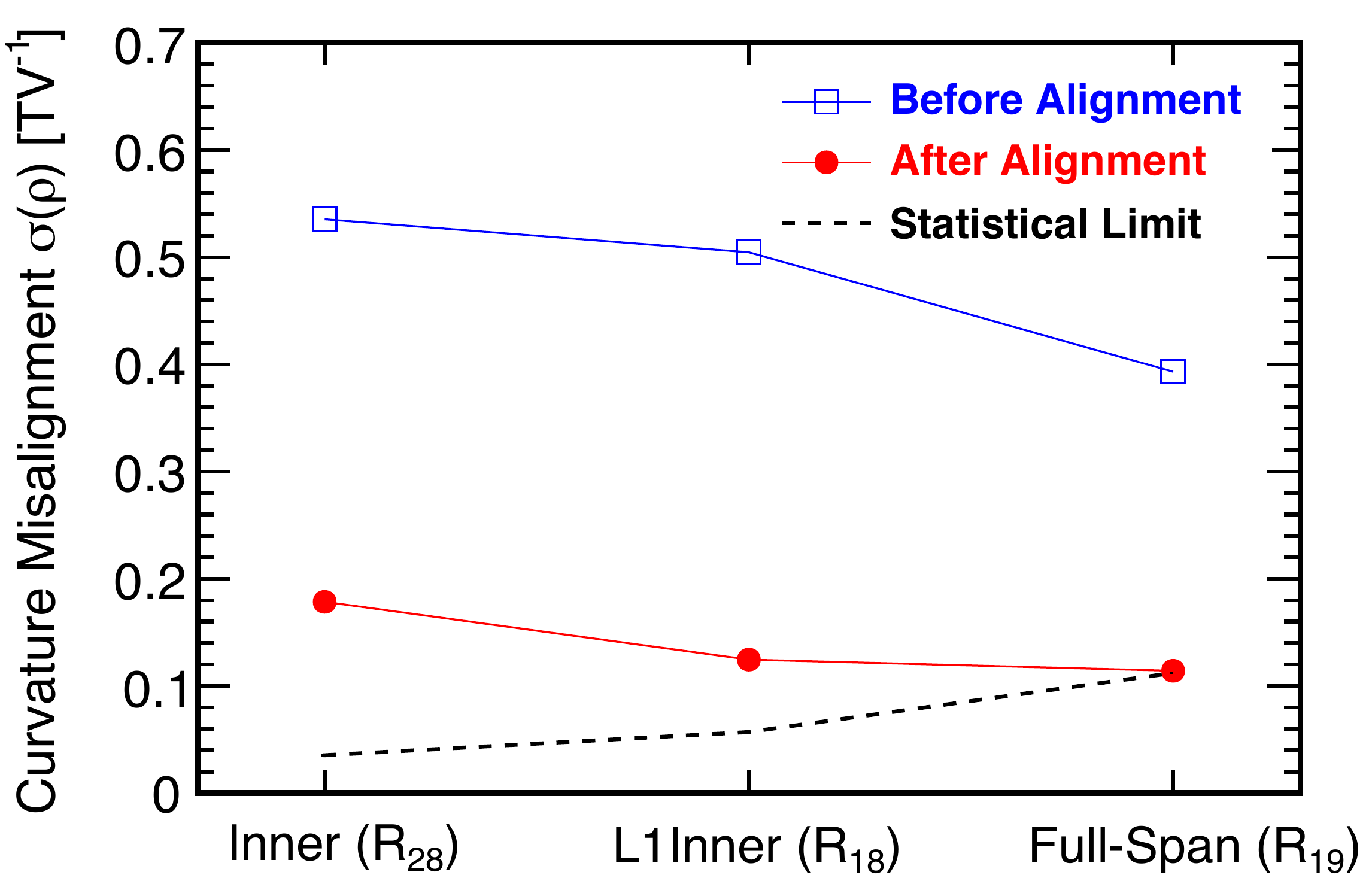}
  \caption{The curvature misalignments, defined as the standard deviations of curvature biases of the 1000 most populated ladder combinations, $\sigma(\rho)$, for the track patterns of the inner tracker ($R_{28}$), L1 and the inner tracker ($R_{18}$), and the full-span tracker ($R_{19}$) before (open squares) and after (full circles) the static alignment. The statistical limits (dashed line) due to the uncertainties arising from the limited number of cosmic-ray events at high rigidities in the curvature bias determination are also shown.}
  \label{misaligncurve}
\end{figure}

To illustrate the full (1st and 2nd) static alignment improvement, Fig.~\ref{curvbaisladder} shows the curvature biases of the 1000 most populated ladder combinations ($\sim$60$\%$ of the total sample) for each track pattern (L2-L8, L1-L8, or L1-L9) before the static alignment (open squares) and after the full static alignment (full circles), which are derived from Eq.(\ref{f:curvealign}).
Figure~\ref{misaligncurve} summarizes the curvature misalignments, defined as the standard deviations of curvature biases of the 1000 ladder combinations, $\sigma(\rho)$, for the individual track patterns, together with the statistical limits (dashed line) due to the uncertainties arising from the limited number of cosmic-ray events at high rigidities in the curvature bias determination.
As seen, with the static alignment approach, the misalignment of the tracker has been greatly reduced for all the track patterns.

\subsubsection{Determination of the total absolute rigidity scale}\label{scaledetermine} 
After the previous 2 rounds of static alignment, the tracker becomes homogeneous, i.e. the relative curvature bias from module combination to module combination has vanished.
However, the whole tracker can have an overall curvature bias, or a shift in the total absolute rigidity scale,
which behaves as a coherent shift in the positions of the tracker layers. 
To determine the total absolute rigidity scale in space, a method using cosmic-ray electrons ($e^{-}$) and positrons ($e^{+}$) events to calibrate the detector has been developed.

Similar method to estimate the curvature bias was used in the CMS experiment~\cite{cms2014alignment}. 
The basic idea is to use the property that the deflection curves of the track trajectories in the magnetic field are mirrored between a charged particle and its anti-particle with the same energy.
When a coherent shift in the tracker layers occurs, the measured absolute inverse rigidity, $|1/R|$, will be shifted by a positive (negative) and by a negative (positive) value for $e^{-}$ and $e^{+}$ respectively.
The rigidity scale shift therefore can be evaluated by comparing the $|1/R|$ distributions between $e^{-}$ and $e^{+}$ events with the same energy measured in the AMS electromagnetic calorimeter detector.
To make full use of the collected cosmic-ray $e^{+}$ and $e^{-}$ events with different energies, an unbinned likelihood method was developed. The detailed description of the method is presented in Ref.~\cite{TRACKScale}.

Using this approach, the total rigidity scale is established with an accuracy of $\pm$1/34 TV$^{-1}$ based on 10 years of AMS data, limited mostly by the available positron statistics.
The estimated small correction for the total curvature bias is converted into position offsets of the individual tracker layers~\cite{TRACKScale}, adding to the layer alignment parameters. 

\subsection{Alignment results}
The results of the static alignment are classified into several aspects shown in the following sections.

\subsubsection{Displacements of the tracker modules during launch}

\begin{figure*}[htpb]
  \includegraphics[width=1.\textwidth]{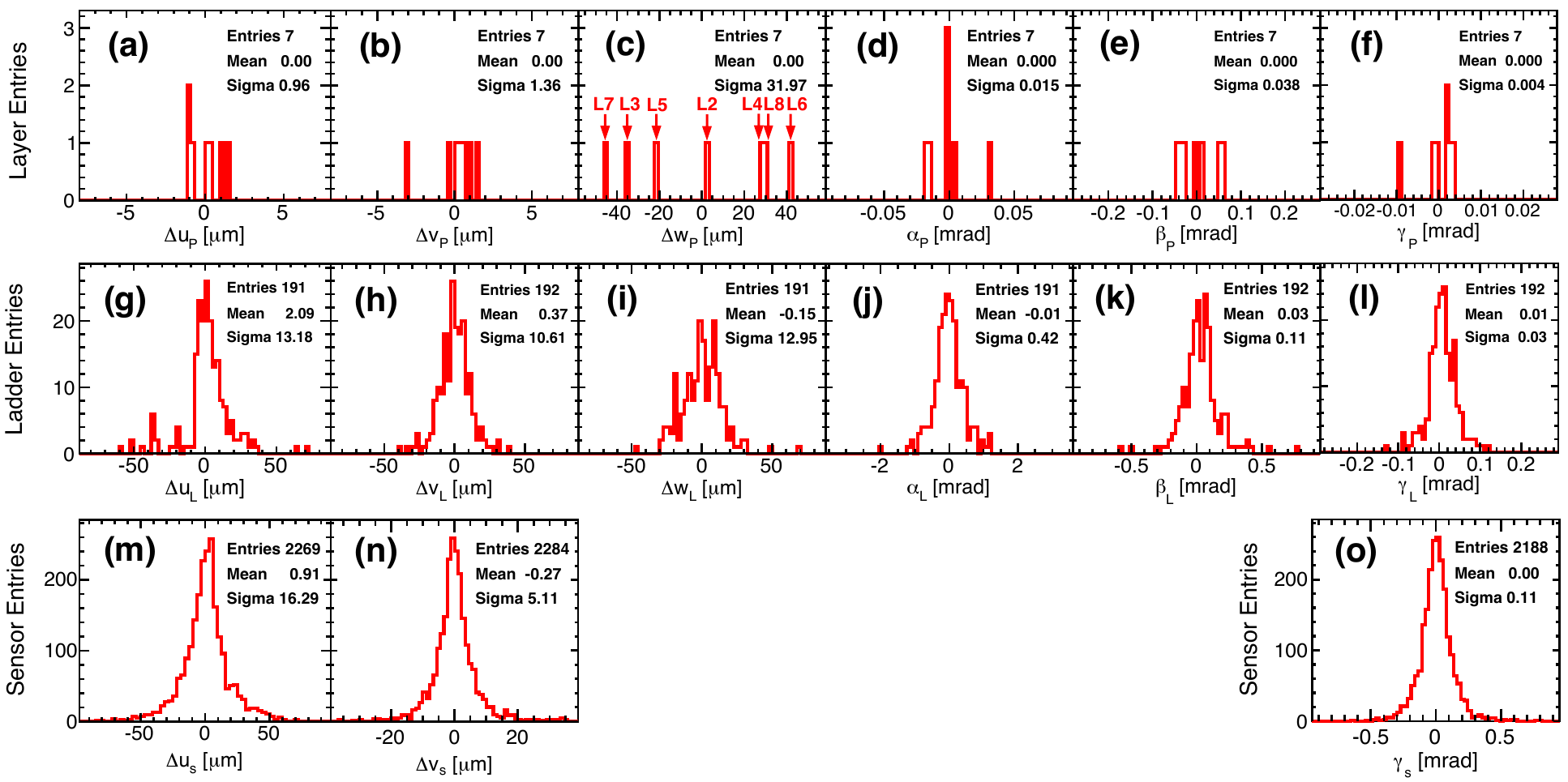}
  \caption{The distributions of the alignment parameters of layers (top row), ladders (middle row), and sensors (bottom row) obtained from the static alignment of all the tracker modules in space. The layer alignment parameters of L1 and L9 are not included in the plots (a-f) as they are dynamically aligned using the position of the inner tracker for the reference.} 
  \label{alignpariss}
\end{figure*}
After the static alignment, we obtain the changes between the positions of the tracker modules in space and those on the ground, which are expressed as the alignment parameters shown in Fig.~\ref{alignpariss}.

As seen in Fig.~\ref{alignpariss} (a-f), the translations of the inner tracker layers are $\sim$1~$\mathrm{\upmu m}$, $\sim$1~$\mathrm{\upmu m}$, and $\sim$32~$\mathrm{\upmu m}$ along the $u_{P}$-, $v_{P}$-, and $w_{P}$-axes ($x$-, $y$-, and $z$-axes) respectively and the rotations are $\sim$0.015~mrad, $\sim$0.04~mrad, and $\sim$0.004~mrad around the $u_{P}$-, $v_{P}$-, and $w_{P}$-axes respectively. 
The translation of $\sim$32~$\mathrm{\upmu m}$ along the $w_{P}$-axis ($z$-axis) can be explained by the outgassing of the support structure, i.e. the foam in the ladder reinforcement frame (see Fig.~\ref{amsladder_fig} (a)), which happened very rapidly under vacuum.
This is confirmed by the fact that the odd and even layers of the inner tracker are shifted in opposite $z$-direction (see Fig.~\ref{alignpariss} (c)), since their ladders are mounted oppositely.
Apart from that, the support structure of the inner tracker planes (the carbon fiber cylinder), exhibits excellent mechanical stability, holding the layers of the inner tracker in place at the micron level through the launch. 

As seen in Fig.~\ref{alignpariss} (g-l), the translations of the ladders are $\sim$13~$\mathrm{\upmu m}$, $\sim$11~$\mathrm{\upmu m}$, and $\sim$13~$\mathrm{\upmu m}$ along the $u_{L}$-, $v_{L}$-, and $w_{L}$-axes respectively and the rotations are $\sim$0.4~mrad, $\sim$0.1~mrad, and $\sim$0.03~mrad around the $u_{L}$-, $v_{L}$-, and $w_{L}$-axes respectively.
The sizable changes of the ladder positions are the major sources of the tracker misalignment in space.

As seen in Fig.~\ref{alignpariss} (m-o), the translations of the sensors are $\sim$16~$\mathrm{\upmu m}$ and $\sim$5~$\mathrm{\upmu m}$ along the $u_{s}$- and $v_{s}$-axes respectively and the rotation is $\sim$0.1~mrad around the $w_{s}$-axis, which are also not small changes.
In particular, the largest translation of $\sim$16~$\mathrm{\upmu m}$ along the $u_{s}$-axis reveals a systematic change of the ladder structure after the launch, that is an increased distance between the adjacent sensors in a ladder. 
The reason might also be related to the deformation of the foam in the ladder reinforcement frame. 

\begin{figure}[htpb]
  \centering
  \includegraphics[width=0.75\textwidth]{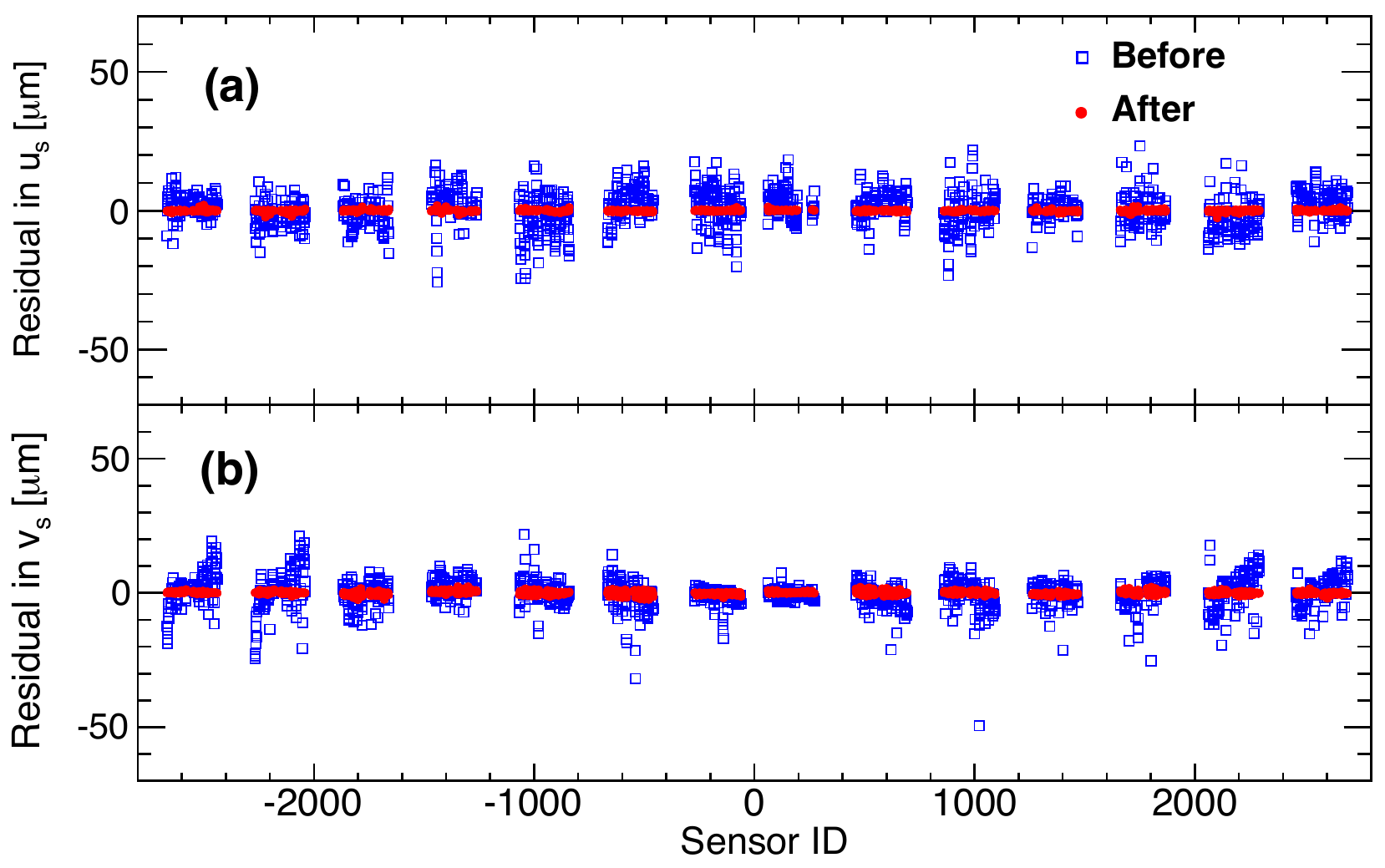}
  \caption{The residual biases of the individual sensors of the inner tracker in (a) the $u_{s}$ direction and (b) the $v_{s}$ direction before (open squares) and after (full circles) the static alignment for a selected cosmic-ray proton sample with rigidity $R>30$~GV based on 10 years of AMS data. A circle or square represents a residual bias of each sensor. The circles or squares of a common group are the sensors from the same half of a tracker layer.}
  \label{alignressensoriss}
\end{figure}
Figure~\ref{alignressensoriss} shows the residual biases of the individual sensors of the inner tracker in the sensor $u_{s}$- and $v_{s}$-directions before and after the static alignment for a selected cosmic-ray proton sample with rigidity $R>30$~GV based on 10 years of AMS data.
Obvious displacements of the tracker modules induced by the launch (before the static alignment) are seen.
After the alignment, there is no bias in the residual of each sensor.

\subsubsection{Stability of the tracker modules in space}
We have also examined the position stability of the inner tracker sensors in space through their residuals over time. 
During the 10 year period, in the microgravity environment, the changes of the sensor positions are found to be very small. 

In order to increase the sensitivity of detecting the tracker movement in space, a similar approach as in section \ref{curvatirealign} is applied to estimate the time dependent rigidity-scale shift of the total tracker, by using that the cosmic-ray flux at high rigidities is constant in time.
The curvature biases, or the rigidity-scale shifts, are measured in 40 time periods of 3 months each 
by fitting the measured event-rate ratios of those periods to the total over 10 years ($n_{i=1-40}$/$n$) with a function similar to Eq.(\ref{f:curvealign}).

\begin{figure}[htpb]
  \centering
  \includegraphics[width=0.7\textwidth]{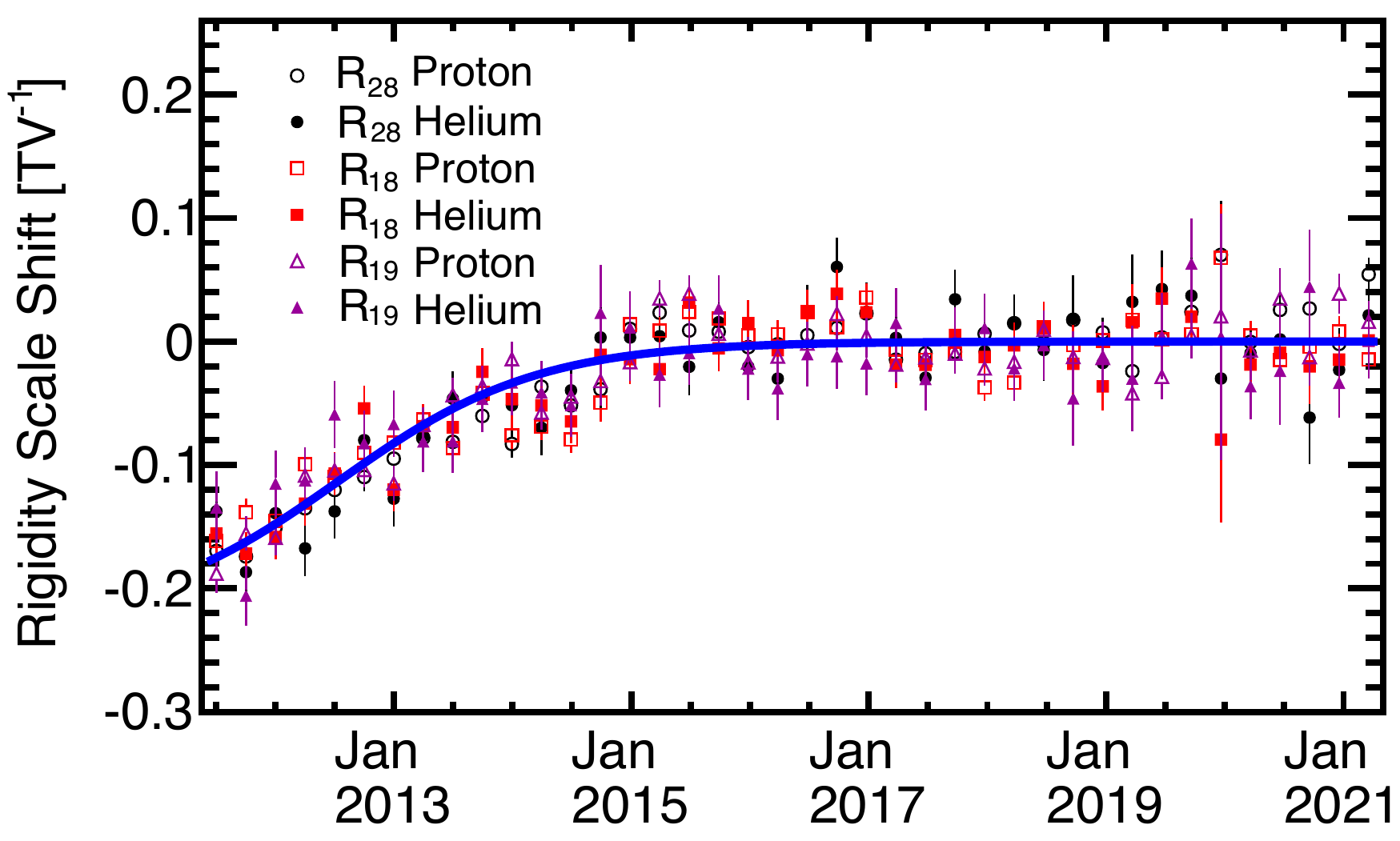}
  \caption{The rigidity-scale shifts as a function of time over 10 years obtained from the event rates of cosmic-ray protons (open symbols) and helium (full symbols) measured using the inner tracker ($R_{28}$, circles), L1 and the inner tracker ($R_{18}$, squares), and the full-span tracker ($R_{19}$, triangles).
The solid curve shows the fit with a logistic function.}
  \label{rigscaletime}
\end{figure}
Figure~\ref{rigscaletime} shows the rigidity-scale shifts as a function of time over 10 years obtained from the event rates of cosmic-ray protons (open symbols) and helium (full symbols) measured using the inner tracker ($R_{28}$, circles), L1 and the inner tracker ($R_{18}$, squares), and the full-span tracker ($R_{19}$, triangles). As seen, the slow shift of the rigidity scale,  or the long-term movement of the inner tracker, is evident before 2015 and progressively decreasing to near zero around 2016.
The amplitude of this movement is fairly small, as the maximum rigidity-scale change of $\sim$0.18~TV$^{-1}$ shown in the figure is equivalent to a displacement of an inner tracker layer of $<$1~$\mathrm{\upmu m}$~\cite{TRACKScale}. 
It is also shown in the figure that the shift of the rigidity measured with the external layers ($R_{18}$ or $R_{19}$) perfectly follows the shift of the rigidity measured with only the inner tracker ($R_{28}$), proving the high stability and reliability of the L1 and L9 dynamic-alignment procedure.

The small correction for the time dependent rigidity-scale shift is converted into position offsets of the individual tracker layers~\cite{TRACKScale}, adding to the layer alignment parameters.
 
\subsubsection{Alignment precision}
\begin{figure}[htpb]
   \centering
  \includegraphics[width=0.65\textwidth]{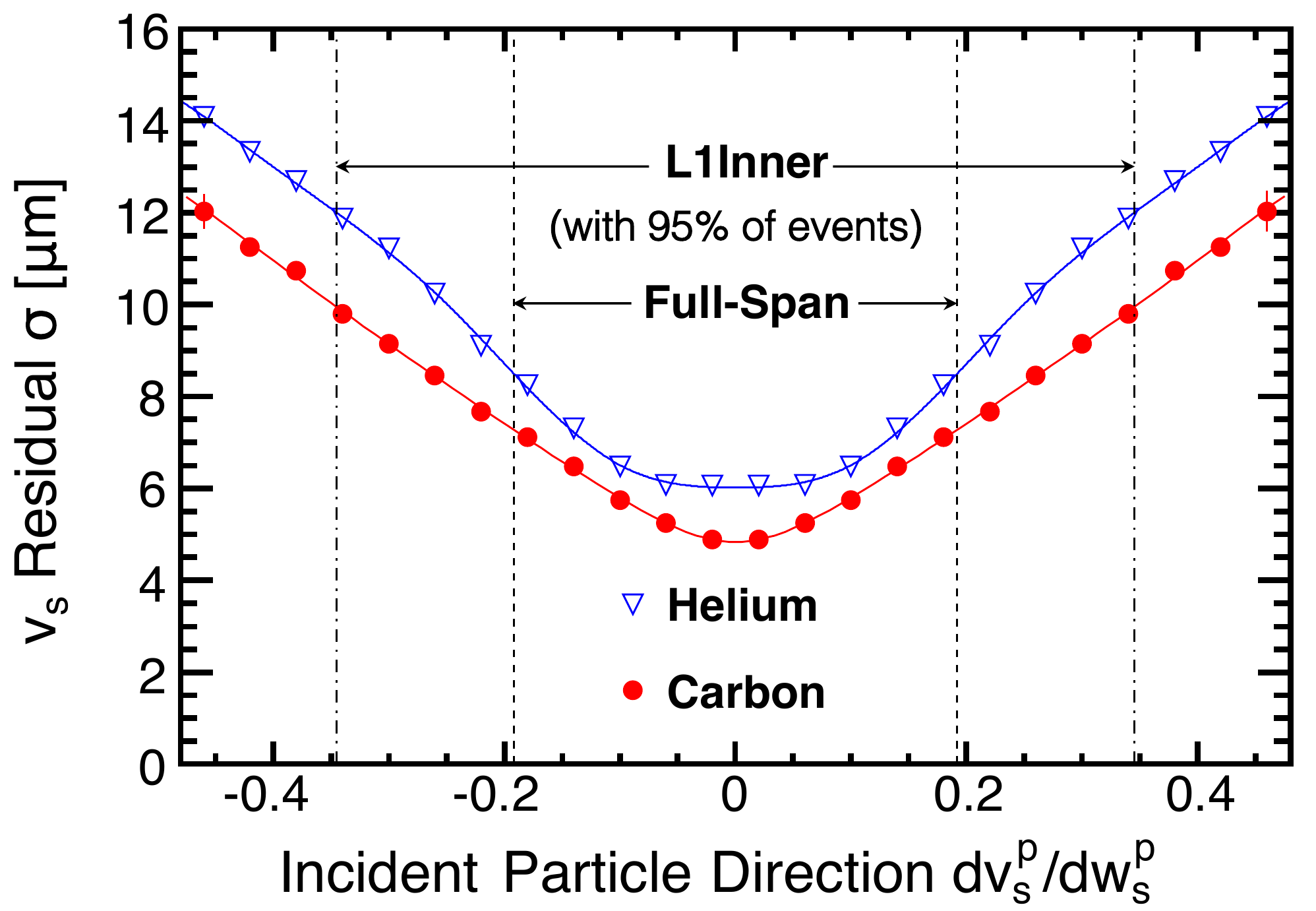}
  \caption{The standard deviation ($\sigma$) of the $v_{s}$ residual (the $v_{s}$ coordinate difference between the measurement from a sensor of L5 and the prediction from the track fit using the other layers),
as functions of the incident particle direction $dv_{s}^{p}/dw_{s}^{p}$ for cosmic-ray helium (triangles) and carbon (full circles) nuclei with rigidities $R>50$~GV. 
The vertical dashed lines indicate the angular boundary of the full-span tracker geometrical acceptance, which includes 95$\%$ of the events.  The vertical dot-dashed lines indicate the same boundary but of the L1-inner tracker acceptance. The intrinsic tracker spatial resolution is predominant in the residual $\sigma$. The average $v_{s}$ coordinate resolutions are 6.5 (7.5)~$\mathrm{\upmu m}$ for helium and 5.1 (5.8)~$\mathrm{\upmu m}$ for carbon in the full-span (L1 and inner) tracker geometry.}
  \label{residualangle}
\end{figure}

After the static alignment, the misalignment in the residual, or incoherent misalignment, is negligible (under a micron as seen in Fig.~\ref{alignressensoriss}) compared with the intrinsic tracker coordinate resolution.
Figure~\ref{residualangle} shows the Gaussian sigma of the $v_{s}$ residual, that is the $v_{s}$ coordinate difference between the measurement from a sensor of L5 and the prediction from the track fit using the other layers, as functions of the incident particle direction in the sensor $v_{s}w_{s}$-plane, $dv_{s}^{p}/dw_{s}^{p}$, for cosmic-ray helium (triangles) and carbon (full circles) nuclei with rigidities $R>50$~GV.
Owing to the precise alignment together with the advanced position finding algorithm~\cite{TRACK2017},
the average $v_s$ or $y$ coordinate resolutions are 6.5 (7.5)~$\mathrm{\upmu m}$ for helium and 5.1 (5.8)~$\mathrm{\upmu m}$ for carbon in the full-span (L1 and inner) tracker geometry. The detailed performance of the AMS tracker coordinate resolutions for all charged particles up to $Q=26$ can be found in Ref.~\cite{TRACK2017}.

Another source of the misalignment in the static alignment is the misalignment of the curvature, or coherent misalignment, which is not visible in the residual and is more crucial.
The curvature misalignment can be split into two parts: (a) the overall curvature bias that will shift the mean of the measured rigidity and (b) the differential curvature bias that will degrade the rigidity resolution. 

The overall curvature bias, or the rigidity scale shift of the total tracker, has been corrected to an accuracy of $\pm$1/34 TV$^{-1}$ by using cosmic-ray electrons and positrons events with the procedure discussed in section \ref{scaledetermine}.

The differential curvature biases for the different combinations of the tracker modules can smear the tracker resolution as shown in the MC study (see Fig.~\ref{rigresomc}).
With the unique alignment approach, most of the smeared rigidity resolution is recovered.
By using the isotropic property of cosmic-ray flux, direct assessment of the misalignment is performed on the data.
As shown in Fig.~\ref{misaligncurve}, after the alignment, the standard deviations of the differential curvature biases among different ladder combinations, are better than 0.18~TV$^{-1}$, 0.125~TV$^{-1}$, and 0.11~TV$^{-1}$ for the rigidities measured using the inner tracker ($R_{28}$), L1 and inner tracker ($R_{18}$), and full-span tracker ($R_{19}$) respectively, which are the misalignments equivalent to additional smearings of the measured position of each layer by less than 0.7~$\mathrm{\upmu m}$, 1.2~$\mathrm{\upmu m}$, and 2.7~$\mathrm{\upmu m}$ for $R_{28}$, $R_{18}$, and $R_{19}$ respectively.
This estimation is based on different ladder combinations and does not include the contribution from the misalignment of the sensors, which cannot be accurately determined from the different sensor combinations due to the limited number of cosmic-ray events per sensor combination at high rigidities.  
Considering that the sensor position change during launch is small, $\sim$5$~\mathrm{\upmu m}$, in the bending direction, based on the MC simulation, we assign an error of $\sim$2$~\mathrm{\upmu m}$ to the sensor misalignment.
So, combining in quadrature, the total differential curvature misalignments equivalent to the position errors of each layer are 2.1~$\mathrm{\upmu m}$, 2.3~$\mathrm{\upmu m}$, and 3.3~$\mathrm{\upmu m}$ for $R_{28}$, $R_{18}$, and $R_{19}$ respectively, which are smaller than both the intrinsic spatial resolution (e.g. 5.1~$\mathrm{\upmu m}$ for carbon nuclei in the full-span geometry) and the alignment errors of the external layers in the dynamic alignment (7.1~$\mathrm{\upmu m}$ for L1 and 7.9~$\mathrm{\upmu m}$ for L9).

\section{Conclusion}
\begin{figure}[htpb]
  \centering
  \includegraphics[width=0.65\textwidth]{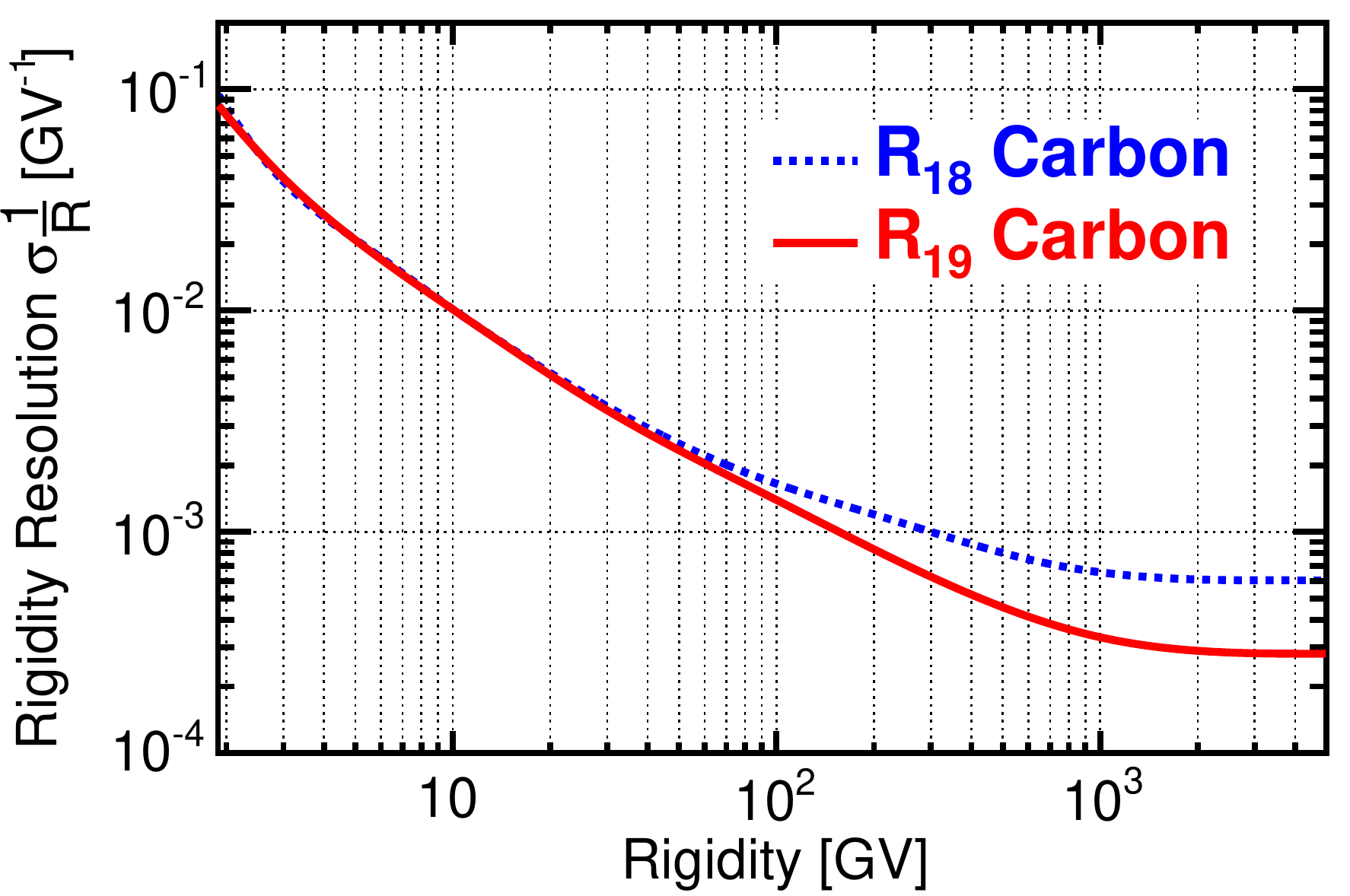}
  \caption{The rigidity resolutions, $\sigma(1/R)$, of L1-inner ($R=R_{18}$) and full-span ($R=R_{19}$) track patterns as functions of the true rigidity for carbon nuclei obtained from MC simulation. The corresponding maximal detectable rigidities, $R^{M}$,  with $R^{M}\sigma(1/R^{M})\equiv1$,
are $R_{18}^{M}=1.6$~TV and $R_{19}^{M}=3.6$~TV.}
  \label{invrigreso}
\end{figure}
Precise alignment of the silicon tracker is invaluable for the success of the AMS mission.
We have presented a series of new methods to align the large permanent magnetic spectrometer for the space experiment, starting from the alignment with the test beam data on the ground through the alignment with the cosmic-ray events in space,
with an ultimate precision of a few microns achieved under harsh conditions.
This allows AMS to accurately measure cosmic rays up to the multi-TV region.
As an example, Fig.~\ref{invrigreso} shows the rigidity resolutions of L1-inner track pattern, $\sigma(1/R_{18})$, and of full-span track pattern, $\sigma(1/R_{19})$, as functions of the true rigidity for carbon nuclei after the full alignment procedure.
The maximal detectable rigidities, $R^{M}$,  with $R^{M}\sigma(1/R^{M})\equiv1$,
are $R_{18}^{M}=1.6$~TV and $R_{19}^{M}=3.6$~TV, correspondingly. 

The developments of the new mathematical alignment algorithms, such as the alignment for the composite detector structure, the alignment for the dynamic system, and the alignment in the presence of the magnetic field, are useful for various HEP experiments equipped with the tracking detectors and particularly valuable for the future spaceborne magnetic spectrometers. 
 
\section*{Acknowledgements}
We acknowledge the continuous support from MIT and its School of Science. We are grateful for the support of the U.S. Department of Energy (DOE), Office of Science. We thank the strong support from CERN IT department.
We thank Dr. Michael Capell for his diligent proofreading of the manuscript. 

\begin{appendix}
\numberwithin{equation}{section}
%\appendixpage
\addappheadtotoc
\stoptocwriting
\section{Coordinate transformation from the local sensor frame to the global tracker frame} \label{appA}
Substituting Eq.(\ref{f:frl}) into Eq.(\ref{f:frp}) gives:
{{\begin{equation}
\begin{split}
\vect{r}_{P}=&\matr{R}^\mathsf{T}_{L}{\Delta}\matr{R}_{L}\bigl[\matr{R}^\mathsf{T}_{s}{\Delta}\matr{R}_{s}(\vect{q}+{\Delta}\vect{q}_{s})+\vect{r}_{0s}+{\Delta}\vect{q}_{L}\bigr]+\vect{r}_{0L} \\
    {\simeq}&\matr{R}^\mathsf{T}_{L}\matr{R}^\mathsf{T}_{s}\bigl[(\matr{R}_{s}{\Delta}\matr{R}_{L}\matr{R}^\mathsf{T}_{s}){\Delta}\matr{R}_{s}\vect{q}+{\Delta}{\vect{q}_{s}}+\matr{R}_{s}{\Delta}\matr{R}_{L}\vect{r}_{0s} 
     +\matr{R}_{s}{\Delta}\vect{q}_{L}\bigr]+\vect{r}_{0L} 
\end{split}
\label{f:frp2}
\end{equation}}}%
Subsequently, substituting Eq.(\ref{f:frp2}) into Eq.(\ref{f:frg}) gives:
{{\begin{align}
\vect{r}_{g}{\simeq}&\matr{R}^\mathsf{T}_{P}\matr{R}^\mathsf{T}_{L}\matr{R}^\mathsf{T}_{s}\bigl[(\matr{R}_{s}\matr{R}_{L}{\Delta}\matr{R}_{P}\matr{R}^\mathsf{T}_{L}\matr{R}^\mathsf{T}_{s})(\matr{R}_{s}{\Delta}\matr{R}_{L}\matr{R}^\mathsf{T}_{s})\Delta\matr{R}_{s}\vect{q}+\Delta\vect{q}_{s}  \nonumber\\
    \quad&+(\matr{R}_{s}\matr{R}_{L}\Delta\matr{R}_{P}\matr{R}^\mathsf{T}_{L}\matr{R}^\mathsf{T}_{s})\matr{R}_{s}\Delta\matr{R}_{L}\vect{r}_{0s}+\matr{R}_{s}\Delta\vect{q}_{L}
    +\matr{R}_{s}\matr{R}_{L}\Delta\matr{R}_{P}\vect{r}_{0L}+\matr{R}_{s}\matr{R}_{L}\Delta\vect{q}_{P}\bigr]+\vect{r}_{0P}%
\intertext{The above equation can be simplified to:}
\vect{r}_{g}{\simeq}&\matr{R}^\mathsf{T}(\vect{q}+\Delta\vect{q})+\vect{r}_{0} \nonumber 
\end{align}}}%
where
{{\begin{align}
\Delta\vect{q}=&\bigl[(\matr{R}_{s}\matr{R}_{L}\Delta\matr{R}_{P}\matr{R}^\mathsf{T}_{L}\matr{R}^\mathsf{T}_{s})(\matr{R}_{s}\Delta\matr{R}_{L}\matr{R}^\mathsf{T}_{s})\Delta\matr{R}_{s}-\matr{E}\bigr]\vect{q}+\Delta\vect{q}_{s}\nonumber \\
    \quad&+\bigl[(\matr{R}_{s}\matr{R}_{L}\Delta\matr{R}_{P}\matr{R}^\mathsf{T}_{L}\matr{R}^\mathsf{T}_{s})\matr{R}_{s}\Delta\matr{R}_{L}-\matr{R}_{s}\bigr]\vect{r}_{0s} \label{f:dq} \\
    &+\matr{R}_{s}\Delta\vect{q}_{L}+\matr{R}_{s}\matr{R}_{L}(\Delta\matr{R}_{P}-\matr{E})\vect{r}_{0L}+\matr{R}_{s}\matr{R}_{L}\Delta\vect{q}_{P}  \nonumber\\
\matr{R}^\mathsf{T}&=\matr{R}^\mathsf{T}_{P}\matr{R}^\mathsf{T}_{L}\matr{R}^\mathsf{T}_{s} \\
\vect{r}_{0}&=\matr{R}_{P}^\mathsf{T}\matr{R}_{L}^\mathsf{T}\vect{r}_{0s}+\matr{R}_{P}^\mathsf{T}\vect{r}_{0L}+\vect{r}_{0P}
\end{align}}}%

\section{${\chi}^2$ minimization and alignment matrix in the global alignment}\label{appB}
Minimization of the ${\chi}^2$ of Eq.(\ref{f:alignchis}) leads to the partial derivative with respect to each ($g$-th) global parameter $\Delta{p}_g$ being zero:
{\begin{equation}
\begin{split}
\frac{\partial {\chi}^{2}}{\partial p_{g}}=&2\sum_{i=1}^{N_{track}}\sum_{j=1}^{n_{meas}}\Bigl({\frac{\partial \vect{\varepsilon}_{ij}}{\partial p_{g}}}\Bigr)^\mathsf{T}\matr{V}_{ij}^{-1}\vect{\varepsilon}_{ij}=0 \\
 {\simeq}&2\sum_{i=1}^{N_{track}}\sum_{j=1}^{n_{meas}}\Bigl({\frac{\partial \vect{\varepsilon}_{ij}}{\partial p_{g}}}\Bigr)^\mathsf{T}\matr{V}_{ij}^{-1}\Bigl[\vect{\varepsilon}_{ij}(\vect{q}_{i}^{0},\vect{p}^{0})
 +\sum_{l'}{\frac{\partial \vect{\varepsilon}_{ij}}{\partial q_{il'}}}{\Delta}q_{il'}+\sum_{g'}{\frac{\partial \vect{\varepsilon}_{ij}}{\partial p_{g'}}}{\Delta}p_{g'}\Bigr] 
 \end{split}
\label{dchisdpg}
\end{equation}}%
where $\vect{\varepsilon}_{ij}{\simeq}\vect{\varepsilon}_{ij}(\vect{q}_{i}^{0},\vect{p}^{0})+\sum_{l'}{\frac{\partial \vect{\varepsilon}_{ij}}{\partial q_{il'}}}{\Delta}q_{il'}+\sum_{g'}{\frac{\partial \vect{\varepsilon}_{ij}}{\partial p_{g'}}}{\Delta}p_{g'}$ depends both on the local track parameters $\Delta\vect{q}_{i}$ and the global alignment parameters $\Delta\vect{p}$, and $\vect{\beta}_{ij}{\simeq}\sum_{l'}{\frac{\partial {\vect\beta}_{ij}}{\partial q_{il'}}}{\Delta}q_{il'}$ as the  intrinsic track property only depends on the local track parameters $\Delta\vect{q}_{i}$.
Eq.(\ref{dchisdpg}) can be further simplified in matrix form as:
{\begin{equation}
\sum_{i=1}^{N_{track}}\vect{d}^{i}=\Bigl(\sum_{i=1}^{N_{track}}\matr{C}^{i}\Bigr){\Delta}\vect{p}+\sum_{i=1}^{N_{track}}\matr{G}^{i}{\Delta}\vect{q}_{i}
\label{f:di}
\end{equation}}%
where $\vect{d}^{i}$ is a vector whose $g$-th element is given by:
{\begin{equation}
d_{g}^{i}=-\sum_{j=1}^{n_{meas}}\Bigl({\frac{\partial \vect{\varepsilon}_{ij}}{\partial p_{g}}}\Bigr)^\mathsf{T}\matr{V}_{ij}^{-1}\vect{\varepsilon}_{ij}(\vect{q}_{i}^{0},\vect{p}^{0})
\label{f:dg}
\end{equation}}%
$\matr{C}^{i}$ is a matrix whose $(g,g')$ entry is given by:
{\begin{equation}
C_{gg'}^{i}=\sum_{j=1}^{n_{meas}}\Bigl({\frac{\partial \vect{\varepsilon}_{ij}}{\partial p_{g}}}\Bigr)^\mathsf{T}\matr{V}_{ij}^{-1}{\frac{\partial \vect{\varepsilon}_{ij}}{\partial p_{g'}}}
\label{f:Cmatrix}
\end{equation}}%
and $\matr{G}^{i}$ is a matrix whose $(g,l')$ entry is given by:
{\begin{equation}
G_{gl'}^{i}=\sum_{j=1}^{n_{meas}}\Bigl({\frac{\partial \vect{\varepsilon}_{ij}}{\partial p_{g}}}\Bigr)^\mathsf{T}\matr{V}_{ij}^{-1}{\frac{\partial \vect{\varepsilon}_{ij}}{\partial q_{il'}}}
\label{f:Gmatrix}
\end{equation}}%
The partial derivatives of the residual with respect to the global alignment parameters, ${\partial \vect{\varepsilon}_{ij}}/{\partial \vect{p}}$, are from Eqs.(\ref{f:resps}) (\ref{f:respl}) (\ref{f:respp}) and with respect to the local track parameters, ${\partial \vect{\varepsilon}_{ij}}/{\partial \vect{q}_{i}}$, are derived from the track fitting algorithm.
In this paper, the track fitting was done with the custom software implementation of the General Broken Lines algorithm~\cite{BGLfitting}.

Minimization of the $\chi^{2}$ of Eq.(\ref{f:alignchis}) leads the partial derivative with respect to each ($l$-th) local track parameter of each ($i$-th) track, $\Delta{q}_{il}$, to equal zero:
{ \begin{equation}
\begin{split}
\frac{\partial {\chi}^{2}}{\partial q_{il}}=&2\sum_{j=1}^{n_{meas}}\Bigl({\frac{\partial \vect{\varepsilon}_{ij}}{\partial q_{il}}}\Bigr)^\mathsf{T}\matr{V}_{ij}^{-1}\vect{\varepsilon}_{ij}+2\sum_{j=2}^{n_{scat}-1}\Bigl({\frac{\partial \vect{\beta}_{ij}}{\partial q_{il}}}\Bigr)^\mathsf{T}\matr{W}_{ij}^{-1}\vect{\beta}_{ij}=0\\
{\simeq}&2\sum_{j=1}^{n_{meas}}\Bigl({\frac{\partial \vect{\varepsilon}_{ij}}{\partial q_{il}}}\Bigr)^\mathsf{T}\matr{V}_{ij}^{-1}\Bigl[\vect{\varepsilon}_{ij}(\vect{q}_{i}^{0},\vect{p}^{0})+\sum_{l'}{\frac{\partial \vect{\varepsilon}_{ij}}{\partial q_{il'}}}{\Delta}q_{il'}
+\sum_{g'}{\frac{\partial \vect{\varepsilon}_{ij}}{\partial p_{g'}}}{\Delta}p_{g'}\Bigr] \\
&+2\sum_{j=2}^{n_{scat}-1}\Bigl({\frac{\partial \vect{\beta}_{ij}}{\partial q_{il}}}\Bigr)^\mathsf{T}\matr{W}_{ij}^{-1}\sum_{l'}{\frac{\partial \vect{\beta}_{ij}}{\partial q_{il'}}}{\Delta}q_{il'}
\end{split}
\label{dchisdqil}
\end{equation}}%
Eq.(\ref{dchisdqil}) can be simplified in matrix form as:
{ \begin{equation}
\vect{b}^{i}=(\matr{G}^{i})^\mathsf{T}{\Delta}\vect{p}+\matrgk{\Gamma}^{i}{\Delta}\vect{q}_{i}
\label{f:bi}
\end{equation}}%
where $\vect{b}^{i}$ is a vector whose $l$-th element is given by:
{ \begin{equation}
b_{l}^{i}=-\sum_{j=1}^{n_{meas}}\Bigl({\frac{\partial \vect{\varepsilon}_{ij}}{\partial q_{il}}}\Bigr)^\mathsf{T}\matr{V}_{ij}^{-1}\vect{\varepsilon}_{ij}(\vect{q}_{i}^{0},\vect{p}^{0})
\label{f:bii}
\end{equation}}%
$(\matr{G}^{i})^\mathsf{T}$ is the transpose of the matrix $\matr{G}^{i}$ which is defined in Eq.(\ref{f:Gmatrix}),
and $\matrgk{\Gamma}^{i}$ is a matrix whose $(l,l')$ entry is given by:
{ \begin{equation}
\Gamma_{ll'}^{i}=\sum_{j=1}^{n_{meas}}\Bigl({\frac{\partial \vect{\varepsilon}_{ij}}{\partial q_{il}}}\Bigr)^\mathsf{T}\matr{V}_{ij}^{-1}\frac{\partial \vect{\varepsilon}_{ij}}{\partial q_{il'}}+\sum_{j=2}^{n_{scat}-1}\Bigl({\frac{\partial \vect{\beta}_{ij}}{\partial q_{il}}}\Bigr)^\mathsf{T}\matr{W}_{ij}^{-1}{\frac{\partial \vect{\beta}_{ij}}{\partial q_{il'}}}
\label{f:gamma}
\end{equation}}%
The partial derivatives of the scattering angle with respect to the local track parameters, ${\partial \vect{\beta}_{ij}}/{\partial \vect{q}_{i}}$, are derived from the track fitting algorithm.

Combining Eq.(\ref{f:di}) and Eq.(\ref{f:bi}), all the global alignment parameters, $\Delta\vect{p}$, and all the local track parameters, $\Delta\vect{q}$, can be solved simultaneously from following matrix equation:
{ \begin{equation*}
\begin{pmatrix}
\sum_{i}\matr{C}^{i} &  \matr{G}^{1} &  \ldots &\matr{G}^{j} & \ldots & \matr{G}^{N} \\
(\matr{G}^{1})^\mathsf{T} & \matrgk{\Gamma}^{1} & \ldots & \matr{0} & \ldots & \matr{0}  \\
\vdots  & \vdots & \ddots & \vdots &\ddots & \vdots\\
(\matr{G}^{j})^\mathsf{T} & \matr{0} & \ldots & \matrgk{\Gamma}^{j} &  \ldots & \matr{0} \\
\vdots & \vdots & \ddots & \vdots & \ddots & \vdots\\
(\matr{G}^{N})^\mathsf{T} & \matr{0} & \ldots& \matr{0} & \ldots & \matrgk{\Gamma}^{N} \\
\end{pmatrix}
\begin{pmatrix}
\Delta\vect{p} \\
\Delta\vect{q}_{1} \\
\vdots\\
\Delta\vect{q}_{j} \\
\vdots\\
\Delta\vect{q}_{N} \\
\end{pmatrix}
=\\
\begin{pmatrix}
\sum_{i}\vect{d}^{i} \\
\vect{b}^{1} \\
\vdots \\
\vect{b}^{j} \\
\vdots \\
\vect{b}^{N} \\
\end{pmatrix}
\end{equation*}}%

\resumetocwriting
\end{appendix}

%\section*{References}

\bibliographystyle{spphys}       % APS-like style for physics
\interlinepenalty=10000
\bibliography{qyantkalign}
\end{document}